\documentstyle[12pt,axodraw,prd,aps]{revtex}
\input{psfig.sty}
\input{axodraw.sty}

\renewcommand{\r}[1]{$(\!\!~\ref{#1})$}                   
\newcommand{\BEA}{\begin{eqnarray}}
\newcommand{\EEA}{\end{eqnarray}}
\newcommand{\BA}{\begin{array}}
\newcommand{\EA}{\end{array}}
\newcommand{\BE}{\begin{equation}}
\newcommand{\EE}{\end{equation}}
\newcommand{\BT}{\begin{tabular}}
\newcommand{\ET}{\end{tabular}}
\newcommand{\BTA}{\begin{tabbing}}
\newcommand{\ETA}{\end{tabbing}}
\newcommand{\BC}{\begin{center}}
\newcommand{\EC}{\end{center}}

\newcommand{\SL}{\\[1mm]}
\newcommand{\NN}{\nonumber\\[4mm]}
\newcommand{\nn}{\nonumber\\}
\newcommand{\LP}{\left}
\newcommand{\RP}{\right}

\newcommand{\nmb}{\nn & & \mbox{}}

\newcommand{\ABS}[1]{\left| #1 \right|}

\newcommand{\PR}{\mbox{{\rm I}\hspace{-0.25 em}{\bf P}}}  

\newcommand{\PD}[2]{\frac{\partial #1}{\partial #2}}      
         
\newcommand{\HA}{\frac{1}{2}}

\def\f{\frac}                                             
\def\P{\partial}

\def\L{{\cal L}}                                          
\def\H{{\cal H}}

\newcommand{\LS}[2]{#1_{#2}}

\newcommand{\TSTB}{\t\s^{m}\bar{\t}}

\newcommand{\tb}{\bar{\t}}

\newcommand{\SWA}{\sin\theta_{\mbox{w}}}
\newcommand{\CWA}{\cos\theta_{\mbox{w}}}
\newcommand{\SWAS}{\sin^{2}\theta_{\mbox{w}}}
\newcommand{\CWAS}{\cos^{2}\theta_{\mbox{w}}}

\def\a{\alpha}
\def\b{\beta}
\def\g{\gamma}

\def\d{\delta}
\def\e{\varepsilon}
\def\t{\theta}

\def\la{\lambda}
\def\La{\Lambda}
\def\s{\sigma}

\newcommand{\be} {\begin{equation}}
\newcommand{\ee} {\end{equation}}
\newcommand{\bea} {\begin{eqnarray}}
\newcommand{\eea} {\end{eqnarray}}

\def\Journal#1#2#3#4{{#1} {\bf #2}, #3 (#4)}
\def\Book#1#2#3{{\bf #1}, {#2} (#3)}
\def\NPB{{\em Nucl. Phys.} B}
\def\PR{{\em Phys. Rev.}}

\begin{document}
\draft
 
\title{Revis\~ao da Constru\c c\~ao de Modelos Supersim\'etricos.}
\author{\bf M. C. Rodriguez}
\address{
Instituto de F\'\i sica Te\'orica\\
Universidade Estadual Paulista\\
Rua Pamplona, 145\\ 
01405-900-- S\~ao Paulo, SP\\
Brazil}
\date{\today}

\maketitle
\begin{abstract}
Foi com base neste estudo que fizemos a constru\c c\~ao da vers\~ao 
supersim\'etrica dos modelos de simetria 
$SU(3)_{C} \otimes SU(3)_{L} \otimes U(1)_{N}$ \cite{susy331}, apresentado 
no final da minha tese de doutorado \cite{mcr1}. Bem como dos estudos 
fenomenol\'ogicos subsequente \cite{mcr}.
\end{abstract}


\section{Introdu\c c\~ao}

As simetrias conhecidas da matriz S em f\'\i sica de part\'\i culas 
s\~ao
\begin{itemize}

\item Invari\^ancia de Poincar\'e, com geradores $P_m$, $M_{mn}$.

\item Simetrias Globais ``internas", relacionadas \`a n\'umeros 
qu\^anticos conservados. Os geradores de tais simetrias s\~ao 
escalares de Lorentz e geram uma \'algebra de Lie que \'e escrita como:
\begin{eqnarray}
\left[ B_{\ell}, B_{k} \right] = i C_{\ell k}^j B_j \quad , \nonumber
\end{eqnarray}
onde os $C_{\ell k}^j$ s\~ao as constantes de estrutura.

\item Simetrias Discretas: C, P e T.
\end{itemize}

Em 1967, Coleman e Mandula \cite{coleman}, mostraram que, sob certas 
suposi\c c\~oes, as simetrias descritas acima s\~ao as 
\'unicas simetrtias poss\'\i veis da matriz S.

Para demonstrar este teorema temos que supor que a \'algebra 
de simetria envolva apenas comutadores. Se supusermos que 
anticomutadores tamb\'em s\~ao geradores de simetria este teorema 
j\'a n\~ao \'e mais v\'alido. Na verdade em  1975 Haag, 
Lopusza\'nski e Sohnius \cite{hls} comprovaram que supersimetria 
\'e a \'unica simetria adicional permitida a matriz S 
sob estas suposi\c c\~oes de introduzir anticomutadores.

Supersimetria foi descoberta independentemente por 
tr\^es grupos de autores \`a saber
\begin{center} 
Yu. Gol'fand \& E. Lichtman      (1971)\\
D. Volkov \& V. Akulov           (1972) \\
J. Wess \& B. Zumino             (1974) 
\end{center}

A motiva\c c\~ao de Gol'fand e Lichtman \cite{1} era o de 
introduzir viola\c c\~ao de paridade em teoria 
qu\^antica de campos. O ponto de partida do paper 
de Volkov e Akulov \cite{2,3} era a quest\~ao de se 
part\'\i culas de Goldstone de spin um meio 
poderiam existir. Wess e Zumino \cite{4} fizeram a 
generaliza\c c\~ao do supergrupo que primeiro 
apareceu no modelo dual de Neveu--Schwarz--Ramond.

Podemos dizer que supersimetria \'e uma simetria 
entre b\'osons e f\'ermions \cite{5} ou mais precisamente \'e 
uma simetria entre estados de spin diferentes. Por 
exemplo uma part\'\i cula de spin $0$ \'e 
transformada em uma part\'\i cula de spin $1/2$ 
sobre uma transforma\c c\~ao de supersimetria.

Devido ao exposto acima conclu\'\i mos que o 
operador $Q$ que gera tais transforma\c c\~oes 
deve ser um espinor, assim $\bar{Q}$ tamb\'em 
\'e um gerador de supersimetria,  e ele 
produz as seguintes trnsforma\c c\~oes
\bea
Q | \mbox{b\'oson} \rangle = | \mbox{f\'ermion} \rangle 
\,\ , \nonumber 
\eea
\bea
Q |\mbox{f\'ermion} \rangle = | \mbox{b\'oson} \rangle 
\,\ . \nonumber
\eea
Assim uma unifica\c c\~ao dos campos de mat\'eria (f\'ermions) com os 
campos de for\c ca (b\'osons) aparece naturalmente.

Nos \'ultimos anos os estudos fenomenol\'ogicos envolvendo Supersimetria 
cresceu bastante e as raz\~oes para isto s\~ao muitas. Al\'em do atrativo de
unificar b\'osons e f\'ermions, outro fato \'e que Supersimetria global fornece 
uma teoria da gravidade menos divergente do que a usual gravita\c c\~ao 
qu\^antica. Outro fato que contribui para aumentar o interesse em Superimetria 
\'e que ela proporciona uma solu\c c\~ao ao problema da hierarquia.

Devido a este enorme interesse aqui pretendo estudar como construir teorias 
Supersim\'etricas. A lagrangiana \'e constru\'\i da passo a passo e com 
bastante detalhe tanto no formalismo de supercampos como em termos de 
componentes.

\section{Nota\c c\~ao}

Aqui neste estudo iremos utilizar a nota\c c\~ao do Livro do Wess-Barger
\cite{wb}, um outro bom livro sobre o assunto 
\'e o Srivastava \cite{sr}  que usam o seguinte tensor
m\'etrica 
\begin{eqnarray}
\eta_{mn} = \left(
\begin{array}{cccc}
-1& 0& 0& 0 \\
0& 1& 0& 0 \\
0& 0& 1& 0 \\
0& 0& 0& 1 \\
\end{array}
\right) \,\ .
\nonumber
\end{eqnarray}
Mas antes de entrarmos em supersimetria, vou rever rapidamente a \'algebra 
espinorial, pois isto ser\'a muito \'util nas pr\'oximas se\c c\~oes j\'a que 
estaremos basicamente trabalhando com espinores de Weyl de duas componentes.

\subsection{\'Algebra Espinorial.}

Na \'algebra espinorial levantamos e abaixamos os \'\i ndices 
espinoriais com a seguinte m\'etrica
\bea
&\epsilon_{ \alpha \beta} &= 
\epsilon_{ \dot{\alpha} \dot{\beta}}= 
\left(
\begin{array}{cc}
0& -1\\
1& 0
\end{array}
\right) \,\ , \nonumber \\ 
&\epsilon^{ \alpha \beta}& = 
\epsilon^{ \dot{\alpha} \dot{\beta}}= 
\left(
\begin{array}{cc}
0& 1\\
-1& 0
\end{array}
\right)
= i \sigma^{2} \,\ .
\label{eq:levidef}
\eea

O levantamento de \'\i ndice sem ponto e o abaixamento de 
\'\i ndice ponto \'e feito da seguinte maneira
\bea
\psi^{ \alpha} &= \epsilon^{ \alpha \beta} \psi _{\beta} \,\ ; \nonumber \\ 
\bar{ \psi}_{ \dot{ \alpha}} &= \epsilon_{ \dot{ \alpha} \dot{ \beta}}
\bar{ \psi}^{ \dot{\beta}} \,\ . 
\nonumber
\eea
Onde fazemos as seguintes identifica\c c\~oes
\be
\bar{\psi}_{ \dot{ \alpha}} \equiv 
(\psi_{ \alpha} )^{\dagger};
\,\ \psi_{ \alpha} \equiv 
(\bar{\psi}_{ \dot{ \alpha}} )^{\dagger} \qquad .
\nonumber
\ee
O ponto sobre o \'\i ndice espinorial representa conjuga\c c\~ao complexa. O 
mesmo papel \'e exercido pela barra sobre o s\'\i mbolo, embora aqui esta 
nota\c c\~ao seja redundante.

Com isto podemos definir a nossa conven\c c\~ao de soma da seguinte maneira
\bea
& \psi \chi & \equiv \psi^{ \alpha} \chi_{ \alpha} = -\psi_{ \alpha}
\chi^{\alpha} \,\ , \nonumber \\  
& \bar{ \psi} \bar{ \chi} & \equiv \bar{ \psi}_{ \dot{ \alpha}} 
\bar{ \chi}^{ \dot{\alpha}} =  -\bar{ \psi}^{ \dot{ \alpha}} 
\bar{ \chi}_{ \dot{\alpha}} \,\ . 
 \nonumber 
\eea

O operador $\epsilon$ ainda satisfaz
\bea
\epsilon_{ \alpha \beta} \epsilon^{\beta \gamma} &= &
\delta_{\alpha}^{\gamma} \,\ ,\nonumber\\
\epsilon _{\dot{ \alpha} \dot{ \beta}} \epsilon ^{\dot{ \beta} \dot{ \gamma}}
&= &\delta _{\dot{ \alpha}}^{\dot{ \gamma}} \,\ .
\nonumber
\eea

As matrizes de Pauli s\~ao
\be
\sigma^{0} = \left(
\begin{array}{cc}
-1& 0\\ 
0& -1
\end{array}
\right), \,\
\sigma^{1} = \left(
\begin{array}{cc}
0& 1\\ 
1& 0
\end{array}
\right), \,\
\sigma^{2} = \left(
\begin{array}{cc}
0& -i\\ 
i& 0
\end{array}
\right), \,\
\sigma^{3} = \left(
\begin{array}{cc}
1& 0 \\ 
0& -1
\end{array}
\right).
\label{pauli}
\ee
J\'a a matriz de Dirac escritas em termos das matrizes de 
Pauli s\~ao dadas pela seguinte equa\c c\~ao:
\be
\gamma^{m} = \left(
\begin{array}{cc}
0& \sigma^{m} \\ 
\bar{ \sigma}^{m}& 0
\end{array}
\right) \,\ ,
\label{gammam}
\ee
j\'a a $\gamma_{5}$ \'e
\be
\gamma^{5} \equiv i \gamma^0 \gamma^1 \gamma^2 \gamma^3 
= \left(
\begin{array}{cc}
1& 0 \\ 
0& -1
\end{array}
\right) \,\ ,
\label{gamma5}
\ee
em todo este estudo estaremos trabalhando nesta representa\c c\~ao quiral.

As matrizes de Pauli $\sigma ^{m}$ e $\bar{ \sigma}^{m}$ s\~ao 
relacionadas pela seguinte equa\c c\~ao
\be
\bar{ \sigma}^{m \dot{ \alpha} \beta} = 
\epsilon ^{ \dot{ \alpha} \dot{ \gamma}} 
\epsilon^{ \beta \delta} 
\sigma^{m}_{ \delta \dot{ \gamma}} \,\ ,
\label{eq:raisecovsigma}
\ee
de onde podemos mostrar que
\bea
\bar{ \sigma}^{m} &= (\sigma^{0},-\vec{ \sigma}) & \,\ .
\nonumber
\eea 

Com base na Eq.(\ref{eq:raisecovsigma}) podemos mostrar 
que estas matrizes satisfazem as seguintes rela\c c\~oes 
de completeza
\bea
tr \sigma^{m} \bar{ \sigma}^{n} &= &-2 \eta^{mn} \,\ ,\nonumber\\
\sigma^{m}_{ \alpha \dot{ \beta}} \bar{ \sigma}_{m}^{ \dot{ \gamma} \delta}
&=& -2\delta _{\alpha}^{\delta}\delta_{ \dot{ \beta}}^{ \dot{ \gamma}}
\,\ .
\label{eq:completeness}
\eea

Ainda podemos definir
\bea
\left( \sigma^{m} \bar{ \sigma}^{n} +
\sigma^{n} \bar{ \sigma}^{m} \right)^{ \beta}_{ \alpha} &=
&-2 \eta^{nm} \delta_{ \alpha}^{ \beta} \,\ ,\nonumber\\
\left( \bar{ \sigma}^{m} \sigma^{n} +
\bar{ \sigma}^{n} \sigma^{m} \right)^{ \dot{ \beta}}_{ \dot{ \alpha}} &=
&-2 \eta^{mn} \delta _{ \dot{ \alpha}}^{ \dot{ \beta}} \,\ , \nonumber \\
{ \sigma^{nm}_{ \alpha}}^{ \beta} &= & 
\frac{1}{4} \left[
\sigma^{n}_{ \alpha \dot{ \gamma}} \bar{ \sigma}^{m \dot{ \gamma} \beta}
- \sigma^{m}_{ \alpha \dot{ \gamma}} 
\bar{ \sigma}^{n \dot{ \gamma} \beta}\right] \,\ , \nonumber\\
{\hbox{$\bar{ \sigma}^{nm \dot{ \alpha}}$}}_{ \dot{ \beta}} &= & 
\frac{1}{4} \left[
\bar{ \sigma}^{n \dot{ \alpha} \gamma} \sigma^{m}_{ \gamma \dot{ \beta}}
- \bar{ \sigma}^{m \dot{ \alpha} \gamma} 
\sigma^{n}_{ \gamma \dot{ \beta}} \right] \,\ .
\label{eq:covsigmarel}
\eea

Algumas identidades \'uteis s\~ao
\bea
\psi^{ \alpha} \psi^{ \beta} &= &
- \frac{1}{2} \epsilon^{ \alpha \beta} \psi \psi \,\ ,\nonumber\\
\psi_{ \alpha} \psi_{ \beta} &= & 
\frac{1}{2} \epsilon_{ \alpha \beta} \psi \psi \,\ ,\nonumber \\
\bar{ \psi}^{ \dot{ \alpha}} \bar{ \psi}^{ \dot{ \beta}} &= & 
\frac{1}{2}\epsilon^{ \dot{ \alpha} \dot{ \beta}} 
\bar{ \psi} \bar{ \psi} \,\ ,\nonumber\\
\bar{ \psi}_{ \dot{ \alpha}} \bar{ \psi}_{ \dot{ \beta}} &= & 
-\frac{1}{2}\epsilon_{ \dot{ \alpha} \dot{ \beta}} 
\bar{ \psi} \bar{ \psi}
\,\ , \nonumber \\
( \theta \phi)( \theta \psi) &= & 
- \frac{1}{2} ( \theta \theta)( \phi \psi) \,\ , \nonumber \\
& & \nonumber\\
( \bar{ \theta} \bar{ \phi})( \bar{ \theta} \bar{ \psi}) &= & 
- \frac{1}{2}( \bar{ \phi} \bar{ \psi})( \bar{ \theta} \bar{ \theta}) \,\ , \nonumber \\
& & \nonumber\\
\phi \sigma^{m}\bar{ \chi} &= & 
-\bar{ \chi} \bar{ \sigma}^{m} \phi \,\ , \nonumber \\
& & \nonumber\\
( \theta \sigma^{m} \bar{ \theta})( \theta \sigma^{n} \bar{ \theta})
&= & -\frac{1}{2} \eta^{mn} ( \theta \theta)( \bar{ \theta} \bar{ \theta}) \,\ .
\label{eq:fierztable}
\eea

At\'e aqui s\'o trabalhamos com espinores de duas componentes, mas as regras 
de Feynman s\~ao escritas em termos de espinores de quatro componentes. Vamos 
agora estudar como passar de espinores de duas para espinores de quatro
componentes.

\subsection{Nota\c c\~ao de Quatro Componentes}
\label{SECT: Four comp not}

Para fazermos isto uma boa ferramenta s\~ao os operadores proje\c c\~ao.

\subsubsection{Os Operadores de Proje\c c\~ao.}

Come\c caremos definido os operadores de proje\c c\~ao da seguinte maneira
\BEA
   L &=& \HA \LP(1-\g_5\RP) \,\ ,
      \label{left proje}\SL
   R &=& \HA \LP(1+\g_5\RP) \,\ .
      \label{right proje}
\EEA

Estes operadores satisfazem as seguintes propiedades
\BEA
  L + R &=& 1 \,\ , \nonumber \\ 
  L L &=& L \,\ , \nonumber \\ 
  R R &=& R \,\ , \nonumber \\ 
  L R &=& R L \;\; = \;\; 0 \,\ , \nonumber \\ 
  L \g^m &=& \g^m R \,\ .
     \label{PL PR equiv zero prop 1}
\EEA

Vamos ver como escrever espinores de quatro componentes em termos de espinores 
de duas componentes. Para isto vamos ver como definir espinores de quatro 
componentes, em termos de espinores de duas componentes.

\subsubsection{Rela\c c\~oes entre espinores de duas e de quatro componentes.}
    \label{Subsect: Connection Between Two- and Four-Component Spinors.}

Come\c caremos por introduzir os seguintes espinores de Weyl de duas componentes 
$\LS{\xi}{\a}$ e $\bar{\eta}^{\dot{\a}}$
\BEA
   \LS{\xi}{\a} &\in & F \,\ , \nn
   \bar{\eta}^{\dot{\a}} &\in & \dot{F}^{*} \,\ , \nonumber
\EEA
onde F e $\dot{F}^{*}$ s\~ao espa\c cos vetoriais e correspondem \`a 
representa\c c\~oes 
inequivalentes do grupo SL(2,C). Vamos construir o espa\c co soma direta 
desses dois espa\c cos da seguinte maneira
\BEA
     D &=& F \oplus \dot{F}^{*} \,\ . \nonumber
\EEA
O espa\c co D \'e uma representa\c c\~ao de dimens\~ao quatro de SL(2,C).
Os elementos de D, s\~ao exatamente os bem conhecidos espinores de Dirac 
de quatro componentes.

Desta maneira um espinor de Dirac que representaremos por , $\Psi$, podem ser 
constru\'\i dos apartir dos espinores de Weyl de duas componentes da seguinte 
maneira
\BEA
   \Psi &=& \LP( \BA{c} \LS{\xi}{\a} \\ \bar{\eta}^{\dot{\a}}  \EA \RP) \,\ .
       \label{Dirac Spinor}
\EEA
Este \'e um espinor de Dirac na representa\c c\~ao quiral.

Um espinor de Majorana que representaremos por , $\la$, tamb\'em \'e um 
espinor de Dirac de quatro componentes mas possui a seguinte condi\c c\~ao 
adicional
\BEA
     \la &=&  \la^{c} \;\; = \;\; C \bar{\la}^{T} \,\ .
     \nonumber
\EEA
Onde C \'e a matriz usual conjuga\c c\~ao de carga. Na representa\c c\~ao 
quiral a matriz conjuga\c c\~ao de carga \'e definida por
\BEA
   C &=& \LP( \BA{cc} 
   -i \sigma^{2} & 0 \\ 0 &i \sigma^{2}  \EA \RP) \,\ 
   = \LP( \BA{cc}
   \epsilon_{\alpha \beta} & 0 \\
   0 & \epsilon^{\dot{\alpha} \dot{\beta}} \EA \RP) \,\ ,
       \label{conj carg}
\EEA
na segunda igualdade foi usada a Eq.(\ref{eq:levidef}).

Enquanto $\bar{\la}$ 
significa para espinores de quatro componentes, o usual espinor adjunto 
$\bar{\la}= \la^{\dagger} \g_{0}$.
Podemos mostrar usando Eq.(\ref{conj carg}) que o espinor de Majorana pode ser 
escrito da seguinte maneira
\BEA
   \Psi^{c} &=& C \bar{\Psi}^{T} \;\; = \;\;
    \LP( \BA{c} \LS{\eta}{\a} \\ \bar{\xi}^{\dot{\a}}  \EA \RP) \,\ ,
       \label{Charge conjugation}
\EEA
ou seja a conjuga\c c\~ao de carga(na representa\c c\~ao quiral) troca 
$\xi$ por $\eta$ e vice-versa. Portanto, podemos facilmente concluir 
que para um espinor de Majorana, $\la$ pode ser escrito como
\BEA
   \la &=& \LP( \BA{c} \LS{\xi}{\a} \\ \bar{\xi}^{\dot{\a}}  \EA \RP) \,\ .
      \label{Majorana Spinor}
\EEA

Usando as Eqs.(\ref{gamma5}), (\ref{left proje}) e (\ref{right proje}) 
obteremos as seguintes express\~oes para os 
operadores proje\c c\~oes
\BEA
      L &=& \LP( \BA{cc}   0 & 0 \\ 0 & 1 \EA \RP) \,\ ,\nn
      R &=& \LP( \BA{cc}   1 & 0 \\ 0 & 0 \EA \RP) \,\ ,\label{svtl1}
\EEA
com isto as componentes quirais de um espinor $\Psi$ s\~ao
\BEA
      \Psi_{R} &=& R \Psi \;\;=\;\;
           \LP( \BA{c} \LS{\xi}{\a} \\ 0  \EA \RP) \,\ ,
          \nn
      \Psi_{L} &=& L \Psi \;\;=\;\;
          \LP( \BA{c} 0 \\ \bar{\eta}^{\dot{\a}}  \EA \RP) \,\ .
          \label{svtl2}
\EEA
O espinor adjunto de Dirac de $\Psi$ usando a Eq.(\ref{gammam}), \'e 
expresso em termos dos espinores de duas componentes da seguinte 
maneira
\BEA
   \bar{\Psi} &=& \Psi^{\dagger} \g_{0}
      \;\; = - \;\;  \LP( \BA{cc}  \eta^{\a} & \bar{\xi}_{\dot{\a}} \EA \RP) 
      \,\ . \label{svtl3}
\EEA

\subsubsection{Rela\c c\~oes \'uteis entre espinores de duas e quatro 
componentes.}

Vamos mostrar algumas rela\c c\~oes \'uteis, que permitem fazer a passagem entre 
espinores de duas componentes para espinores de quatro 
componente, e vice-versa. Considere 
os espinores, $\Psi_{1}(x)$ e $\Psi_{2}(x)$, definidos na
Eq.(\ref{Dirac Spinor}). Usando as Eqs.(\ref{gammam}), (\ref{gamma5}), 
(\ref{eq:fierztable}), (\ref{svtl1}), (\ref{svtl2}) e (\ref{svtl3}), podemos facilmente mostrar as 
seguintes igualdades
\BEA
   \bar{\Psi}_{1}\Psi_{2} &=&- \eta_{1}\xi_{2} - \bar{\xi}_{1}\bar{\eta}_{2} \,\ ,
             \nonumber \SL
   \bar{\Psi}_{1}\g^{m}\Psi_{2}
           &=& - \bar{\xi}_{1}\bar{\s}^{m}\xi_{2}
              + \bar{\eta}_{2}\bar{\s}^{m}\eta_{1} \,\ ,
              \nonumber \SL
   \bar{\Psi}_{1}\g_{5}\Psi_{2}
           &=& - \eta_{1}\xi_{2}+ \bar{\xi}_{1}\bar{\eta}_{2} \,\ ,
              \nonumber \SL
   \bar{\Psi}_{1}\g^{m}\g_{5}\Psi_{2}
           &=& - \bar{\xi}_{1}\bar{\s}^{m}\xi_{2}
          + \eta_{1}\s^{m}\bar{\eta}_{2} \,\ ,
              \nonumber \SL
   \bar{\Psi}_{1}\g^{m}\P_{m}\Psi_{2}
           &=&  -\eta_{1}\s^{m}\P_{m}\bar{\eta}_{2}
          - \bar{\xi}_{1}\bar{\s}^{m}\P_{m}\xi_{2} \nn
           &=&  -\bar{\eta}_{2}\bar{\s}^{m}\P_{m}\eta_{1}
          - \bar{\xi}_{1}\bar{\s}^{m}\P_{m}\xi_{2}
              +\P_{m}\LP(\bar{\eta}_{2}\bar{\s}^{m}\eta_{1} \RP) \,\ ,
              \nonumber \SL
    \bar{\Psi}_1 L \Psi_2 &=& -\bar{\xi}_1\bar{\eta}_2 \,\ ,
       \nonumber  \SL
    \bar{\Psi}_1 R \Psi_2 &=& -\eta_{1}\xi_{2} \,\ ,
       \nonumber \SL
    \bar{\Psi}_1\g^{m} L \Psi_2
                             &=&-\eta_1{\s}^{m}\bar{\eta}_2 \nn
        &=& \bar{\eta}_2\bar{\s}^{m}\eta_1  \,\ ,
        \nonumber \SL
    \bar{\Psi}_1\g^{m} R \Psi_2 &=& -\bar{\xi}_1\bar{\s}^{m}\xi_2 \,\ ,
       \nonumber \SL
       \bar{\Psi}_1\g^{m} L\P_{m} \Psi_2
        &=& -\eta_1\s^{m}\P_{m}\bar{\eta}_2 \nn
        &=& -\bar{\eta}_2\bar{\s}^{m}\P_{m}\eta_1
             +\P_{m}\LP(\bar{\eta}_2\bar{\s}^{m}\eta_1\RP) \,\ ,
        \nonumber \SL
    \bar{\Psi}_1\g^{m} R\P_{m} \Psi_2
        &=& -\bar{\xi}_1\bar{\s}^{m}\P_{m}\xi_2 \,\ .
        \label{svtl4}
\EEA
Com estas igualdades podemos converter todas as lagrangianas escritas na forma de 
duas componentes na forma de quatro componentes satisfazendo as nossas conven\c c\~oes.

Com isto finalizamos nossa revis\~ao da \'algebra espinorial. Nosso pr\'oximo 
passo ser\'a definir e ver algumas consequ\^encias da \'algebra supersim\'etrica. 

\section{\'Algebra Supersim\'etrica.}

A \'algebra de supersimetria, considerando apenas um 
gerador de supersimetria em quatro dimens\~oes \'e 
\footnote{Para o caso mais geral veja o livro do Wess-Barger ou Srivastava.}
\bea
& \{ Q_{ \alpha}, \bar{Q}_{ \dot{\beta}} \}  &=
2\sigma ^{m}_{\alpha \dot{\beta}}P_{m} \,\ , \nonumber \\
& \{ Q_{\alpha},Q_{\beta} \}  &=
\{ \bar{Q}_{ \dot{\alpha}}, \bar{Q}_{\dot{\beta}} \}  = 0 
\,\ , \nonumber \\
& [Q_{\alpha},P_m]  &= [\bar{Q}_{ \dot{\alpha}},P_m] = 0 
\,\ , \nonumber \\
& [Q_{\alpha},M^{mn}]  &= \frac{1}{2}
{ \hbox{$ \sigma^{mn} _{ \alpha}$}}^{ \beta}Q_{\beta} 
\,\ , \nonumber \\
& [\bar{Q}^{ \dot{\alpha}}, M^{mn}]  &= \frac{1}{2}
{\hbox{$ \bar{ \sigma}^{mn \dot{\alpha}}$}}_{ \dot{\beta}}
\bar{Q}^{ \dot{\beta}} \,\ , \nonumber \\
& [P_m, P_n]  &=  0 \,\ , \nonumber \\
& [P_m,M^{nq}]  &= \delta_{m}^{n}P^p - \delta_{m}^{p}P^n 
\,\ , \nonumber \\
& [M^{mn},M^{pq}] &= \eta ^{mp}M^{nq} - \eta ^{mq}M^{np}  
\,\ . \label{eq:nonesusyalg}
\eea

\subsubsection{Consequ\^encias desta \'algebra.}

Algumas consequ\^encias desta \'algebra s\~ao
\begin{itemize}
\item[{\bf 1}] Cada supermultipleto 
\footnote{supermultipleto cont\'em estados bos\^onicos 
e fermi\^onicos, conforme definiremos mais adiante.} 
cont\'em o mesmo n\'umero de grau  de f\'ermions e b\'osons.
\item[{\bf 2}] As massas de todos os estados em um supermultipleto s\~ao 
degenerados, e as massas dos b\'osons e f\'ermions s\~ao iguais.
\item[{\bf 3}]Em uma teoria supersim\'etrica qualquer estado tem energia 
positiva definida.

\end{itemize}

Para mostrarmos o primeiro ponto, considere o operador $(-1)^{2s}$ 
onde $s$ \'e o momento angular de spin. Pelo teorema de 
spin-estat\'\i stica, 
este operador tem auto-valor $+1$ atuando em um estado bos\^onico e 
auto-valor $-1$ atuando em um estado fermi\^onico. Portanto este operador 
\'e exatamente proporcional ao n\'umero de b\'osons $n_B$ menos o n\'umero 
de f\'ermions $n_F$, ou seja
\bea
Tr[(-1)^{2s} P_{m} \sigma^{m}_{ \a \dot{ \b}}] 
\propto n_{B}-n_{F} \,\ .
\label{beleza}
\eea

Por outro lado o operador $(-1)^{2s}$ 
deve anticomutar com qualquer operador fermi\^onico, e
em particular com $Q$ e $\bar{Q}$. Podemos tomar o tra\c co deste operador e 
o resultado \'e o seguinte
\bea
Tr[(-1)^{2s} P_{m} \sigma^{m}_{ \a \dot{ \b}}]
&\propto&
Tr[(-1)^{2s} \left( Q_{ \a} \bar{Q}_{ \dot{ \b}}
+ \bar{Q}_{ \dot{ \b}} Q_{ \a} \right) ]
\nonumber\\
&=&
Tr[(-1)^{2s} Q_{ \a} \bar{Q}_{ \dot{ \b}}
- \bar{Q}_{ \dot{ \b}} (-1)^{2s} Q_{ \a}]
\nonumber\\
&=&
Tr[(-1)^{2s} Q_{ \a} \bar{Q}_{ \dot{ \b}}]-
Tr[ \bar{Q}_{ \dot{ \b}} (-1)^{2s}  Q_{ \a}]
\nonumber\\
&=&Tr[(-1)^{2s} Q_{ \a} \bar{Q}_{ \dot{ \b}}] -
Tr[(-1)^{2s} Q_{ \a} \bar{Q}_{ \dot{ \b}}]
\nonumber \\
&=& 0 \,\ .
\label{chuchu}
\eea

Comparando Eq.(\ref{beleza}) com Eq.(\ref{chuchu}) conclu\'\i mos facilmente 
que
\bea
n_{B}=n_{F} \,\ .
\eea

O segundo ponto resulta do fato de $P^{2}$ ser um operador de Casimir 
da teoria, pois $[P,Q]=0$.

J\'a o terceiro ponto vem da seguinte rela\c c\~ao de anti-comuta\c c\~ao
$ \{ Q_{ \alpha}, \bar{Q}_{ \dot{ \a}} \}  = 
2\sigma ^{m}_{\alpha \dot{ \a}}P_{m}$ da \'algebra supersim\'etrica. 
Usando a primeira equa\c c\~ao da Eq.(\ref{eq:completeness}) 
resulta que 
\bea
\bar{ \s}^{n \a \dot{ \a}} \{ Q_{ \alpha}, \bar{Q}_{ \dot{ \a}} \}=
-4P^{n} \,\ .
\eea
Assim a Hamiltoniana de uma teoria supersim\'etrica \'e escrita da seguinte 
maneira($\bar{\sigma}^0=\sigma^0=-I$)
\bea
H= P_0 = \frac{1}{4} \left( Q_1 \bar{Q}_1 + \bar{Q}_1 Q_1 + 
Q_2 \bar{Q}_2 + \bar{Q}_2 Q_2 \right) \,\ .
\label{energia}
\eea
Isto implica que $H$ \'e um operador positivo e definido no 
espa\c co de Hilbert
\bea
\langle \psi |H|\psi \rangle \; \geq 0 \; 
\quad \forall \psi\ \,\ . \nonumber 
\eea

Se o v\'acuo $|0 \rangle$ \'e supersim\'etrico, ent\~ao 
$Q_{ \alpha}|0 \rangle=\bar{Q}_{ \dot{ \alpha}}|0 \rangle = 0$ e 
Eq.(\ref{energia}) implica que $E_{vac} \equiv \langle 0|H|0 \rangle = 0$. 
Por outro lado se o v\'acuo n\~ao \'e supersim\'etrico, ou seja existe 
no m\'\i nimo um dos geradores de supersimetria que n\~ao aniquila o v\'acuo, 
ent\~ao Eq.(\ref{energia}) implica $E_{vac}>0$.

Os geradores de supersimetria $Q$ e $\bar{Q}$ tamb\'em 
comutam com todos os outros geradores de transforma\c c\~ao de gauge. 
Portanto part\'\i culas no mesmo supermultipleto devem 
estar tamb\'em na mesma representa\c c\~ao do grupo de 
gauge, e assim devem ter a mesma carga el\'etrica e todos 
os outros n\'umeros qu\^anticos tamb\'em ser\~ao iguais. Antes de vermos uma 
representa\c c\~ao destes operadores $Q$ e $\bar{Q}$, vamos introduzir o 
super-espa\c co onde todo o formalismo de supersimetria pode ser expresso 
de uma maneira econ\^omica, compacta e extremamente elegante.

\section{Super-Espa\c co}

Uma formula\c c\~ao mais elegante das transforma\c c\~oes 
supersim\'etricas \'e encontrada no Super-espa\c co. Super-espa\c co 
\'e o espa\c co Euclideano(Minkowski) normal completado pela adi\c c\~ao 
de duas novas coordenadas, que s\~ao grassmanianas, isto \'e 
anti-comutante, ou seja
\bea
\{ \t^{\a}, \t^{ \b} \} = \,\ \{ \t^{\a}, \bar{ \t}^{ \dot{ \b}} \} = \,\ 
\{ \bar{ \t}^{ \dot{ \a}}, \bar{ \t}^{ \dot{ \b}} \} =0 \,\ .
\label{superespaco}
\eea
As vari\'aveis $\t$ e $\bar{ \t}$ t\^em dimens\~oes de $E^{-1/2}$, isto 
ficar\'a claro quando estivermos construindo o supercampo quiral. Com a 
introdu\c c\~ao destas novas vari\'aveis espinoriais, n\'os necessitamos 
aumentar a dimens\~ao do espa\c co-tempo: temos que passar de 4 para 8. Um 
ponto no Super-Espa\c co se denota por $z^a=(x^a,\theta^{\alpha},
\bar{\theta}_{\dot{\alpha}})$.

Neste Super-espa\c co podemos representar uma transforma\c c\~ao 
supersim\'etrica como uma transforma\c c\~ao sobre pontos, de maneira 
an\'aloga ao que acontece com os operadores 
$P_{m}$ e $M_{mn}$ que geram transla\c c\~oes e rota\c c\~oes no espa\c co 
Euclideano(Minkowski).

Para acharmos uma representa\c c\~ao dos geradores de supersimetria e 
construir lagrangianas, n\'os temos que saber como calcular derivadas e 
integrais neste Super-espa\c co, e \'e isto o que faremos a seguir.

\subsection{Derivadas no Super-Espa\c co}

Devido as suas propiedades de anticomuta\c c\~ao, $\theta$ e $\bar{ \theta}$ 
n\~ao podem variar 
continuamente, logo elas t\^em de ser objetos discretos.

Devido a isto definir a diferencia\c c\~ao com rela\c c\~ao as 
vari\'aveis de Grassman da maneira usual, como a taxa de duas 
varia\c c\~oes infinitesimais, n\~ao faz o menor sentido.

Por\'em, podemos, formalmente definir diferencia\c c\~ao como:
\bea
\partial_{ \a} \t^{ \b}&=& \d^{ \a}_{ \b} \,\ , \nonumber \\
\bar{\partial}^{ \dot{ \a}} \bar{ \t}_{ \dot{ \b}}&=& 
\d^{ \dot{ \b}}_{ \dot{\a}} \,\ , \nonumber
\eea
onde
\bea
\partial_{ \a} & \equiv& \frac{ \partial}{ \partial \t^{ \a}} \,\ , \nonumber \\
\bar{\partial}^{ \dot{ \a}} & \equiv & 
\frac{ \partial}{ \partial \bar{ \t}_{ \dot{ \a}}} \,\ . \nonumber
\eea

Algumas propiedades destas derivadas s\~ao
\bea
\partial^{ \a} \theta^{ \beta} &=& -\epsilon^{ \a \beta} \,\ , \nonumber \\
\bar{\partial}^{ \dot{ \a}} \bar{ \theta}^{ \dot{ \beta}} &=& 
-\epsilon^{ \dot{ \a} \dot{ \beta}} \,\ , \nonumber \\ 
\partial_{ \a} ( \theta \theta) &=& 2 \theta_{\alpha} \,\ , \nonumber \\
\bar{\partial}^{ \dot{ \a}} ( \bar{ \theta} \bar{ \theta}) &=& 
-2 \bar{ \theta}_{ \dot{ \a}} \,\ , \nonumber\\
\partial^2 ( \theta \theta) &=& 4 \,\ , \nonumber \\
\bar{ \partial}^2 ( \bar{ \theta} \bar{ \theta}) &=& 4 \,\ ,\nonumber
\label{eq:supdrels}
\eea
com $\partial^{2}= \partial^{ \a} \partial_{ \a}$ e 
$\bar{ \partial}^2= \bar{ \partial}_{ \dot{ \a}} 
\bar{ \partial}^{ \dot{ \a}}$.

\subsection{Integral do Super-Espa\c co.}

A integral de Berezin \cite{berezin} para um \'unico par\^ametro de Grassman 
$\t$ \'e definida como
\bea
\int d \t \t&=& 1 \,\ , \nonumber \\
\int d \t c &=& 0 \,\ .
\label{intgr1}
\eea
Com $c$ sendo uma constante em rela\c c\~ao a vari\'avel $\theta$.

J\'a para uma fun\c c\~ao arbitr\'aria de um \'unico par\^ametro $\t$, tem a 
seguinte expans\~ao, exata, de Taylor
\be
f( \t)=a+b \t \,\ , 
\label{svtl5}
\ee
ent\~ao das Eqs.(\ref{intgr1}) e (\ref{svtl5}) podemos escrever
\be
\int d \t f( \t)=b \,\ ,
\ee
ou seja 
{\bf a integra\c c\~ao de Berezin \'e equivalente a deriva\c c\~ao}
\be
\frac{df(\t)}{d \t}=b= \int d \t f( \t) \,\ .
\ee

No caso do Super-espa\c co com apenas um gerador de supersimetria, de 
coordenadas $\t, \bar{ \t}$, usaremos as seguintes conven\c c\~oes
\bea
d^2 \theta &=& -\frac{1}{4} d \theta^{ \alpha}\,\ d \theta ^{\beta}\,
\epsilon_{ \alpha \beta} \quad ,\nonumber\\
d^2 \bar{ \theta} &=& -\frac{1}{4} d \bar{ \theta}_{ \dot{ \alpha}} \,\ 
d \bar{ \theta}_{ \dot{ \beta}} \,\
\epsilon^{ \dot{ \alpha} \dot{ \beta}} \quad ,\\
d^4 \theta &=& d^2 \theta \,\ d^2 \bar{ \theta} \quad .\nonumber
\label{eq:intsuperdefs}
\eea
Usando esta nota\c c\~ao, temos as seguintes identidades
\bea
\int d^2 \theta \,\ \theta \,\ \theta &=& 1\,\ ,\nonumber \\
\int d^2 \bar{ \theta} \,\ \bar{ \theta} \,\ \bar{ \theta} &=& 1
\,\ , \nonumber \\
\int d^4 \theta \,\ \theta \,\ \theta \bar{ \theta} \,\ \bar{ \theta} &=& 1
\,\ , \nonumber \\ 
\int d^2 \theta \,\ \theta^{ \alpha}&=&0 \,\ , \nonumber \\
\int d^2 \bar{ \theta} \,\ \bar{ \theta}_{ \dot{ \alpha}}&=&0 \,\ , \nonumber \\
\int d^2 \theta c= \int d^2 \bar{ \theta} c &=&0 \,\ .
\label{eq:intidents}
\eea

\section{Supercampos}

Supercamos proporcionam uma descri\c c\~ao elegante e compacta das 
representa\c c\~oes de supersimetria no Super-espa\c co. 
Definiremos uma transforma\c c\~ao em um supercampo da seguinte maneira
\bea
\delta_{ \xi} \Phi=( \xi Q+ \bar{ \xi} \bar{Q}) \Phi \,\ , 
\label{svtl6}
\eea
onde $Q$ e $\bar{Q}$ s\~ao os geradores de supersimetria.
Que podem ser escritos 
no Super-espa\c co da seguinte maneira
\bea
Q_{ \alpha} &=& \partial_{ \alpha} - i\sigma ^m_{ \alpha \dot{\beta}}
\bar{ \theta} ^{ \dot{ \beta}}\partial _m \quad ,\nonumber\\
\bar{Q}_{ \dot{ \alpha}} &=& \partial_{ \dot{ \alpha}} - i \theta ^{\beta} 
\sigma^m_{ \beta \dot{ \alpha}} \partial _m \quad .\nonumber
\eea
Para a\c c\~oes mais geral, temos que introduzir derivadas covariante do 
Super-espa\c co que chamaremos de $D_{ \alpha}$ e $\bar{D}_{ \dot{ \alpha}}$ 
e que satisfazem a seguinte \'algebra
\bea 
\{ D_{ \a}, Q_{ \b} \}= 
\{ D_{ \a}, \bar{Q}_{ \dot{ \b}} \}=
\{ \bar{D}_{ \dot{ \a}},Q_{ \b} \}=
\{ \bar{D}_{ \dot{ \a}}, \bar{Q}_{ \dot{ \b}} \}=0 \,\ .
\eea

Fazendo estes c\'alculo podemos mostrar
\bea
D_{ \alpha} &=& \partial_{ \alpha} + i \sigma^m_{ \alpha \dot{ \beta}}
\bar{ \theta}^{ \dot{ \beta}} \partial_m \,\ , \nonumber\\
\bar{D}_{ \dot{ \alpha}} &=&- \partial_{ \dot{ \alpha}} - i \theta ^{\beta} 
\sigma^m_{ \beta \dot{ \alpha}} \partial _m \quad .
\label{svetlana}
\eea

Agora nos resta apenas definir o supercampo: Um supercampo \'e simplesmente 
uma fun\c c\~ao de $z^a \equiv 
(x^a,\theta^{\alpha},\bar{\theta}_{\dot{\alpha}})$, que pode ser escrita da 
seguinte maneira geral:
\begin{eqnarray}
\Phi(z)&=&f_{00}(x)+\theta^{\alpha}f_{10 \alpha}(x)+
\bar{\theta}_{\dot{\alpha}} 
f^{\dot{\alpha}}_{01}(x)+\theta \theta f_{20}(x)+\bar{\theta}\bar{\theta} 
f_{02}(x)+\theta^{\alpha}f_{11 \alpha \dot{\alpha}}(x) \bar{\theta}^{\dot{\alpha}}  \nonumber \\ &+& 
\theta \theta \bar{\theta}_{\dot{\alpha}} 
f^{\dot{\alpha}}_{21}(x)+\bar{\theta} \bar{\theta}\theta^{\alpha}
f_{12 \alpha}(x)+\theta \theta \bar{\theta} \bar{\theta}f_{22}(x),
\label{supercampodefinition}
\end{eqnarray}
onde
\begin{equation}
\theta \theta= \theta^{\alpha} \theta_{\alpha}=\epsilon^{\alpha \beta} 
\theta_{\beta} \theta_{\alpha}=\epsilon^{12} \theta_2 \theta_1+ 
\epsilon^{21} \theta_1 \theta_2=-2 \theta_1 \theta_2,
\end{equation}
devido ao fato de que $\theta_{\alpha} (\bar{\theta}_{\dot{\alpha}})$ s\~ao 
vari\'aveis anticomutantes, os termos do tipo $\theta \theta 
\theta^{\alpha}$ e $\bar{\theta} \bar{\theta} \bar{ \theta}_{ \dot{ \alpha}}$ 
s\~ao nulos. Al\'em disto os $f_{00}, f_{20}, f_{02}$ e $f_{22}$ s\~ao campos 
escalares, enquanto que $f_{01}, f_{10}, f_{12}$ e $f_{21}$ s\~ao campos 
espinoriais; $f_{11}$ \'e um campo vetorial.

Tendo definido este supercampo geral, iremos agora discutir os dois 
Supercampos de interesse.

\subsection{Supercampo Quiral}

Um supercampo quiral \'e definido por
\be
\bar{D}_{ \dot{ \alpha}} \Phi = 0 \qquad .
\label{eq:chsupdef}
\ee
Acharemos a solu\c c\~ao mais geral para a Eq.(\ref{eq:chsupdef}).
Para isto definiremos uma nova coordenada bos\^onica $y^m$ definida 
no Super-espa\c co por
\be
y^m = x^m + i \theta \sigma^m \bar{ \theta} \,\ .
\label{eq:ysdef}
\ee
Usando a Eq.(\ref{svetlana}) podemos mostrar que
\bea
\bar{D}_{ \dot{ \alpha}} \,\ y^m &=& 0 \,\ ,\nonumber \\
\bar{D}_{ \dot{ \alpha}} \,\ \theta^{ \alpha} &=& 0 \,\ ,
\label{eq:obviousders}
\eea
logo qualquer fun\c c\~ao $\Phi(y, \theta)$ de $y^m$ e $\theta$
(mas n\~ao de $\bar{ \theta}$) satisfaz
\be
\bar{D}_{ \dot{ \alpha}} \,\Phi(y, \theta)= 0 \qquad .
\label{eq:itschiman}
\ee
Com isto se pode mostrar que sua expans\~ao nesta nova coordenada 
\'e a seguinte \footnote{Esta expans\~ao \'e exata, pois na Eq.
(\ref{superespaco}) termos com mais que tr\^es $\theta$ desaparecem.}
\be
\Phi(y, \theta) = A(y) + \sqrt{2} \theta \psi(y) +\theta \theta F(y) \,\ ,
\label{eq:mostgenchone}
\ee
onde $A(y)$, $F(y)$ s\~ao campos escalares complexos de spin $0$, enquanto
$\psi_{\alpha}(y)$ \'e um espinor de Weyl complexo de spin $1/2$. Os tr\^es 
termos do supercampo $\Phi$ t\^em dimens\~ao de $E$. Lembremos que a 
dimens\~ao de um campo escalar \'e de $E$, enquanto a de um espinor \'e de 
$E^{3/2}$, e o campo $F$, ver sua equa\c c\~ao de movimento, tem dimens\~ao 
de $E^{2}$, as derivadas possuem dimens\~ao de $E$.

Tamb\'em podemos escrever este supercampo em termos das coordenadas 
do Super-espa\c co e neste caso sua expans\~ao de Taylor \'e dada por
\bea
\Phi(x, \theta, \bar{ \theta}) &=& A(x) + \sqrt{2} \theta \psi (x) +
\theta \theta F(x) \nonumber\\
&+&i \theta \sigma^m \bar{ \theta} \partial_m A(x) -
\frac{i}{ \sqrt{2}}( \theta \theta ) \partial_m \psi (x) \sigma^m \bar{ \theta}
\nonumber \\  
&+&
\frac{1}{4} (\theta \theta)( \bar{ \theta} \bar{ \theta} ) \Box A(x) \quad .
\label{eq:fullchiexp}
\eea
Uma transforma\c c\~ao de supersimetria, dada pela Eq.(\ref{svtl6}), produz neste supercampo as seguintes 
varia\c c\~oes
\bea
\delta_{ \xi} A &=& \sqrt{2} \xi \psi  \,\ (\mbox{b\'oson 
$\rightarrow$ f\'ermion}) 
\,\ , \nonumber\\
\delta_{ \xi} \psi &=& \sqrt{2} \xi F +i \sqrt{2} \sigma^m \bar{ \xi}
\partial_m A \,\ (\mbox{f\'ermion $\rightarrow$ b\'oson}) \,\ , \nonumber \\
\delta_{ \xi} F &=&i \sqrt{2} \bar{ \xi} \bar{ \sigma}^{m} \partial_m \psi
\,\ (\mbox{F $\rightarrow$ derivada total}) \,\ .   
\eea

De maneira an\'aloga podemos definir um supercampo antiquiral por
\bea
D_{ \alpha} \bar{ \Phi} &=& 0 \,\ ,\nonumber\\
\bar{ \Phi} &=& \bar{ \Phi}(\bar{y},\bar{ \theta}) \,\ ;
\,\ \bar{y} = x^m - i \theta \sigma^m \bar{ \theta} \,\ .
\eea
De maneira an\'aloga ao que fizemos no caso do supercampo quiral, 
podemos escrever
\be
\bar{ \Phi}( \bar{y}, \theta) = \bar{A}( \bar{y}) + 
\sqrt{2} \bar{ \theta} \bar{ \psi}( \bar{y}) + 
\bar{ \theta} \bar{ \theta} \bar{F}( \bar{y}) \,\ , 
\ee
e 
\bea
\bar{ \Phi}(x, \theta, \bar{ \theta}) &=& \bar{A}(x) + 
\sqrt{2} \bar{ \theta} \bar{ \psi} (x) +
\bar{ \theta} \bar{ \theta} \bar{F}(x) \nonumber\\
&-&i \theta \sigma^m \bar{ \theta} \partial_m \bar{A}(x) +
\frac{i}{ \sqrt{2}}( \bar{ \theta} \bar{ \theta} ) \theta \sigma^m \partial_m 
\bar{ \psi} (x)
\nonumber \\  
&+&
\frac{1}{4} (\theta \theta)( \bar{ \theta} \bar{ \theta} ) \Box \bar{A}(x)
\quad .
\eea

Produtos de supercampos quirais $\Phi_1$,$\Phi_2 \ldots \Phi_n$ 
s\~ao novamente supercampos quirais, e igualmente para o seu conjugado.
Ent\~ao no caso de dois supercampos quirais podemos escrever
\bea
\Phi_i(y, \theta) \Phi_j(y, \theta) &=& A_i(y)A_j(y) + \sqrt{2} \theta 
\left[ \psi_i(y)A_j(y)+A_i(y) \psi_j(y) \right] \nonumber \\
&+& \theta \theta \left[ A_i(y)F_j(y)+F_i(y)A_j(y)- \psi_i(y) \psi_j(y) 
\right] \,\ .
\label{bi}
\eea
O \'ultimo termo em $\theta \theta$ na expans\~ao acima se parece com um 
termo de 
massa dos f\'ermions! J\'a no caso de tr\^es supercampos quirais teremos
\bea
\Phi_i(y, \theta) \Phi_j(y, \theta) \Phi_k(y, \theta) &=& A_i(y)A_j(y)A_k(y)
\nonumber \\
&+& 
\sqrt{2} \theta 
\left[ \psi_i(y)A_j(y)A_k(y)+ A_i(y) \psi_j(y)A_k(y) 
+A_i(y)A_j(y) \psi_k(y) \right] \nonumber \\
&+& \theta \theta \left[ F_i(y)A_j(y)A_k(y)+ A_i(y)F_j(y)A_k(y) + 
A_i(y)A_j(y)F_k(y) \right. \nonumber \\
&-& \left. \psi_i(y) \psi_j(y)A_k(y)- A_i(y) \psi_j(y) \psi_k(y) 
- \psi_i(y)A_j(y) \psi_k(y) \right] \,\ .
\nonumber \\
\label{tri}
\eea
Repare que os tr\^es \'ultimos termos na equa\c c\~ao acima descreve intera\c c\~oes 
de Yukawa entre um escalar e dois f\'ermions, no modelo padr\~ao tais intera\c c\~oes 
geram as massas dos quarks e dos l\'eptons.
As componentes $\theta \theta$ das Eqs.(\ref{bi}) e (\ref{tri}) s\~ao 
independentes da base em que s\~ao calculadas \cite{wb}.

Mas o produto de $\bar{ \Phi} \Phi$, por\'em, n\~ao \'e um supercampo quiral
\bea
\bar{ \Phi}_{i}(x, \theta, \bar{ \theta}) \Phi_{j}(x, \theta, \bar{ \theta})&=& 
\bar{A}_i(x)A_j(x)+ \sqrt{2} \theta \psi_j(x) \bar{A}(x)+ \sqrt{2} \bar{ \theta} 
\bar{ \psi}_i(x)A_j(x) \nonumber \\
&+& \theta \theta \bar{A}_i(x)F_j(x)+ \bar{ \theta} \bar{ \theta} 
\bar{F}_i(x)A_j(x) \nonumber \\
&+& \theta^{ \alpha} \bar{ \theta}^{ \dot{ \alpha}} 
\left[ i \sigma^m_{ \alpha \dot{ \alpha}} \left(  
\bar{A}_i(x) \partial_mA_j(x)- \partial_m \bar{A}_i(x)A_j(x) \right)
-2 \bar{ \psi}_{i \dot{ \alpha}}(x) \psi_{j \alpha}(x) \right] \nonumber \\
&+& \theta \theta \bar{ \theta}^{ \dot{ \alpha}}
\left[ \frac{i}{ \sqrt{2}} \sigma^m_{ \alpha \dot{ \alpha}} \left(  
\bar{A}_i(x) \partial_m \psi_j^{ \alpha}(x)- \partial_m \bar{A}_i(x) 
\psi_j^{ \alpha}(x) \right)
-\sqrt{2}F_j(x) \bar{ \psi}_{i \dot{ \alpha}}(x) \right] \nonumber \\
&+& 
\bar{ \theta} \bar{ \theta} \theta^{ \alpha} 
\left[ \frac{-i}{ \sqrt{2}} \sigma^m_{ \alpha \dot{ \alpha}} \left(  
\bar{ \psi}^{ \dot{ \alpha}}_i(x) \partial_m A_j(x)- 
\partial_m \bar{ \psi}^{ \dot{ \alpha}}_i(x) A_j(x) \right)
+\sqrt{2} \bar{F}_i(x) \psi_{j \alpha}(x) \right] \nonumber \\
&+& \theta \theta \bar{ \theta} \bar{ \theta} \left[ 
\bar{F}_i(x)F_j(x)+ \frac{1}{4} \bar{A}_i(x) \Box A_j(x)+ \frac{1}{4}\Box 
\bar{A}_i(x) A_j(x) \right. \nonumber \\
&+& \left. 
\frac{i}{2} \partial_m \bar{ \psi}_i(x) \bar{ \sigma}^m \psi_j(x) 
\frac{i}{2} \bar{ \psi}_i(x) \bar{ \sigma}^m \partial_m \psi_j(x) 
- \frac{1}{2} \partial_m \bar{A}_i(x) \partial^m A_j(x) \right] \,\ . \nonumber \\
\label{WB1}
\eea
O termo proporcional a $\theta \theta \bar{\theta} \bar{\theta}$ cont\'em 
termos de energia cin\'etica para $A$ bem como para 
$\Psi$! Repare que os campos $F$ n\~ao se propagam. O campo $F$ \'e 
introduzido para restabelecer a regra da igualdade dos graus de liberdade 
fermi\^onico e bos\^onico em uma teoria Supersimetrica. Lembremos que 
$\Psi$ \'e um campo fermi\^onico que tem $4$ graus de liberdade, j\'a o campo 
escalar $A$ tem apenas $2$ graus de liberdade, dai a necessidade de 
introduzir o campo $F$, para que tenhamos $4$ graus de liberdades 
bos\^onicos. O campo $F$ \'e chamado na literatura de campo auxiliar.

Este supercampo quiral descreve part\'\i culas de spin $0$ e de spin $1/2$, tais como o 
Higgs, os l\'eptons e quarks do modelo padr\~ao. Por\'em, ainda necessitamos descrever 
part\'\i culas de spin $1$, que s\~ao os b\'osons de gauge do modelo padr\~ao.
Para isto  precisamos introduzir o supercampo vetorial, e \'e isto que faremos
a seguir.

\subsection{Supercampo Real.}

Estes supercampos s\~ao definidos por
\be
V(x, \theta, \bar{ \theta}) = V^{\dagger}(x, \theta, \bar{ \theta}) \,\ ,
\label{definicao}
\ee
e tem a seguinte expans\~ao de Taylor em pot\^encia de $\theta^{ \alpha}$, 
$\bar{\theta}^{ \dot{ \alpha}}$:
\bea
V(x, \theta, \bar{ \theta})&=&C(x)+ 
i \theta \chi(x)-i \bar{ \theta} \bar{ \chi}(x)
\nonumber \\ 
&+& \frac{i}{2} \theta \theta \left[ M(x)+iN(x) \right]- \frac{i}{2}
\bar{ \theta} \bar{ \theta} \left[ M(x)-iN(x) \right]- \theta \sigma^m 
\bar{ \theta}A_m(x) \nonumber \\
&+&i \theta \theta \bar{ \theta} \left[ \bar{ \lambda}(x)+ 
\frac{i}{2} \bar{ \sigma}^m \partial_m \chi(x) \right]- 
i \bar{ \theta} \bar{ \theta} \theta \left[ \lambda(x)+ 
\frac{i}{2} \sigma^m \partial_m \bar{ \chi}(x) \right] \nonumber \\
&+& \frac{1}{2} \theta \theta \bar{ \theta} \bar{ \theta} 
\left[D(x)+ \frac{1}{2} \Box C(x) \right] \,\ .
\label{abertura}
\eea
As componentes $C,D,M,N,A_m$ devem ser reais para que Eq.(\ref{abertura}) 
satisfa\c ca \newline Eq.(\ref{definicao}).

Na literatura existe um gauge especial, chamado gauge de Wess-Zumino, 
onde esse supercampo \'e escrito da seguinte maneira 
\be
V_{WZ}(x, \theta, \bar{ \theta}) = -\theta \sigma^m \bar{ \theta} A_m(x) +
i( \theta \theta ) \bar{ \theta} \bar{ \lambda}(x) - 
i(\bar{ \theta} \bar{ \theta} )\theta \lambda(x)
+ \frac{1}{2} (\theta \theta )( \bar{ \theta} \bar{ \theta} )D(x) \,\ ,
\label{eq:vwzdef}
\ee
onde $A_{m}$ \'e um b\'oson de gauge de spin um, $\lambda$ \'e um 
f\'ermion de Weyl de spin meio enquanto $D$ \'e um campo escalar 
real de spin zero. Neste gauge podemos escrever ainda
\bea
V^2_{WZ}&=&- \frac{1}{2} \theta \theta \bar{ \theta} \bar{ \theta}A_mA^m
\,\ , \nonumber \\
V^n_{WZ}&=&0 \,\ ,
\label{oksana}
\eea
para $n \geq 3$.
Uma transforma\c c\~ao infinitesimal, dada pela Eq.(\ref{svtl6}), neste supercampo 
produz as seguintes transforma\c c\~oes
\bea
\delta_{ \xi} A_m &=&i \xi \sigma^m \bar{ \lambda}+ 
i \bar{ \xi} \sigma^m \lambda \,\ (\mbox{b\'oson $\rightarrow$ f\'ermion})
\,\ , \nonumber \\
\delta_{ \xi} \lambda &=& F_{mn} ( \sigma^{mn} \xi)+i \xi D 
\,\ (\mbox{f\'ermion $\rightarrow$ b\'oson})\,\ ,\\
\delta_{ \xi} \bar{ \lambda} &=& F_{mn} ( \bar{ \sigma^{mn}} \bar{ \xi})- 
i \bar{ \xi} D \,\ (\mbox{f\'ermion $\rightarrow$ b\'oson})
\,\ , \nonumber \\ 
\delta_{ \xi} D &=&\bar{ \xi} \bar{ \sigma}^m \partial_m \lambda - 
\xi \sigma^m \partial_m \bar{ \lambda} 
(\mbox{D $\rightarrow$ derivada total})
\,\ , \nonumber 
\eea 
onde$F_{mn}= \partial_m A_n- \partial_n A_m$.

Agora vamos construir o campo de for\c ca supersim\'etrico, no caso abeliano, 
ele \'e definido da seguinte maneira
\bea
W_{ \alpha} &=& -\frac{1}{4} ( \bar{D} \bar{D})D_{ \alpha} 
V(x, \theta, \bar{ \theta}) \,\ ,\nonumber\\
\bar{W}_{ \dot{ \alpha}} &=& -\frac{1}{4} (DD) 
\bar{D}_{ \dot{ \alpha}} V(x, \theta, \bar{ \theta}) \,\ .
\label{eq:simplewdef}
\eea
Uma defini\c c\~ao equivalente \'e dada para o caso n\~ao abeliano, 
que \'e a seguinte
\bea
W_{ \alpha} &=& -\frac{1}{4} ( \bar{D} \bar{D})e^{-V}D_{ \alpha} 
e^{V} \,\ ,\nonumber\\
\bar{W}_{ \dot{ \alpha}} &=& -\frac{1}{4} (DD)e^{-V} 
\bar{D}_{ \dot{ \alpha}} e^{V} \,\ .
\label{eq:nonawdef}
\eea

Se abrirmos em componentes a Eqs.(\ref{eq:simplewdef}) e 
(\ref{eq:nonawdef}) o resultado ser\'a
\bea
W_{ \alpha} &=& -i \lambda_{ \alpha} + \theta_{ \alpha}D
- \frac{i}{2}( \sigma^m \bar{ \sigma}^n \theta)_{\alpha}
F_{mn} \nonumber \\
&+&( \theta \theta) \sigma^m_{ \alpha \dot{ \beta}} \partial_m
\bar{ \lambda}^{\dot{ \beta}} \,\ ,
\label{eq:abelwcom}\\
\bar{W}_{ \dot{ \alpha}} &=& i\bar{ \lambda} _{\dot{ \alpha}} +
\bar{ \theta}_{ \dot{ \alpha}}D+ 
\frac{i}{2}( \bar{ \sigma}^m \sigma^n \bar{ \theta})_{ \dot{ \alpha}} 
F_{mn}\nonumber\\
&-&( \bar{ \theta} \bar{ \theta})
\hbox{$ \bar{ \sigma}^m_{ \dot{ \alpha}}$}^{\beta} \partial_m
\lambda _{\beta} \,\ .\nonumber
\eea
No caso abeliano $F_{mn}= \partial_m A_n- \partial_n A_m$ e no 
caso n\~ao abeliano \'e \newline 
$F_{mn}= \partial_m A_n- \partial_n A_m-gt^{abc}A_{m}^{b}A_{n}^{c}$. Este 
campo de for\c ca \'e um supercampo quiral, pois pode-se mostrar que
\BEA
\bar{D}_{ \dot{ \beta}}W_{ \a}&=&0 \,\ , \nonumber \\
D_{ \a} \bar{W}_{ \dot{ \beta}}&=&0 \,\ , \nonumber \\
\EEA
e esta \'e a defini\c c\~ao de supercampo quiral, conforme j\'a discutimos 
anteriormente.

\section{A\c c\~oes Supersim\'etricas}

Uma vez tendo introduzidos os supercampos e analisados algumas de suas 
expans\~oes em componentes, iremos agora 
ver como construir a\c c\~oes supersim\'etricas usando 
os supercampos definidos na \'ultima se\c c\~ao.

\subsection{A\c c\~ao com o Supercampo Quiral}

A a\c c\~ao mais simples que podemos construir \'e
\be
\int d^{4}x \int d^{4} \theta \bar{ \Phi} \Phi +
\int d^{4}x \left[d^{2} \theta (\frac{1}{2} m \Phi^2 + 
\frac{1}{3} g \Phi ^3 ) +{\rm h.c.} \right] \,\ ,
\label{aq}
\ee
onde $\Phi$ \'e um supercampo quiral. Mudan\c ca de base de $y$ para 
$x$ n\~ao muda a a\c c\~ao. O segundo termo \'e o superpotencial 
\footnote{Na realidade a forma mais geral para o superpotencial \'e
$W= \lambda \Phi+1/2m \Phi^{2}+1/3g \Phi^{3}$, como o primeiro termo 
n\~ao \'e importante para o que faremos a seguir n\'os n\~ao iremos 
analisar ele.} da teoria. Nesta a\c c\~ao paramos em $\Phi^3$ porque, 
s\'o podemos ter termos 
escalares proporcionais a $A^{2}$ e $A^{3}$, pois 
termos com pot\^encias maiores que 3 geram diverg\^encias 
quadr\'aticas \`a n\'\i vel de dois loops.

Mas da Eq.(\ref{WB1}) sabemos que
\be
\int d^{4} \theta \bar{ \Phi} \Phi  =
\frac{1}{4} \Box \bar{A}A +\frac{1}{4} \bar{A} \Box A + \bar{F}F
-\frac{1}{2} \partial^m \bar{A} \partial_m A  
+ \frac{i}{2} \partial_m \bar{ \psi}_i \bar{ \sigma}^m \psi_j
-\frac{i}{2} \bar{ \psi}_i \bar{ \sigma}^m \partial_m \psi_j
\,\ ,
\label{eq:ppdag}
\ee
mas podemos escrever este termo da seguinte forma
\be
\int d^{4} \theta \bar{ \Phi} \Phi  =
\bar{A} \Box A + \bar{F}F+i \partial_m \bar{ \psi}_i \bar{ \sigma}^m \psi_j
\,\ .
\label{quiral1}
\ee

J\'a da Eq.(\ref{bi}) e Eq.(\ref{tri}) vem que
\be
\int d^{2} \theta \Phi_i \Phi_j=A_iF_j+A_jF_i- \psi_i \psi_j \,\ ,
\ee
\be
\int d^{2} \theta \Phi_i \Phi_j \Phi_k=F_iA_jA_k+A_iF_jA_k+A_iA_jF_k- 
\psi_i \psi_j A_k- A_i \psi_j \psi_k - \psi_i A_j \psi_k
\,\ ,
\ee
com isto podemos reescrever o superpotencial da seguinte maneira
\be
\int d^{2} \theta \left[ \frac{m}{2} \Phi^2 + \frac{g}{3} \Phi^3 \right]
= m \left(AF- \frac{1}{2} \psi \psi \right) + g(A^2F- \psi \psi A) \quad ,
\ee
assim a nossa a\c c\~ao da Eq.(\ref{aq}) em termos de componentes torna-se
\be
\int d^{4}x \left \{ \bar{F}F+ \bar{A} \Box A+i \partial_m \bar{ \psi}_i 
\bar{ \sigma}^m \psi_j
+ \left[ m \left( AF- \frac{1}{2} \psi \psi \right) + g(A^2F- \psi \psi A)+ 
{\rm h.c} \right] \right \} \,\ .
\label{aq2} 
\ee

Notamos que nesta a\c c\~ao n\~ao cont\'em derivadas atuando 
em $F(x)$, isto significa que $F(x)$ \'e um {\bf campo auxiliar} 
que pode ser eliminado resolvendo suas equa\c c\~oes de 
movimento que s\~ao dadas pelas equa\c c\~oes de Euler-Lagrange 
\BEA
     \PD{\L}{\phi} - \P_{m}\PD{\L}{(\P_{m}\phi)} &=& 0 \,\ ,  \nonumber
       \label{Euler-Lagrange Equation}
\EEA
onde $\phi$ \'e {\em qualquer} (inclusive os hermitianos conjugados) campo de
Minkowski. Formalmente campos auxiliares s\~ao definidos como campos que n\~ao
possuem termos cin\'eticos. Assim, esta defini\c c\~ao nos diz que as 
equa\c c\~oes Euler-Lagrange para os campos auxiliares \'e simplesmente 
$\PD{\L}{\phi}=0$. Usando esta equa\c c\~ao simplificada obtemos
\bea
\frac{ \delta {\cal L}}{ \delta F} &=&
\bar{F} + mA + gA^2 = 0 \,\ ,\nonumber\\
\frac{ \delta {\cal L}}{ \delta \bar{F}} &=&
F + m \bar{A} + g(\bar{A})^2 = 0 \,\ .
\label{aq3}
\eea
Usando Eq.(\ref{aq2}) e Eq.(\ref{aq3}) podemos escrever a nossa a\c c\~ao
original da seguinte 
 maneira
\be
\int d^{4}x \left[  \bar{A} \Box A+i \partial_m \bar{ \psi}_i 
\bar{ \sigma}^m \psi_j - (\frac{1}{2} m \psi \psi +g \psi \psi A+h.c) - 
V_{F}(A, \bar{A}) \right] \,\ ,
\ee
onde o potencial escalar $V_{F}(A, \bar{A})$ \'e dado por
\be
V_{F}(A, \bar{A}) = \vert F \vert^2 = \bar{F}F=
[m \bar{A} + g( \bar{A})^2][mA + gA^2] \,\ ,
\ee
e descreve o termo de massa dos escalares e as intera\c c\~oes dos escalares. 
Repare que o campo escalar $A$ e o f\'ermion $\psi$ tem a mesma massa $m$. O 
acoplamento entre dois f\'ermions e um escalar \'e o mesmo que entre 
quatro escalares, e \'e dado por $g$.

A introdu\c c\~ao do superpotencial $W$ facilita bastante escrever a a\c c\~ao 
precedente e ela \'e escrita da seguinte maneira
\begin{equation}
\int d^{4}x \left[  \bar{A} \Box A+i \partial_m \bar{ \psi}_i 
\bar{ \sigma}^m \psi_j - \frac{1}{2} \sum_{ij} \left( 
\frac{ \partial^{2} W}{ \partial A_{i} \partial A_{j}}+hc \right) - 
\left| \frac{\partial W}{\partial A_{j}} \right|^{2} \right] \,\ ,
\ee
onde agora
\begin{eqnarray}
\frac{ \partial W}{ \partial A_{i}}&=&m_{ij}A_{j}+g_{ijk} 
A_{j}A_{k} \equiv V_{F}(A, \bar{A}) \nonumber \\
\frac{ \partial^{2} W}{ \partial A_{i} \partial A_{j}}&=&m_{ij}+2g_{ijk}A_{k}, 
\end{eqnarray}
onde $A_{i}$ \'e um campo escalar. Devemos ressaltar que ap\'os substituir 
as Eqs.(\ref{aq3}) o superpotencial $W$ ser\'a fun\c c\~ao apenas dos campos
escalares $A_{i}$.

\subsection{Intera\c c\~oes Invariante de Gauge.}

A a\c c\~ao que vamos estudar agora \'e
\be
\int d^{4}x \int d^{4} \theta \bar{ \Phi} e^{gV} \Phi \,\ ,
\ee
onde $g$ \'e a constante de acoplamento, real, de algum grupo. 
Podemos escrever no gauge de Wess-Zumino o seguinte
\be
\bar{ \Phi} e^{gV_{WZ}} \Phi= \bar{ \Phi} \Phi+
g \bar{ \Phi} V_{WZ}  \Phi+
\frac{g^{2}}{2} \bar{ \Phi} V_{WZ}^{2}  \Phi \,\ ,
\label{akulov}
\ee
esta expans\~ao \'e exata neste gauge, ver Eq.(\ref{oksana}).

O primeiro termo j\'a conhecemos est\'a na Eq.(\ref{quiral1}), j\'a 
os outros dois usando 
Eqs.(\ref{eq:vwzdef}) e (\ref{oksana}) podemos mostrar que
\bea
\bar{ \Phi}V_{WZ}&=&- \theta \sigma^{m} \bar{ \theta} \bar{A} A_{m}+
i \theta \theta \bar{ \theta} \bar{A} \lambda - 
\sqrt{2} \bar{ \theta} \bar{ \psi} \theta \sigma^{m} \bar{ \theta}A_{m} + 
\frac{1}{2} \theta \theta \bar{ \theta} \bar{ \theta}( \bar{A}D-i 
\partial_{n} \bar{A}A^{n}-i \sqrt{2} \bar{ \psi} \bar{ \lambda}) 
\nonumber \\
\bar{ \Phi}V^{2}_{WZ}&=&- \frac{1}{2} \theta \theta \bar{ \theta} 
\bar{ \theta} \bar{A} A_{m}A^{m} \,\ .
\eea

Fazendo a outra parte das contas encontramos
\bea
\int d^{4} \theta \bar{ \Phi}V_{WZ} \Phi&=& \frac{1}{2}[ A \bar{A}D-iA 
\partial_{n} \bar{A}A^{n}-i \sqrt{2}A \bar{ \psi} \bar{ \lambda}+
i \sqrt{2} \bar{A} \lambda \psi+A_{m} \bar{ \psi} \bar{ \sigma}^{m} \psi+i 
\bar{A}A^{n} \partial_{n}A] \nonumber \\
\int d^{4} \theta \bar{ \Phi}V^{2}_{WZ} \Phi&=&- \frac{1}{4} \bar{A} 
A^{m}A_{m}A \,\ ,
\label{ig}
\eea
substituindo Eq.(\ref{quiral1}) e Eq.(\ref{ig}) na Eq.(\ref{akulov}) obteremos
\bea
\int d^{4}x \int d^{2} \theta \bar{ \Phi} e^{gV_{WZ}} \Phi&=& \bar{F}F+ 
A \Box \bar{A}+i \partial_m \bar{ \psi}_i \bar{ \sigma}^m \psi_j \nonumber \\
&+&gA^n \left( \frac{1}{2} \bar{ \psi} \bar{ \sigma}^{n} \psi+ \frac{i}{2} 
\bar{A} \partial_n A- \frac{i}{2} \partial_n \bar{A} A \right) \nonumber \\
&-& \frac{ig}{ \sqrt{2}}(A \bar{ \lambda} \bar{ \psi}- \bar{A} \lambda \psi)+ 
\frac{1}{2} \left( gD- \frac{1}{2} g^{2} A_n A^n \right) \bar{A}A \,\ . \nonumber \\ 
\label{d1}
\eea
Repare que esta parte n\~ao apenas descreve as intera\c c\~oes dos campos de 
mat\'eria com os campos de gauge, dado na segunda linha, bem como a 
intera\c c\~ao escalar b\'oson de gauge. Tamb\'em temos 
as intera\c c\~oes de Yukawa entre os f\'ermions, $\Psi$, sf\'ermions, 
$A$, e os gauginos $\lambda$, conforme nos diz o primeiro termo da 
terceira linha da equa\c c\~ao 
acima.

\subsection{Teoria de  Yang-Mills supersim\'etrica}

A lagrangiana de Yang-Mills supersim\'etrica \'e
\bea
&&\frac{1}{4} \int d^{4}x \int d^{2} \theta \,\ \left( W^{\alpha}W_{\alpha} 
+ \bar{W}_{ \dot{ \alpha}} \bar{W}^{ \dot{ \alpha}} \right) \nonumber \\
&& = \int d^{4}x \left[ -\frac{1}{4} F_{mn}F^{mn}
-i \lambda \sigma^m {\cal D}_{m} \bar{ \lambda} + \frac{1}{2} D^2 \right]
\quad ,
\label{d2}
\eea
com ${\cal D}_{m} \lambda^{a}= \partial_{m} \lambda^{a}-gf^{abc}A^{b}_{m} 
\lambda^{c}$ sendo a derivada covariante usual, ou 
seja $f$ \'e a constante de estrutura da \'algebra de Lie e $g$ \'e o 
acoplamento de gauge. Onde percebemos que esta a\c c\~ao cont\'em as 
a\c c\~oes de Maxwell e de Dirac para campos livres, bem como acopla os 
gauginos aos campos de gauge.

Como antes $D$ tamb\'em \'e um campo auxiliar e 
pode ser removido usando suas equa\c c\~oes de movimento. Das Eq.(\ref{d1}) e 
Eq.(\ref{d2}) vemos que a lagrangiana dos $D$ termos \'e dada por
\bea
{\cal L}_{D}= \frac{1}{2}D^{2}+ \frac{1}{2} g \bar{A} AD \,\ ,
\eea
que resulta na seguinte equa\c c\~ao de movimento para os campos $D$
\bea
D=- \frac{g}{2} \bar{A}A \,\ ,
\eea
no caso n\~ao abeliano onde $V=T_{a}V^{a}$ pode-se mostrar que a equa\c c\~ao 
de movimento dada acima \'e modificada para
\bea
D_{a}=-g \bar{A} T_{a} A \,\ ,
\eea
e o potencial escalar neste caso \'e dado por
\bea
V_{D}(\bar{A},A)=- {\cal L}_{D}= \frac{1}{2}D^{2} \,\ .
\eea

\section{O Potencial Escalar}

Ao contr\'ario do modelo padr\~ao, onde o potencial escalar \'e arbitr\'ario 
e \'e definido apenas pela exig\^encia da invari\^ancia de gauge. No caso de 
teorias supersim\'etricas o potencial escalar \'e completamente definido pelo 
superpotencial, e consiste das contribui\c c\~oes dos termos $D$ e $F$ que 
discutimos na \'ultima se\c c\~ao. O potencial escalar completo de uma teoria 
supersim\'etrica \'e a soma dessas duas contribui\c c\~oes, ou seja
\bea
V(\bar{A},A)&=& V_{D}(\bar{A},A)+V_{F}(\bar{A},A) \nonumber \\
&=&|F|^{2}+ \frac{1}{2}D^{2} \,\ ,
\label{joia}
\eea
e percebemos que $V(\bar{A},A) \geq 0$, pois s\~ao quadr\'aticos neste campos.

{\em Dessa maneira, a forma da lagrangiana \'e praticamente fixada pela exig\^encia 
de uma simetria. As \'unicas liberdades deste tipo de teorias s\~ao o
conte\'udo de campos, os valores dos acoplamentos de gauge $g$, os
acoplamentos de Yukawa $g_{ijk}$ e as massas das part\'\i culas do modelo}.
Com isto agora j\'a temos quase todas as ferramentas necess\'arias para 
construir um modelo supersim\'etrico real\'\i stico, 
falta apenas ver como quebrar
supersimetria. Se supersimetria n\~ao fosse quebrada, os f\'ermions e os 
seus superparceiros bos\^onicos deveriam ser degenerados na massa. No espectro 
do modelo padr\~ao claramente n\~ao satisfaz esta condi\c c\~ao 
j\'a que nenhum parceiro supersim\'etrico foi encontrado, pois n\~ao existe um 
sel\'etron com massa de $511 KeV$, nem um sm\'uon de $106 MeV$ e etc. De tal modo 
que se supersimetria \'e realizada pela natureza, ela deve ser quebrada.

Antes de ver como quebrar supersimetria devemos mencionar que nenhuma das 
part\'\i culas j\'a conhecidas seja o parceiro supersim\'etrico de outra, 
porque o superparceiro deve diferir de meio no spin, enquanto tendo a mesma propiedade 
de transforma\c c\~ao sobre o grupo de gauge bem como sobre qualquer simetria global 
da teoria. Mas antes de tratarmos disto vamos mostrar como supersimetria 
resolve o problema da hierarquia.
 
\section{Problema da Hierarquia}

Para enetendermos este problema considere a seguinte lagrangiana (n\~ao 
supersim\'etrica) de um campo escalar complexo $A$ e um f\'ermion de 
Weyl $\chi$
\BEA
\L= & - & \partial_m \bar{A} \partial^m A -i \bar{ \chi} \bar{ \sigma}^m
\partial_m \chi - \frac{1}{2} \,m_f\, (\chi\chi + \bar{ \chi} \bar{ \chi})- 
 m^2_b\, \bar{A} A \nonumber \\
& - & \;Y\,(A\chi\chi + \bar{A} \bar{ \chi} \bar{ \chi}) \;-\; 
\lambda \, (\bar{A} A)^2\ .
\label{toy model}
\EEA
Se $m_{b}=m_{f}$ e $Y= \lambda$ esta lagrangiana seria supersim\'etrica 
conforme j\'a foi visto na se\c c\~ao 6.1, 
mas agora vamos considerar o caso n\~ao supersim\'etrico.

Neste caso $\L$ possui uma simetria quiral para $m_{f}=0$ dada por
\BEA
A  \rightarrow  e^{-2 i \alpha}\, A\ , \qquad
\chi  \rightarrow e^{i \alpha}\, \chi \ .
\EEA
Esta simetria proibe a gera\c c\~ao de massa para o f\'ermion por 
corre\c c\~oes qu\^anticas. Para $m_{f} \neq 0$ a massa do f\'ermion recebe 
corre\c c\~oes radiativas, mas todos os poss\'\i veis diagramas t\^em que 
ter inser\c c\~oes de massa, podem ser visto na Fig.1. A corre\c c\~ao 
de massa que \'e proporcional a 
$m_{f}$ \'e dada por
\BEA
\delta m_f \sim Y^2 m_f \ln {m_f^2\over \Lambda^2}\ , 
\EEA
onde $\Lambda$ \'e o cutoff ultravioleta. Portanto a massa de um f\'ermion 
quiral n\~ao recebe grandes corre\c c\~oes radiativas se a massa ``nua" 
for pequena.

J\'a no caso de campos escalares a situa\c c\~ao \'e diferente. Os diagramas 
que d\~ao corre\c c\~oes de um loop \`a $m_{b}$ est\~ao mostrados nas Fig.2 e 
Fig.3. Ambos diagramas s\~ao divergente quadraticamente, mas eles t\^em 
sinais opostos devido ao fato que no segundo diagrama f\'ermions est\~ao no 
loop enquanto que no primeiro n\'os temos b\'osons rodando no loop. A 
corre\c c\~ao na massa neste caso \'e dada por
\BEA
\delta m_b^2 \sim  (\lambda  -  ~Y^2)\, \Lambda^2\ .
\EEA
Assim em teorias n\~ao supersim\'etricas campos escalares recebem grandes 
corre\c c\~oes de massa (mesmo que a massa ``nua" seja zero) e massas 
pequenas para os escalares n\~ao parece ser ``natural". Poi a massa 
``natural" para qualquer part\'\i cula escalar \'e $\Lambda$, que \'e da 
ordem da massa de Planck cujo valor \'e $M_{plank} \simeq 10^{19}$GeV.

Uma das possibilidades para se evitar este problema \'e por considerar supersimetria 
global e associar o Higgs a um supermultipleto quiral. A simetria quiral que 
proibe massa ao f\'ermion, higgsino, tamb\'em ir\'a proibir correspondente 
termo de massa ao Higgs, j\'a que se supersimetria \'e v\'alida temos que 
ter $m_f=m_b$. Se esta simetria quiral junto com supersimetria permanece 
at\'e uma escala $O(1TeV)$, as 
corre\c c\~oes radiativas ir\~ao induzir massa escalar da ordem de $O(M_W)$ 
o que explicaria porque $M_{H}<<<M_{plank}$ e assim resolvendo este problema. 
Devemos fazer a quebra de supersimetria sem introduzir diverg\^encias 
quadr\'aticas para resolver o problema da hierarquia.

\section{Quebra Espont\^anea de Supersimetria.}

De uma perspectiva 
te\'orica, esperamos que supersimetria seja uma simetria que \'e quebrada
espontaneamente. Em outras palavras, o modelo deve ter uma lagrangiana que 
\'e invariante sobre supersimetria, mas um v\'acuo que n\~ao \'e. Desta 
maneira, supersimetria \'e escondida a baixas energias em uma maneira
an\'aloga a simetria eletrofraca no modelo padr\~ao {\bf SM}.

Da Eq.(\ref{energia}) 
aprendemos que se os geradores de supersimetria aniquilam o v\'acuo, a 
hamiltoniana $H$ tamb\'em aniquila o v\'acuo. Isto por sua vez implica que 
o potencial escalar $V(\bar{A},A)$ de uma teoria supersim\'etrica tem um
estado supersim\'etrico fundamental tem  que ser zero neste
m\'\i nimo \bea
\label{Vmin}
E_{vac} = 0\quad \Rightarrow\quad 
\langle  V \rangle \equiv
V(\bar{A},A) |_{\rm min} = 0\,\ .
\eea
A forma geral do potencial escalar 
$V(\bar{A},A)=  F^i \bar{F}^i + \frac{1}{2} D^a D^a$ foi mostrada na
Eq.(\ref{joia}). Como  $V(\bar{A},A)$ \'e positivo n\'os facilmente 
conclu\'\i mos da Eq.(\ref{Vmin}) que em um estado supersim\'etrico
fundamental
\bea
\langle F^i\rangle \equiv F^i |_{\rm min} = 0 \quad {\rm e} \quad 
\langle  D^a \rangle \equiv D^a |_{\rm min} = 0 \,\ ,
\eea
tem que ser v\'alida. O oposto tamb\'em \'e verdadeiro
\bea
\langle F^i\rangle \neq 0 \quad {\rm ou}\quad
\langle  D^a \rangle \neq 0\quad
\Rightarrow\quad
V|_{min} >0\quad \Rightarrow\quad
 Q_{\alpha} |0>\ \neq 0 \,\ , \nonumber \\ 
\eea
e supersimetria \'e espont\^aneamente quebrada. Assim 
$\langle F^i\rangle$ e $\langle D^a\rangle$ s\~ao os par\^ametros que comandam
a quebra de supersimetria quando s\~ao diferentes de zero.

Potenciais espec\'\i ficos que resultam em termos $D$ e $F$ diferentes de zero 
foram constru\'\i dos \cite{ORa,FIl}. Por exemplo, o modelo de O'Raifeartaigh 
\cite{ORa} tr\^es supercampos quirais s\~ao necess\'arios para quebrar 
supersimetria. O espectro de massa dos seis b\'osons reais e os tr\^es 
f\'ermions de Weyl do modelo s\~ao
\bea
\mbox{\rm B\'osons:} \quad && (0,0,m^2,m^2, m^2 \pm 2\lambda g) \,\ , \nonumber \\
\mbox{\rm F\'ermions:} \quad && (0,m^2,m^2) \,\ . \nonumber
\eea

J\'a no caso do modelo de Fayet e Iliopoulos \cite{FIl}, eles observaram que 
a componente $\theta \theta \bar{ \theta} \bar{ \theta}$ do supercampo
vetorial \'e invariante tanto por invari\^ancia de gauge quanto por 
supersimetria. Este modelo descreve dois f\'ermions de massa 
$\sqrt{m^{2}+(1/2)e^{2}v^{2}}$, um campo vetorial e um campo escalar, cada de 
massa $\sqrt{(1/2)e^{2}v^{2}}$, um campo escalar complexo de massa 
$\sqrt{m^{2}}$, e um espinor de goldstone n\~ao massivo.

Repare que nos dois modelos descrito acima vale a seguinte rela\c c\~ao 
\bea \label{bosferm}
\sum_{\rm bosons} M_{b}^2 
= 2 \sum_{\rm fermions} M_{f}^2\ ,
\eea
que \'e muito ruim para fenomenologia. Muitos modelos de
quebra de  supersimetria j\'a foram propostos, mas ainda n\~ao existe consenso
em como  isto deve ser feito.

Por\'em, de um ponto de
vista pr\'atico, \'e extremamente \'util simplesmente parametrizar nossa 
ignor\^ancia desses resultado por introduzir termos extras que quebrem 
supersimetria explicitamente na lagrangiana. O acoplamento extra de quebra 
de supersimetria deve ser ``soft" para manter naturalmente uma hierarquia 
entre a escala eletrofraca e a escala de Planck, conforme discutimos na 
\'ultima se\c c\~ao.

A filosofia atr\'as destes termos ``softs" \'e a seguinte: existe um setor que 
quebra supersimetria espontaneamente em uma escala de energia maior que a escala 
eletrofraca. A quebra de supersimetria \'e comunicada de alguma maneira(atrav\'es 
de intera\c c\~oes de gauge ou atrav\'es da intera\c c\~ao gravitacional) aos campos 
do modelo e como resultado os termos ``softs" aparecem. Uma implementa\c c\~ao desta 
id\'eia \'e a de quebrar supersimetria espontaneamente em um ``setor escondido", que \'e 
um setor que n\~ao interage com as part\'\i culas do modelo padr\~ao (setor vis\'\i vel) 
exceto atrav\'es de uma teoria de supergravidade que serve de intermedi\'ario para esta 
quebra, para mais detalhe veja a pr\'oxima sub-se\c c\~ao.

Em 1982 L. Girardello e M. T. Grisaru \cite{10} mostraram que o seguinte termo 
que quebra supersimetria
\BEA
{\cal L}_{soft}=- \frac{1}{2} (M_{ \lambda} \lambda^{a} \lambda^{a}+hc) 
-m^{2} \bar{A}A+b^{ij}A_{i}A_{j}+a^{ijk}A_{i}A_{j}A_{k} \,\ ,
\label{soft}
\EEA
\'e livre de diverg\^encias quadr\'aticas nas corre\c c\~oes qu\^anticas 
para a massa do escalar da teoria. Pode ser mostrado que a aus\^encia de 
diverg\^encias quadr\'aticas em teorias supersim\'etricas estabiliza a 
massa do Higgs e assim a escala eletrofraca. O primeiro termo desta 
lagrangiana \'e um termo de massa para os gauginos, o segundo \'e o termo 
de massa para os escalares e os termos $b^{ij}$ e $a^{ijk}$ t\^em a mesma
forma  dos termos proporcionais a $M^{ij}$ e $y^{ijk}$ no superpotencial,
assim eles  ser\~ao permitidos pela invari\^ancia de gauge se e somente se um
termo  correspondente existe no superpotencial.

\subsection{Modelos de Quebra de SUSY.}

N\'os introduzimos acima quebra expl\'\i cita de SUSY devido ao fato de n\~ao 
conhecermos ainda o mecanismo de quebra de SUSY. Se SUSY \'e quebrada 
espont\^aneamente ent\~ao existe um f\'ermion de Goldstone chamado de 
{\it goldstino}. Quando supersimetria \'e global o goldstino n\~ao tem massa. 
Por\'em em supersimetria local (supergravidade) o goldstino \'e ``comido"  
pelo gravitino $(\tilde{g}_{3/2})$ que desta maneira adquire uma massa 
$m_{3/2}$ \cite{superHiggs}. Isto \'e conhecido na literatura como o 
{\rm Mecanismo de Super-Higgs} e \'e completamente an\'alogo ao mecanismo de 
Higgs das teorias de Gauge.

Modelos que apresentam quebra espont\^anea de SUSY assumem que SUSY \'e quebrada
num setor `hidden'' ou ``secluded'' que \'e completamente neutro com 
rela\c c\~ao
ao grupo de Gauge do SM. A informa\c c\~ao da quebra de SUSY \'e ent\~ao
informada ao setor ``visible'', que cont\'em o MSSM, por algum mecanismo. Doi 
cen\'arios deste tipo de quebra tem sido estudado em detalhe: quebra de SUSY 
gravity--mediated e gauge--mediated.

No caso de {\it quebra de SUSY gravity--mediated}, a quebra de SUSY \'e
transmitida ao MSSM pela intera\c c\~ao gravitacional \cite{gravmed}. A quebra 
ocorre a ${\cal O}(10^{10})$ GeV ou acima, e o gravitino ganha uma massa da
ordem da escala eletro-fraca. O modelo de estrutura mais simples deste tipo de 
mecanismo \'e o {\it minimal supergravity model} (mSUGRA) \cite{12,msugra}. 
Como mSUGRA possui apenas cinco par\^ametros livres (o SM possui 18) ele \'e 
altamente preditivo e por isto mais usado nas buscas experimentais. Por\'em, 
devemos ter em mente, que ele \'e bem restritivo.

Os modelos {Gauge--mediated} (GMSB) \cite{gaugemed} tem um setor ``secluded'' 
onde SUSY \'e quebrada e um setor ``messenger'' consistindo de part\'\i culas 
com n\'umeros qu\^anticos de $SU(3) \otimes SU(2) \otimes U(1)$. Os 
``messengers'' se acoplam diretamente ao setor ``secluded'' e a quebra de SUSY 
\'e transmitida ao MSSM atrav\'es de trocas virtuais dos ``messengers''. Uma 
caracter\'\i stica b\'asica deste tipos de modelos \'e que a escala de quebra 
de SUSY ocorre a escalas bem abaixos do que no caso gravity--mediated, a escala
\'e da ${\cal O}(10^4\mbox{--}10^5)$~GeV. Por\'em o alcance da massa do
gravitino vai de eV at\'e KeV.

Uma discuss\~ao mais detalhada da quebra de SUSY est\'a al\'em deste estudo.
Para pessoas interessadas em mSUGRA veja \cite{DPF95,snowmass}.  
Uma revis\~ao da quebra de SUSY gauge--mediated \'e dada em \cite{giudice}.

\section{A Invari\^ancia R.}

Esta nova simetria \'e uma simetria global que n\~ao comuta com supersimetria,
foi introduzida em 1975 por P. Fayet \cite{11} e \'e chamada de simetria $R$. 
Exigir \`a invari\^ancia R da a\c c\~ao coloca restri\c c\~oes na forma dos termos 
de intera\c c\~ao, al\'em disto muitas teorias supersim\'etricas interessantes 
s\~ao invariantes por R e assim possuem uma simetria $U(1)_{R}$ adicional.

O conceito de simetria $R$ pode ser melhor entendida no Super-espa\c co.
Chamaremos os geradores desta simetria por ${\bf R}$, este operador gera a 
seguinte transforma\c c\~ao nas vari\'aveis de 
grassman
\BEA
{\bf R} \t \rightarrow e^{-i \a} \t \,\ ,
\EEA
assim $\t$ tem carga $R( \t)=1$. Quando atua nos supercampos quirais ela produz
\BEA
   {\bf R} \Phi(x,\t,\tb) &=& e^{2 i n_{\Phi}\a}\Phi(x, e^{-i\a}\t,e^{i\a}\tb ) \,\ ,
          \nonumber \\
   {\bf R} \bar{ \Phi}(x,\t,\tb) &=&
          e^{-2 i n_{\Phi}\a}\bar{ \Phi}(x, e^{-i\a}\t, e^{i\a}\tb ) \,\ ,
          \label{The R-Invariance prop 2}
\EEA
onde $2n_{ \Phi}$ \'e a carga quiral e no vetorial
\BEA
   {\bf R} V(x,\t,\tb) &=& V(x, e^{-i\a}\t, e^{i\a}\tb ) \,\ .
         \label{The R-Invariance prop 3}
\EEA

Em termos de componentes, as transforma\c c\~oes acima para o multipleto 
quiral s\~ao
\BEA
   \LP.  \BA{lcr}
          A       &\longrightarrow&       e^{2in_{\Phi}\a} A \\
          \psi    &\longrightarrow&       e^{2i\LP(n_{\Phi}-\HA\RP)\a} \psi\\
          F       &\longrightarrow&       e^{2i\LP(n_{\Phi}-1\RP)\a} F
             \EA \RP\} \,\ ,
           \label{The R-Invariance prop 4}
\EEA
e para o multipleto vetorial temos
\BEA
   \LP.  \BA{lcr}
          A_{m} &\longrightarrow&       A_{m} \\
          \la     &\longrightarrow&       e^{i\a} \la\\
          D       &\longrightarrow&       D
          \EA \RP\} \,\ .
          \label{The R-Invariance prop 5}
\EEA

Para produtos de termos de supercampos quirais, pode-se mostrar que
~\cite{FAYET}
\BEA
  {\bf R} \prod_{a}\;\Phi_{a}(x,\t,\tb)
    &=& e^{2i\sum_{a}n_{a}\a}\,\prod_{a}\Phi(x, e^{-i\a}\t,e^{i\a}\tb ) \,\ ,
      \nonumber
\EEA
e os seguintes termos de supercampos s\~ao todos invariantes por paridade R
\BEA
   & &   \int d^{4}\t\; \bar{ \Phi}(x,\t,\tb) \Phi(x,\t,\tb) \,\ ,
              \nonumber \\
   & &   \int d^{4}\t\; \bar{ \Phi}(x,\t,\tb) e^{V(x,\t,\tb)}
                   \Phi(x,\t,\tb) \,\ ,
                   \nonumber \\
   & &   \int d^{4}\t\; \prod_{a}\Phi_{a}(x,\t,\tb) \,\ ,
              \hspace{2cm} \mbox{se  } \sum_{a} n_{a} = 1 \,\ . 
              \nonumber 
\EEA

\section{O Modelo Padr\~ao Supersim\'etrico M\'\i nimo.} 

As diferentes classes de extens\~oes supersim\'etricas do {\em modelo padr\~ao}
{\bf SM} s\~ao divididas naturalmente em duas classes. A 
primeira \'e o {\em modelo padr\~ao supersim\'etrico m\'\i nimo} {\bf MSSM} 
\cite{INO82a}---\cite{LI89} que
cont\'em o n\'umero m\'\i nimo  de campos e par\^ametros exigidos para
construir um modelo  real\'\i stico de quarks e l\'eptons. A segunda classe 
conhecido como {\em modelo padr\~ao supersim\'etrico n\~ao m\'\i nimo} 
(NMSSM)~\cite{NMSSM}. Este modelo cont\'em  mais par\^ametros e campos 
sem qualquer aumento da predictividade do modelo e por isto n\~ao ser\~ao 
aqui considerados. A respeito do {\bf MSSM} existem exelentes revis\~oes tais 
como \cite{12} e \cite{kane}---\cite{HAB79} nas quais este estudo est\'a baseado.

\subsection{Ingredientes do Modelo.}

Os ingredientes do {\bf MSSM} podem ser definidos pelos 
seguintes pontos:
\begin{enumerate}
  \item
      O grupo de Gauge \'e: $SU(3)_{C}\otimes SU(2)_{L}\otimes U(1)_{Y}$.
  \item
      O conte\'udo m\'\i nimo de part\'\i culas, sendo de tr\^es fam\'\i lias 
        de quarks e l\'eptons, doze b\'osons de gauge (definidos da maneira 
        usual), dois dubletos de Higgs e, claramente, todos os seus 
        parceiros  supersim\'etricos.
\begin{enumerate}	
  \item Introduz\'\i mos juntos com os b\'osons de Gauge os ``{\sc gauginos}'' 
  que s\~ao f\'ermios, ou seja s\~ao part\'\i culas de spin $1/2$.
  
  \item Os l\'eptons e quarks t\^em como parceiros supersim\'etricos os 
  ``{\sc sl\'eptons}" e ``{\sc squarks}", que s\~ao part\'\i culas escalares. 
  Como temos que ter um parceiro supersim\'etrico para cada grau de liberdade, 
  dois campos bos\^onicos s\~ao necess\'arios para cada campo fermi\^onico do 
  {\bf SM}. Eles s\~ao conhecidos como estados ``esquerdo" e ``direito" e s\~ao 
  denotados por $\tilde{l}_L$, $\tilde{l}_R$, $\tilde{q}_L$ e $\tilde{q}_R$. 
  Devemos aqui ressaltar que $L$ e $R$ n\~ao significa a helicidade das 
  spart\'\i culas (eles s\~ao part\'\i culas de spin$0$) mas a dos seus
  superparceiros 
  
  \item J\'a no caso dos dois dubletos de Higgs temos que acresentar seus
  parceiros que s\~ao f\'ermions e s\~ao conhecidos por {\sc higgsinos}.
\end{enumerate}      
   \item Termos de quebra {\em soft} para parametrizar a quebra de 
supersimetria.
   \item Simetria R uma simetria discreta exata.
\end{enumerate}

Se constru\'\i ssemos uma teoria baseada apenas nos tr\^es primeiros pontos 
descritos acima, surgiria uma teoria que viola n\'umero lept\^onico e 
bari\^onico, ent\~ao devido a isto temos que exigir uma simetria discreta a 
mais.

A simetria R \'e introduzida para eliminar estes termos. A paridade R de um 
estado est\'a relacionado ao seu spin ($S$), n\'umero bari\^onico ($B$) e ao 
n\'umero lept\^onico ($L$) da seguinte maneira
\BEA
    R_{p} &=& \LP(-1\RP)^{2S+3B+L}.
\EEA

Uma consequ\^encia imediata da express\~ao acima \'e que todas as 
part\'\i culas do {\bf SM} (incluindo os b\'osons de Higgs) s\~ao 
de paridade R par, enquanto seus superparceiros s\~ao de paridade R 
\'\i mpar. Como resultado as ``novas" part\'\i culas supersim\'etricas podem 
ser produzidas apenas em pares, e qualquer de seu decaimento tem que conter um 
n\'umero \'\i mpar de part\'\i culas supersim\'etricas. Isto implica que a
part\'\i cula supersim\'etrica mais leve, conhecida como 
``Lighest Supersymmetric Particle" ({\em LSP}), tem que ser
est\'avel, pois ela n\~ao tem canal de decaimento permitido. Isto proporciona 
sinais caracter\'\i sticos para a produ\c c\~ao das spart\'\i culas. O 
argumento \'e o seguinte. Como os {\em LSP} s\~ao est\'aveis, alguns deles devem 
ter sobrevividos a era do Big Bang. Se os {\em LSP} sentem as intera\c c\~oes 
eletromagn\'eticas ou fortes, muitos ou a maioria dessas rel\'\i quias 
cosmol\'ogicas devem ter sido confinados para formar ``is\'otopos ex\'oticos". 
Buscas \cite{L1} a esses ``ex\'oticos" levam a v\'\i nculos muito fortes para 
sua abund\^ancia, que exclue todos os modelos com part\'\i culas est\'aveis 
carregadas ou mesmo que interagem fortemente a menos que suas massas 
ultrapasse v\'arios $TeV$ \cite{L2}. Assim no contexto do {\bf MSSM} isto 
significa que o {\em LSP} deve ser neutro, logo o {\em LSP} se parece como um 
neutrino pesado nas buscas experimentais.

Supersimetria (SUSY) em sua vers\~ao local inclue a gravidade: a teoria
resultante \'e conhecida como {\sf supergravidade}. O modelo ent\~ao inclue 
tamb\'em o gr\'aviton (spin--2) e seu parceiro fermi\^onico o gravitino 
(spin--$\textstyle \frac{3}{2}$)

\subsection{Supercampos do {\bf MSSM}.}

As primeiras vers\~oes do {\bf MSSM} foram constru\'\i das nos 
anos 70 \cite{11}. 
Neste trabalho iremos
promover todos os campos fermi\^onicos do {\bf SM} a supercampos quirais, um
para cada gera\c c\~ao. Iremos representar estes supercampos por
$\hat{l}(x,\t,\tb)$, no caso do l\'epton carregado, e $\hat{\nu}_{l}(x,\t,\tb)$
para o neutrino deste l\'epton e conven\c c\~ao an\'aloga para as outras 
part\'\i culas. Os \'\i ndices das gera\c c\~oes ser\~ao
suprimidos \footnote{Aqui iremos seguir a conven\c c\~ao padr\~ao de que todo
supermultipleto quiral ser\~ao definidos em termos de espinores de Weyl 
de m\~ao esquerda, assim que os conjugados dos l\'eptons de m\~ao esquerda 
ir\~ao aparecer para representar os l\'eptons de m\~ao
direita. Soma sobre o \'\i ndice de gera\c c\~ao est\'a subentendida por todo
este estudo se nada for dito para indicar o contr\'ario.}.  

Os supercampos dos l\'eptons de m\~ao esquerda (para cada 
gera\c c\~ao) s\~ao \footnote{Usaremos chapeu ($\hat{}$) para indicar
supercampos.} 
\BEA
   \hat{L}(x,\t,\tb) &=&
      \LP( \BA{c} \hat{\nu}_{l}(x,\t,\tb) \\
                  \hat{e}(x,\t,\tb)          \EA \RP)_{L} \,\ ,
             \nonumber \\
   \hat{R}(x,\t,\tb) &=& \hat{e}_{R}(x,\t,\tb) \,\ ,
\EEA
j\'a os supercampos dos quarks s\~ao escritos da seguinte maneira
\BEA
   \hat{Q}(x,\t,\tb) &=&
      \LP( \BA{c} \hat{u}_{i}(x,\t,\tb) \\
                  \hat{d}_{i}(x,\t,\tb)          \EA \RP)_{L} \,\ ,
             \nonumber \\
   \hat{U}(x,\t,\tb) &=& \hat{u}_{R}(x,\t,\tb) \,\ , \nonumber \\
   \hat{D}(x,\t,\tb) &=& \hat{d}_{R}(x,\t,\tb) \,\ .
\EEA

J\'a no setor de Higgs deste modelo necessitamos de dois dubletos
\footnote{Dois dubletos de higgs s\~ao necess\'arios para evitar anomalias
de gauge que surgem dos higgsinos de spin $1/2$ e para gerar massas a todos os 
quarks e l\'eptons.}, 
que definiremos como
\BEA
   \hat{H}_{1}(x,\t,\tb) &=&
       \LP( \BA{c} \hat{H}_{1}^{1}(x,\t,\tb) \\
                   \hat{H}_{1}^{2}(x,\t,\tb)     \EA \RP),
               \nonumber
\EEA
e
\BEA
   \hat{H}_{2}(x,\t,\tb)  &=&
         \LP( \BA{c} \hat{H}_{2}^{1}(x,\t,\tb) \\
                     \hat{H}_{2}^{2}(x,\t,\tb)    \EA \RP).
                \label{The Lagrangian for Supersymmetric QFD prop 5}
\EEA
Observe que os \'\i ndices superiores desses supercampos, digamos 
$\hat{H}_{1}^{2}(x,\t,\tb)$, \'e um \'\i ndice de $SU(2)$ que toma as 
seguintes s\'eries de valores $\{1,2\}$. A mesma coisa acontece para 
$\hat{L}(x,\t,\tb)$ e $\hat{Q}(x,\t,\tb)$. Os n\'umeros qu\^anticos de 
cada supercampo est\'a na tabela~\ref{Table Superfields}, na 
tabela~\ref{tab:mssm} mostramos o conte\'udo de part\'\i culas do modelo.

Como a nossa teoria possui uma invari\^ancia de gauge 
$SU(3)_C \otimes SU(2)_L \otimes U(1)_N$. 
Isto significa que a teoria tem doze campos de gauge, sendo um 
$\hat{v}(x,\t,\tb)$ para o grupo de gauge $U(1)$, tr\^es $\hat{V}^{i}(x,\t,\tb)$ 
($i=1,2,3$) para $SU(2)$ e oito $\hat{V}^{a}_c(x,\t,\tb)$ 
($i=1, \cdots, 8$) para $SU(3)$. Como no {\bf SM} colocaremos este b\'osons de 
gauge na representa\c c\~ao adjunta do grupo, e usaremos a seguinte 
nota\c c\~ao
\BEA
   \hat{V}'(x,\t,\tb)       &=& {\bf Y} \hat{v}(x,\t,\tb) \,\ , \SL
           \nonumber \\
   \hat{V}(x,\t,\tb)    &=& {\bf T}^{i} \hat{V}^{i}(x,\t,\tb) \,\ ,
                  \hspace{1cm}       i = 1,2,3 \,\ , \SL
           \nonumber \\
   \hat{V}_c(x,\t,\tb)    &=& {\bf T}^{a} \hat{V}^{a}_c(x,\t,\tb) \,\ ,
                  \hspace{1cm}       a = 1, \cdots, 8 \,\ ..
           \label{The Lagrangian for Supersymmetric QFD prop 7}
\EEA
Onde ${\bf Y}$ e ${\bf T}^{i} \equiv {\bf \sigma^{i}/2}$, 
${\bf T}^{a} \equiv {\bf \lambda^{a}/2}$,  s\~ao os geradores de
$U(1)$, $SU(2)$ e $SU(3)$ respectivamente.

\begin{table}[tbh]
 \BC
 \begin{tabular}{|l|c|c|r|} \hline
  Tipo de        &  Supercampos
       & \multicolumn{2}{|c|}{N\'umeros Qu\^anticos} \\ \cline{3-4}
  Multipleto             &                               &  $SU(2)$    & $U(1)$
                 \\ \hline \hline

  Mat\'eria           &  $\hat{L}(x,\t,\tb)$          &   dubleto   &  $-1/2$ \\
                   &  $\hat{R}(x,\t,\tb)$          &   singleto     &  $1$ \\
		   &  $\hat{Q}(x,\t,\tb)$          &   dubleto   &  $1/6$ \\
                   &  $\hat{U}(x,\t,\tb)$          &   singleto     &  $-2/3$ \\
		   &  $\hat{D}(x,\t,\tb)$          &   singleto     &  $1/3$ \\
                   &  $\hat{H}_{1}(x,\t,\tb) $     &   dubleto      &  $-1/2$  \\
                   &  $\hat{H}_{2}(x,\t,\tb) $     &   dubleto      &  $1/2$
                 \\  \hline
  Gauge            &  $\hat{v}(x,\t,\tb)$         &  singleto      &  0 \\
                   &  $\hat{V}^{i}(x,\t,\tb)$      &  tripleto      &  0 \\
		   &  $\hat{V}^{a}_{c}(x,\t,\tb)$      &  octeto      &  0
                  \\ \hline
  \end{tabular}
     \caption{A nota\c c\~ao e os n\'umeros qu\^anticos de cada supercampo 
no {\bf MSSM}. Todos os \'\i ndices das gera\c c\~oes foram suprimidos.}
     \label{Table Superfields}
 \EC
\end{table}

\begin{table}[t] \center
\renewcommand{\arraystretch}{1.5}
\begin{tabular}{|l|cc|cc|}
\hline
Supercampo & Part\'\i cula & Spin & Superparceiro & Spin \\
\hline
\hline
\quad $\hat{v}$ (U(1)) & $V_m$   & 1 & $\lambda_{B}\,\,$ & $\frac{1}{2}$ \\
\quad $\hat{V}^i$ (SU(2)) & $V^i_m$ & 1 & $\lambda^i_A$  & $\frac{1}{2}$ \\
\quad $\hat{V}^a_c (SU(3))$ & $G^a_m$ & 1 & $\tilde{ g}^a$ & $\frac{1}{2}$ \\
\hline
\quad $\hat{Q}$   & $(u,\,d)_L^{}$ & $\frac{1}{2}$ & 
$(\tilde{ u}_L^{},\,\tilde{ d}_L^{})$ & 0 \\
\quad $\hat{U}$ & $\bar u_R^{}$  & $\frac{1}{2}$ & $\tilde{ u}_R^*$ & 0 \\
\quad $\hat{D}$ & $\bar d_R^{}$  & $\frac{1}{2}$ & $\tilde{ d}_R^*$ & 0 \\
\hline
\quad $\hat{L}$   & $(\nu,\,e)_L^{}$ & $\frac{1}{2}$ & 
$(\tilde{ \nu}_L^{},\,\tilde{ e}_L^{})$ & 0 \\
\quad $\hat{R}$ & $\bar e_R^{}$    & $\frac{1}{2}$ & $\tilde{ e}_R^*$ & 0 \\
\hline
\quad $\hat{H}_1^{}$ & $(H_1^0,\, H_1^-)$ & 0 
    & $(\tilde{ H}_1^0,\, \tilde{ H}_1^-)$    & $\frac{1}{2}$ \\
\quad $\hat{H}_2^{}$ & $(H_2^+,\, H_2^0)$ & 0 
    & $(\tilde{ H}_2^+,\, \tilde{ H}_2^0)$    & $\frac{1}{2}$ \\
\hline
\end{tabular}
\renewcommand{\arraystretch}{1}
\caption{Conte\'udo de Part\'\i culas do MSSM.}
\label{tab:mssm}
\end{table}

\subsection{Lagrangiana do {\bf MSSM}.}

Na lagrangiana que iremos construir, iremos assumir que a nossa teoria pode
ser  vista como sendo um limite de baixa energia de uma teoria de 
supergravidade, ou seja ela deve  ser invariante por supersimetria e pelo 
grupo de gauge. Na se\c c\~ao seis j\'a discutimos como deve ser a forma da 
lagrangiana para ser invariante por  supersimetria. Na pr\'oxima se\c c\~ao 
mostraremos que a lagrangiana que iremos  escrever a seguir tamb\'em \'e 
invariante pelo grupo de gauge.

Nossa lagrangiana tem que ter a seguinte forma 
\BEA
   \L_{MSSM} &=& \L_{SUSY} + \L_{soft} \,\ .
      \label{The Lagrangian for Supersymmetric QFD prop 1}
\EEA 
Onde $\L_{SUSY}$ \'e a parte supersim\'etrica da lagrangiana, enquanto 
$\L_{Soft}$ quebra explicitamente supersimetria conforme discutido na 
se\c c\~ao nove.

\subsubsection{O Termo Supersim\'etrico $\L_{SUSY}$.}

O termo supersim\'etrico iremos dividir da seguinte 
maneira
\BEA
   \L_{SUSY} &=&   \L_{lepton}
                  + \L_{Quarks}
                  + \L_{Gauge}
                  + \L_{Higgs} \,\ ,
        \label{The Supersymmetric Term prop 1}
\EEA
onde
\BEA
  \L_{lepton}
     &=& \int d^{4}\t\;\LP[\,\hat{ \bar{L}}e^{2g\hat{V}-g'\hat{V}'}
                           \hat{L} +
                           \hat{ \bar{R}}e^{g'\hat{V}'}
                           \hat{R}\,\RP] \,\ , \nonumber \\
  \L_{Quarks}
     &=& \int d^{4}\t\;\LP[\,\hat{ \bar{Q}}e^{2g_s\hat{V}^a_c+2g\hat{V}+(g'/3)\hat{V}'}
                           \hat{Q} +
                           \hat{ \bar{U}}e^{2g_s\hat{V}^a_c-(4g'/3)\hat{V}'}
                           \hat{U} +
			   \hat{ \bar{D}}e^{2g_s\hat{V}^a_c+(2g'/3)\hat{V}'}
                           \hat{D}\,\RP] \,\ , \nonumber \\
        \label{The Supersymmetric Term prop 2}  \SL	
  \L_{Gauge}
     &=&  \f{1}{4} \int  d^{2}\t\;
         \LP[ \sum_{a=1}^{8} W^{a \alpha}_{s}W_{s \alpha}^{a}+
	 \sum_{i=1}^{3} W^{i \alpha}W_{ \alpha}^{i}+ 
 W^{ \prime \alpha}W_{ \alpha}^{ \prime}\,
         + h.c. \RP]\,,
        \label{The Supersymmetric Term prop 3}
\EEA
e finalmente
\BEA
  \L_{Higgs}
     &=&  \int d^{4}\t\;\LP[\,\hat{ \bar{H}}_{1}e^{2g\hat{V}-g'\hat{V}'}     
                      \hat{H}_{1}                          +
\hat{ \bar{H}}_{2}e^{2g\hat{V}+g'\hat{V}'}                           
\hat{H}_{2}                +  W+ \bar{W} \RP]\!.\hspace{2mm}
        \label{The Supersymmetric Term prop 4}
\EEA
Aqui $g_s$, $g$ e $g'$ s\~ao os acoplamentos de gauge para $SU(3)$, $SU(2)$
e $U(1)$ respectivamente e o fator $2$ que aparece relacionado com a constante
de  acoplamento $g$, s\~ao introduzido por conveni\^encia. Com esta escolha o 
campo de for\c ca $V_{mn}^{i}$ contido em $\LS{W}{\a}$ corresponde ao do 
{\bf SM}. $\LS{W}{s \a}$, $\LS{W}{\a}$ e $\LS{W}{\a}'$ s\~ao os $SU(3)$, $SU(2)$ e
$U(1)$ campos de for\c cas definidos por
\BEA
   \LS{W^a}{s \a}     &=&  -\f{1}{8g_s}\,\bar{D}\bar{D}e^{-2g_s\hat{V}_c}
   \LS{D}{\a}e^{2g\hat{V}_c} \,\ ,
           \nonumber \\
   \LS{W}{\a}     &=&  -\f{1}{8g}\,\bar{D}\bar{D}e^{-2g\hat{V}}\LS{D}{\a}
 e^{2g\hat{V}} \,\ ,
           \nonumber \\
   \LS{W}{\a}'    &=&  -\f{1}{4}\,D D \bar{D}_{\a} \hat{V}' \,\ .
\EEA
Al\'em disto,
$W \equiv W[\hat{L},\hat{R},\hat{Q},\hat{U},\hat{D},\hat{H}_{1},\hat{H}_{2}]$
\'e o superpotencial da teoria que iremos discutir a 
seguir.

\subsubsection{O Superpotencial.}

Lembraremos que o superpotencial pode ser no m\'aximo c\'ubico nos supercampos
para garantir que a teoria seja renormaliz\'avel. O superpotencial do 
{\bf MSSM} tem a seguinte forma 
\BEA
    W &=&  W_{H} + W_{Y}+W_{NR} \,\ , \label{caos}
\EEA
com a parte do Higgs $W_{H}$ dada por
\BEA
   W_{H} &=& \mu\; \e_{\alpha \beta}\hat{H}_{1}^{i}\hat{H}_{2}^{j} \,\ , \nonumber
\EEA
e a correspondente parte de Yukawa $W_{Y}$ tem a seguinte forma 
\BEA
     W_{Y}[\hat{L},\hat{R},\hat{Q},\hat{U},\hat{D},\hat{H}_{1}, \hat{H}_{2}]
             &=& \e_{\alpha \beta}\LP[\,f\hat{H}^{i}_{1}\hat{L}^{j}\hat{R}
               +f_{1}\hat{H}^{i}_{1}\hat{Q}^{j}\hat{D}
               +f_{2}\hat{H}^{j}_{2}\hat{Q}^{i}\hat{U}\,\RP] \,\ . \nonumber
\EEA
Onde  $\mu$ \'e um par\^ametro de massa e $\e_{\alpha \beta}$ \'e um tensor
anti-sim\'etrico definido por 
\BEA
  \e &=& \LP( \BA{rr}   0 & 1 \\
                       -1 & 0     \EA \RP) \,\ .
\EEA
Alem disto, $f$, $f_{1}$ e $f_{2}$ s\~ao acoplamentos de Yukawa e s\~ao 
matrizes $3 \times 3$ no espa\c co das fam\'\i lias. Ou seja precisamos de 
$H_{1}$ e $H_{2}$ para gerar massas a todos os f\'ermions carregados do modelo. 
Repare que os neutrinos permanecem sem massa.

J\'a $W_{NR}$ \'e a parte do superpotencial que n\~ao conserva 
paridade R e tem a seguinte forma
\BEA
W_{NR}[\hat{L},\hat{R},\hat{Q},\hat{U},\hat{D}, \hat{H}_{2}] 
             &=& \mu_{1} \hat{L} \hat{H}_{2}+
\lambda \epsilon_{\alpha \beta} \hat{L}^{i} \hat{L}^{j} \hat{R}+ 
\lambda^{1} \epsilon_{\alpha \beta} \hat{L}^{i} \hat{Q}^{j} \hat{D}+
\lambda^{2} \epsilon_{\alpha \beta} \hat{L}^{i} \hat{H}_{2}^{j}+
\lambda^{3} \hat{U} \hat{D} \hat{D} \,\ . \nonumber
\EEA
Os primeiros tr\^es termos, que cont\^em tr\^es supercampos, na equa\c c\~ao 
acima n\~ao conservam o n\'umero lept\^onico $L$, enquanto o segundo e o 
quarto termo n\~ao conservam o n\'umero bari\^onico $B$.
Dentro deste modelo s\'o podemos quebrar paridade R somente se qualquer 
$L$ ou $B$ n\~ao for conservado. Se ambos $L$ e $B$ forem quebrados, teremos 
r\'apidas taxas para os decaimentos dos n\'ucleons e isto \'e inaceit\'avel 
fenomenologicamente. \'E devido a isto que em modelos com quebra de paridade 
$R$, apenas alguns dos coeficientes s\~ao diferentes de zero. Devemos contudo 
enfatizar que existem fortes v\'\i nculos nos acoplamentos $\mu_{1}$, 
$\lambda$, $\lambda^{1}$, $\lambda^{2}$ e $\lambda^{3}$. Mas neste estudo 
n\'os n\~ao consideraremos quebra desta simetria.

Lembre-se que se impomos invari\^ancia sobre o grupo de gauge em 
todas as intera\c c\~oes no {\bf SM}, achamos que todas os 
termos de dimens\~ao quatro ou menor automaticamente preservam 
tanto n\'umero lept\^onico como o n\'umero bari\^onico. Aqui iremos 
considerar conserva\c c\~ao da paridade R, por isto n\~ao 
iremos tratar com este termo neste estudo.

\subsubsection{O Termo de Quebra Soft $\L_{Soft}$.}

O termo soft \'e escrito da seguinte maneira:
\BEA
   \L_{Soft} &=& \L_{SMT} + \L_{GMT}+ \L_{INT} \,\ ,
        \label{The Soft SUSY-Breaking Term prop 2aaa}
\EEA
onde o termo de massa escalar $\L_{SMT}$ \'e
 escrito da seguinte maneira
\BEA
   \L_{SMT} &=& -\int d^{4}\t\; \LP[\,
            M_{L}^{2}\;\tilde{ \bar{L}}\tilde{L}
          + M_{R}^{2}\tilde{ \bar{R}}\tilde{R}
	  + M_{Q}^{2}\;\tilde{ \bar{Q}}\tilde{Q}
	  \RP.
\nmb  \hspace{1.7cm} \LP.
          + M_{U}^{2}\tilde{ \bar{U}}\tilde{U}
	  + M_{D}^{2}\tilde{ \bar{D}}\tilde{D}
          + M_{1}^{2} \bar{H}_{1}H_{1} \RP.
\nmb  \hspace{1.7cm} \LP.
          + M_{2}^{2} \bar{H}_{2}H_{2}
        - M_{12}^{2}\e_{ij}\LP(H_{1}^{i}H_{2}^{j}+h.c.\RP) \right] \,\ ,
      \label{burro}
\EEA
mas a invari\^ancia $SU(2)$ exige mesmo par\^ametros de quebras para cada 
dubleto de m\~ao esquerda dos sf\'ermions, ou seja 
$m^{2}_{ \tilde{l}}= m^{2}_{ \tilde{ \nu}}$. Muitos
autores incluem a parte $W_{H}$ na parte de quebra soft. Isto \'e devido ao 
fato que a constante de acoplamento $\mu$ tem dimens\~ao de massa como os 
outros termos de quebra soft apresentados acima. O termo de massa do gaugino
\'e 
\BEA
   \L_{GMT} &=&-   \HA \int d^{4}\t\; \LP[
              \LP(\,M_3\; \sum_{a=1}^{8} \tilde{g}^{a} \tilde{g}^{a}
	      + M\; \sum_{i=1}^{3}\; \lambda^{i}_{A} \lambda^{i}_{A}
            + M' \;   \lambda_{B} \lambda_{B}\,\RP)
            + h.c. \RP] \,\ .
            \label{The Soft SUSY-Breaking Term prop 3}
\EEA

J\'a o termo $\L_{INT}$ \'e o seguinte
\BEA
\L_{INT} = \LP( A_E H_1\tilde{L}\tilde{R}+A_D H_1\tilde{Q}\tilde{D}+
A_U H_2\tilde{Q}\tilde{U}+h.c. \RP) \,\ .
\EEA

\subsubsection{Conclus\~ao}

Para conluir esta se\c c\~ao, vamos reunir todos os nossos resultados para a 
lagrangiana $\L_{MSSM}$, em termos dos supercampos. O resultado \'e

\BEA
  \L_{MSSM} &=&   \L_{SUSY} + \L_{Soft} \NN
      &=& \int d^{4}\t\;\LP\{\,\hat{ \bar{L}}e^{2g\hat{V}+g'\hat{V}'}
                           \hat{L} +
                           \hat{ \bar{R}}e^{g'\hat{V}'}
                           \hat{R} \RP.   \nn
      & & \mbox{}\hspace{1.3cm}
          +\hat{ \bar{H}}_{1}e^{2g\hat{V}+g'\hat{V}'}
                           \hat{H}_{1}
                         + \hat{ \bar{H}}_{2}e^{2g\hat{V}+g'\hat{V}'}
                           \hat{H}_{2}\nn
      & & \mbox{}\hspace{1.3cm}
               + \mu \e_{ij} \hat{H}_{1}^{i} \hat{H}_{2}^{j}+f \e_{ij}
               \hat{H}_{1}^{i}\hat{L}^{j}\hat{R}+h.c. \nn
      & & \mbox{}\hspace{1.3cm}
          -\LP[\,
            M_{L}^{2}\,\tilde{ \bar{L}}\tilde{L}
          + M_{R}^{2}\,\tilde{ \bar{R}}\tilde{R}
	  + M_{Q}^{2}\;\tilde{ \bar{Q}}\tilde{Q}
	  \RP.
\nmb  \hspace{1.7cm} \LP.
          + M_{U}^{2}\tilde{ \bar{U}}\tilde{U}
	  + M_{D}^{2}\tilde{ \bar{D}}\tilde{D}
          + M_{1}^{2}\, \bar{H}_{1}H_{1} \RP.
      \nmb \hspace{2.1cm}  \LP.
          + M_{2}^{2} \bar{H}_{2}H_{2}
          - M_{12}^{2}\e_{ij}\LP(H_{1}^{i}H_{2}^{j}+h.c.\RP)
             \RP.  
\nmb \hspace{1.7cm} \left.
	  + \LP( A_E H_1\tilde{L}\tilde{R}+A_D H_1\tilde{Q}\tilde{D}+
A_U H_2\tilde{Q}\tilde{U}+h.c. \RP) \right]
          \nmb\hspace{1.3cm} \LP.
       - \HA\LP[\LP(\,M_3\, \sum_{a=1}^{8} \tilde{g}^{a} \tilde{g}^{a}
       +M\, \sum_{i=1}^{3} 
       \lambda^{i}_{A} \lambda^{i}_{A}
            + M' \,   \lambda_{B} \lambda_{B}\,\RP)
            + h.c.\,\RP] \RP \} \nonumber \\ 
	    & & \mbox{} \hspace{1.3cm}
          + \int d^{2} \theta \f{1}{4}
         \LP[\LP(\, \sum_{a=1}^{8} W^{a \alpha}_{s}W_{s \alpha}^{a}+
	 \sum_{i=1}^{3}
                W^{i \alpha}W^{i}_{ \alpha}+
                W^{ \prime \alpha}W^{ \prime}_{ \alpha} 
              \RP) + h.c.\,\RP]  \,\ . \nonumber \\
\EEA

\section{A Invari\^ancia de Gauge de $\L_{MSSM}$.}

As transforma\c c\~oes de gauge do supercampos s\~ao definidas por
\BEA \LP. \BA{lclcl}
  \Phi'(x,\t,\tb) &=& e^{-ig\Lambda(x,\t,\tb)}\Phi(x,\t,\tb),
  \hspace{0.5cm} & & \bar{D}_{ \dot{ \alpha}}\Lambda = 0  \\*[2mm]
  \bar{\Phi}^{'}(x,\t,\tb) &=& \bar{ \Phi}(x,\t,\tb)e^{ig
   \bar{ \Lambda}(x,\t,\tb)}
   \hspace{0.5cm}& & \LS{D}{\a}\bar{ \Lambda} = 0 \\*[2mm]
   e^{gV'} &=& e^{-ig\bar{\Lambda}}\,e^{gV}\,
         e^{ig\Lambda}  & & \EA \RP\} \,\ .
         \label{Non-Abelian Gauge Transformations prop 6}
\EEA
e as do campo de for\c ca $\LS{W}{\a}$ por
\BEA
   \LS{W}{\a} \rightarrow \LS{W}{\a}'
    &=&  e^{-ig\Lambda}\,\LS{W}{\a}\,e^{ig\Lambda} \,\ .
       \label{Supersymmetric Field Strengths prop 8}
\EEA

Vamos come\c car mostrando a invari\^ancia por $SU(2)$ e depois mostraremos a 
invari\^ancia por $U(1)$. Qualquer outro termo a mais adicionada a lagrangiana 
discutida na \'ultima se\c c\~ao ou n\~ao \'e invariante por $SU(2)$ ou por 
$U(1)$.

\subsection{A Invari\^ancia por $SU(2)$.}

Como $[\hat{V},\hat{V}']=[\hat{\La},\hat{V}']=0$, o termo 
$\hat{ \bar{L}}e^{2g\hat{V}+g'\hat{V}'}\hat{L}$ pode ser mostrado que \'e 
invariante por $SU(2)$, da seguinte maneira
\BEA
  \hat{ \bar{L}}e^{2g \hat{V}}e^{g'\hat{V}^{ \prime}} \hat{L}
 & \longrightarrow&
    \hat{ \bar{L}} e^{2ig \hat{ \bar{ \Lambda}}}
    e^{-2ig \hat{ \bar{ \Lambda}}}
    e^{2g \hat{V}} e^{2ig \hat{ \Lambda}}
    e^{g'\hat{V}'}  e^{-2ig \hat{ \Lambda}}
    \hat{L}  \nn
   &=& \hat{ \bar{L}}e^{2g \hat{V}+g'\hat{V}'}\hat{L} \,\ .
\EEA
A invari\^ancia dos termos cin\'eticos de $\hat{R}$,
$\hat{H}_{1}$ or $\hat{H}_{2}$ s\~ao mostradas da mesma maneira.

J\'a a invari\^ancia do termo cin\'etico dos b\'osons de gauge \'e mostrado da
seguinte maneira 
\BEA
   W^{\a\;a}W_{\a}^{a} = Tr \LP(W^{\a}W_{\a}\RP)
      &\longrightarrow& Tr \LP(
          e^{-2ig\hat{\La}} W^{\a} e^{2ig\hat{\La}}
          e^{-2ig\hat{\La}} W_{\a} e^{2ig\hat{\La}} \RP)\nn
      &=& Tr \LP(W^{\a}W_{\a}\RP)\nn
      &=& W^{\a\;a}W_{\a}^{a} \,\ .
\EEA
Onde utilizamos a propiedade cicl\'\i ca do tra\c co. A invari\^ancia de 
$W^{'\,\a}W_{\a}'$ \'e trivial pois $W_{\a}'$ \'e um singleto sobre este grupo.

Para demonstrarmos a invari\^ancia do superpotencial W, n\'os come\c caremos por 
$W_{H}$,
\BEA
  W_{H} = \mu\e^{ij}\;\hat{H}_{1}^{i}\hat{H}_{2}^{j}
     &\longrightarrow&
       \mu \e^{ij}\; \LP[e^{-2ig\hat{\La}}\hat{H}_{1}\RP]^{i}\,
                     \LP[e^{-2ig\hat{\La}}\hat{H}_{2}\RP]^{j}
                     \hspace{0.4cm} i,j = 1,2 \nn
     &=& \mu \e^{ij}\,{\cal U}^{ik}{\cal U}^{jl} \;
              \hat{H}_{1}^{k}\hat{H}_{2}^{l} \,\ ,
              \hspace{1.4cm} {\cal U} = e^{-2ig\hat{\La}} \,\ . \nonumber \\
\EEA
Para que $W_{H}$ seja invariante devemos ter
\BEA
      \e^{kl} &=& \e^{ij}\,{\cal U}^{ik}{\cal U}^{jl} \,\ .
\EEA
Esta rela\c c\~ao na verdade \'e v\'alida como iremos mostrar agora.
A matriz ${\cal U}= e^{-2ig\hat{\La}}$ \'e obviamente uma matriz
$2\times 2$, e seu determinante \'e
\BEA
   \det {\cal U} =  e^{-2igTr\LP(\hat{\La}\RP)} = 1 \,\ ,
\EEA
como $Tr\LP(\hat{\La}\RP)\equiv Tr\LP(T^{a}\hat{\La}^{a}\RP) = 0$.
Portanto ${\cal U}$ \'e uma matriz de $SU(2)$.
Logo ${\cal U}$, como qualquer matriz de $SU(2)$, pode ser escrita como
\BEA
    {\cal U} &=& \LP( \BA{cc}
          \hat{A}               &   \hat{B} \\
          - \hat{ \bar{B}}   &   \hat{ \bar{A}} \EA \RP) \,\ ,
\EEA
com
\BEA
    \hat{ \bar{A}}\hat{A}+\hat{ \bar{B}}\hat{B} &=& 1 \,\ .
\EEA
Onde $\hat{A}$ e $\hat{B}$ s\~ao funcionais dos supercampos quirais
$\hat{\La}^{a}$.  Sua depend\^encia nestes supercampos n\~ao tem nenhuma
import\^ancia para o que faremos a seguir, por isto n\~ao iremos nos 
preocupar com esta depend\^encia.

Portanto
\BEA
   \e^{ij}\,{\cal U}^{ik}{\cal U}^{jl}
        &=& \LP[{\cal U}^{T}\e\,{\cal U}\RP]^{kl} \nn
        &=& \LP(\BA{cc}
               0       & \hat{ \bar{A}}\hat{A}+\hat{ \bar{B}}\hat{B} \\
               -\LP(\hat{ \bar{A}}\hat{A}+\hat{ \hat{B}}\hat{B}\RP) & 0
             \EA \RP)^{kl} \nn
        &=& \LP( \BA{rr}
              0  & 1 \\
              -1 & 0     \EA \RP)^{kl}\nn
        &=& \e^{kl} \,\ ,
\EEA
e $W_{H}$ \'e invariante de gauge sobre $SU(2)$.

A invari\^ancia de $W_{Y}$ \'e mostrada de maneira an\'aloga pois
$\hat{H}_{1}$ e $\hat{L}$ s\~ao ambos dubletos sob SU(2),
enquanto que $\hat{R}$ \'e um singleto sob este grupo.
Assim o superpotencial
$W= W_{H}+W_{Y}$ \'e invariante sob $SU(2)$.

Assim a lagrangiana total $\L_{MSSM}$ \'e invariante sob SU(2) como deveria de
ser.

\subsection{A Invari\^ancia $U(1)$.}

Muitas das invari\^ancias mostradas acima s\~ao facilmente generalizadas para
$U(1)$ com as substitui\c c\~oes $2g\rightarrow g'$,
$T^{a}\hat{\La}^{a}\rightarrow Y\hat{\la}'=\hat{\La}'$.

A \'unica invari\^ancia que \'e necess\'aria checar novamente, \'e a que 
cont\'em dois ou
mais supercampos quirais. Tais termos aparecem apenas no superpotencial W, e a
invari\^ancia \'e provada da seguinte maneira
\BEA
   W_{H} = \mu\e^{ij}\;\hat{H}_{1}^{i}\hat{H}_{2}^{j}
         &\longrightarrow &
                 \mu\e^{ij}\; e^{-ig'\LP(Y_{H_{1}}+Y_{H_{2}}\RP) \hat{\la}'}\,
                 \hat{H}_{1}^{i}\hat{H}_{2}^{j} \nn
          &=&    W_{H} \,\ ,
\EEA
e
\BEA
   W_{Y} =
        f\e^{ij}\;\hat{H}_{1}^{i}\hat{L}^{j}\hat{R}
      &\longrightarrow &
           f\e^{ij}\; e^{-ig'\LP(Y_{H_{1}}+Y_{L}+Y_{R}\RP)\hat{\la}'}
            \hat{H}_{1}^{i}\hat{L}^{j}\hat{R} \nn
      &=&   W_{Y} \,\ ,
\EEA
pois $Y_{H_{1}}+Y_{H_{2}} = 0$ e $Y_{H_{1}}+Y_{L}+Y_{R} = 0$
de acordo com a Tab.~\ref{Table Superfields}. Portanto nossa teoria tamb\'em
\'e invariante por $U(1)$.

Isto completa a prova de toda a invari\^ancia da teoria pelo grupo 
$SU(2) \otimes U(1)$.

\section{Expans\~ao em Componentes}

Aqui nesta se\c c\~ao vamos abrir em componentes todos os supercampos e a
nossa lagrangiana.

A nota\c c\~ao utilizada em todo este estudo \'e chapeu ($\hat{\, \,}$) 
para indicar supercampos enquanto  til 
($\tilde{\, \,}$) representa o parceiro supersim\'etrico das part\'\i culas do
{\bf SM}. Os sub-\'\i ndices L and R nos campos fermi\^onicos, significa campo
de m\~ao esquerda e direita respectivamente. Quando estes 
sub-\'\i ndices aparecem em um campo bos\^onico, por
exemplo em $\tilde{L}_{L}$, ele apenas indica um campo particular e n\~ao tem
nada haver com campo de m\~ao esquerda ou direita.

\subsection{Supercampos.}

Na se\c c\~ao anterior, n\'os arranjamos um dos supercampos para estar 
na representa\c c\~ao 
dubleto($\hat{L}$) de $SU(2)$ e o outro em um singleto ($\hat{R}$). 
Estes supercampos quirais t\^em a seguinte expans\~ao em componentes 
\footnote{Usamos til ($\tilde{}$) para indicar os parceiros supersim\'etricos 
das part\'\i culas usuais do SM.}
\BEA
  \hat{L}(x,\t,\tb) &=&  \LP( \BA{l}   \hat{\nu}_{l}(x,\t,\tb) \\
                                       \hat{l}(x,\t,\tb)   \EA \RP)_{L} \nn
             &=&    \tilde{L}(x)
                 +  i\;\TSTB \;\P_{m}\tilde{L}(x)
+ \f{1}{4}\;\t\t\;\tb\tb\;\P^{m}\P_{m}\tilde{L}(x) \nn
             & & \mbox{}
                 +  \sqrt{2}\;\t L(x)
                 +  \f{i}{\sqrt{2}}\;\t\t\;\tb\bar{\s}^{m}\P_{m}L(x)
                 +  \t\t\;F_{L}(x) \,\ ,
                 \label{Component Field Expansion prop 1} \SL
  \hat{R}(x,\t,\tb)
      &=&     \hat{l}_{R}(x) \nn
       &=&    \tilde{R}(x)
            + i\,\TSTB\;\P_{m} \tilde{R}(x)
            + \f{1}{4} \; \t\t\;\tb\tb\;\P^{m}\P_{m} \tilde{R}(x)\nn
       & & \mbox{}
            + \sqrt{2}\;\t R(x)
            + \f{i}{\sqrt{2}}\;\t\t\,\tb\bar{\s}^{m}\P_{m}R(x)
            + \t\t\;F_{R}(x) \,\ .
            \label{Component Field Expansion prop 2}
\EEA
para os quarks teremos
\BEA
  \hat{Q}(x,\t,\tb) &=&  \LP( \BA{l}   \hat{u}(x,\t,\tb) \\
                                       \hat{d}(x,\t,\tb)   \EA \RP)_{L} \nn
             &=&    \tilde{Q}(x)
                 +  i\;\TSTB \;\P_{m}\tilde{Q}(x)
+ \f{1}{4}\;\t\t\;\tb\tb\;\P^{m}\P_{m}\tilde{Q}(x) \nn
             & & \mbox{}
                 +  \sqrt{2}\;\t Q(x)
                 +  \f{i}{\sqrt{2}}\;\t\t\;\tb\bar{\s}^{m}\P_{m}Q(x)
                 +  \t\t\;F_{Q}(x) \,\ ,
                 \label{Component QField Expansion prop 1} \SL
  \hat{U}(x,\t,\tb)
      &=&     \hat{u}_{R}(x) \nn
       &=&    \tilde{U}(x)
            + i\,\TSTB\;\P_{m} \tilde{U}(x)
            + \f{1}{4} \; \t\t\;\tb\tb\;\P^{m}\P_{m} \tilde{U}(x)\nn
       & & \mbox{}
            + \sqrt{2}\;\t U(x)
            + \f{i}{\sqrt{2}}\;\t\t\,\tb\bar{\s}^{m}\P_{m}U(x)
            + \t\t\;F_{U}(x) \,\ \SL
	\hat{D}(x,\t,\tb)
      &=&     \hat{d}_{R}(x) \nn
       &=&    \tilde{D}(x)
            + i\,\TSTB\;\P_{m} \tilde{D}(x)
            + \f{1}{4} \; \t\t\;\tb\tb\;\P^{m}\P_{m} \tilde{D}(x)\nn
       & & \mbox{}
            + \sqrt{2}\;\t D(x)
            + \f{i}{\sqrt{2}}\;\t\t\,\tb\bar{\s}^{m}\P_{m}D(x)
            + \t\t\;F_{D}(x) \,\ .
            \label{Component QField Expansion prop 2}
\EEA
\begin{table}[tbh]
 \BC
 \begin{tabular}{|l|c|r|r|} \hline
  Nome do Campo       & S\'\i mbolo              &  Spin   &  Carga  \\ \hline
\hline

  L\'eptons         & $L^{1}   $              &  $1/2$  &    0      \\
                  & $L^{2}   $              &  $1/2$  &  $-1$     \\
                  & $R       $              &  $1/2$  &   $1$     \\
  Sl\'eptons        & $\tilde{L}^{1}$     &    $0$  &   $0$     \\
                  & $\tilde{L}^{2}$     &    $0$  &  $-1$     \\
                  & $\tilde{R}$         &    $0$  &  $ 1$     \\
    Quarks         & $Q^{1}   $              &  $1/2$  &    2/3      \\
                  & $Q^{2}   $              &  $1/2$  &  $-1/3$     \\
                  & $U       $              &  $1/2$  &   $-2/3$     \\
		  & $D       $              &  $1/2$  &   $1/3$     \\
    SQuarks         & $\tilde{Q}^{1}   $          &  $1/2$  &    2/3      \\
                  & $\tilde{Q}^{2}   $            &  $1/2$  &  $-1/3$     \\
                  & $\tilde{U}       $            &  $1/2$  &   $-2/3$     \\
		  & $\tilde{D}       $            &  $1/2$  &   $1/3$     \\
  B\'osons de Higgs & $H^{1}_{1}$         &    $0$  &  $ 0$     \\
                  & $H^{1}_{2}$         &    $0$  &  $-1$     \\
                  & $H^{2}_{1}$         &    $0$  &   $1$     \\
                  & $H^{2}_{2}$         &    $0$  &   $0$     \\
  Higgsinos       & $\psi_{H_{1}}^{1}$  &  $1/2$  &   $0$     \\
                  & $\psi_{H_{1}}^{2}$  &  $1/2$  &  $-1$     \\
                  & $\psi_{H_{2}}^{1}$  &  $1/2$  &   $1$     \\
                  & $\psi_{H_{2}}^{2}$  &  $1/2$  &   $0$     \\
  B\'osons de Gauge & $V^{i}_{m}$       &    $1$  &    --     \\
                  & $V_{m}$          &    $1$  &    --     \\
  Gauginos        & $\la^{i}_{A}$           &  $1/2$  &    --     \\
                  & $\la_{B}$              &  $1/2$  &    --     \\ \hline
\end{tabular}
\EC
  \caption{Um resumo de todos os campos do {\bf SM} e seus superparceiros no
{\bf MSSM}. Os n\'umeros qu\^anticos dos v\'arios campos tamb\'em est\~ao
listados. Todos os campos fermi\^onicos est\~ao dados em termos de espinores
de Weyl de duas componentes.}            
\label{Table:  Total Component fields}
\end{table}
\vspace{0.5cm}

As componentes dos campos est\~ao definidas por
\BEA
  \BA{lllll}
     \tilde{L}(x) = \LP( \BA{c}  \tilde{\nu}(x) \\ \tilde{e}_{L}(x)\EA
\RP)
             & \hspace{0.4cm} &
     L(x)   = \LP( \BA{c}  \nu(x) \\ e(x) \EA \RP)_{L}
            &   \hspace{0.4cm} &
     F_{L}(x) = \LP( \BA{c}  f^{\nu}_{L}(x) \\ f^{l}_{L}(x) \EA \RP) \,\ , 
     \EA \nonumber \\
  \label{Component Field Expansion prop 3}
\EEA
e\footnote{A rela\c c\~ao $\tilde{ \bar{R}}$  e $\tilde{ \bar{L}}$ criam
sl\'eptons de carga oposta.}
\BEA
  \BA{lllll}
      \tilde{R}(x) = \tilde{ e}^{*}_{R}(x)
            & \hspace{0.6cm} &
     R(x)   = \bar{e}_{R}(x)
            &   \hspace{0.6cm} &
     F_{R}(x) = f^{l}_{R}(x) \,\ .
    \EA
  \label{Component Field Expansion prop 4}
\EEA
j\'a para os quarks
\BEA
  \BA{lllll}
     \tilde{Q}(x) = \LP( \BA{c}  \tilde{u}_{L}(x) \\ \tilde{d}_{L}(x)\EA
\RP)
             & \hspace{0.4cm} &
     Q(x)   = \LP( \BA{c}  u(x) \\ d(x) \EA \RP)_{L}
            &   \hspace{0.4cm} &
     F_{Q}(x) = \LP( \BA{c}  f^{u}_{L}(x) \\ f^{d}_{L}(x) \EA \RP) \,\ , 
     \EA \nonumber \\
  \label{Component QField Expansion prop 3}
\EEA
e
\BEA
  \BA{lllll}
      \tilde{U}(x) = \tilde{ u}^{*}_{R}(x)
            & \hspace{0.6cm} &
     U(x)   = \bar{u}_{R}(x)
            &   \hspace{0.6cm} &
     F_{U}(x) = f^{u}_{R}(x) \,\ .
    \EA
  \label{Component QField Expansion prop 4}
\EEA
e
\BEA
  \BA{lllll}
      \tilde{D}(x) = \tilde{ d}^{*}_{R}(x)
            & \hspace{0.6cm} &
     D(x)   = \bar{d}_{R}(x)
            &   \hspace{0.6cm} &
     F_{D}(x) = f^{d}_{R}(x) \,\ .
    \EA
  \label{Component QField Expansion prop 4}
\EEA

\begin{table}[tbh]
 \BC
 \begin{tabular}{|l|c|r|r|} \hline
  Nome do Campo       & S\'\i mbolo              &  Spin   &  Carga  \\ \hline
\hline

Campo Auxiliar do L\'epton   & $f^{\nu}_{L}$    &    $0$  &   $0$     \\
                         & $f^{l}_{L}$  &    $0$  &  $-1$     \\
                         & $f^{l}_{R}$  &    $0$  &   $1$     \\
Campo Auxiliar do Quark   & $f^{u}_{L}$    &    $0$  &   $2/3$     \\
                         & $f^{d}_{L}$  &    $0$  &  $-1/3$     \\
                         & $f^{u}_{R}$  &    $0$  &   $-2/3$     \\
			 & $f^{d}_{R}$  &    $0$  &   $1/3$     \\
Campo Auxiliar do Higgs   & $f_{1}^{1}$  &    $0$  &   $0$     \\
                         & $f_{1}^{2}$  &    $0$  &  $-1$     \\
                         & $f_{2}^{1}$  &    $0$  &   $1$     \\
                         & $f_{2}^{2}$  &    $0$  &   $0$     \\
Campo Auxiliar de Gauge   & $D^{i}$      &    $1$  &   --     \\
                         & $D$         &    $1$  &   --     \\  \hline
\end{tabular}
\EC
  \caption{Os campos auxiliares deste modelo e seus n\'umeros qu\^anticos.}
          \label{Table:  Auxiliary Component fields}
\end{table}
\vspace{0.7cm}
Para os dois dubletos dos Higgs temos
\BEA
  \hat{H}_{1}(x,\t,\tb) &=& \LP( \BA{c}   \hat{H}_{1}^{1}(x,\t,\tb) \\
                                          \hat{H}_{1}^{2}(x,\t,\tb) \EA \RP)\nn
             &=&    H_{1}(x)
                 +  i\;\TSTB \;\P_{m}H_{1}(x)
                 +  \f{1}{4}\;\t\t\;\tb\tb\;\P^{m}\P_{m}H_{1}(x) \nn
             & & \mbox{}
                 +  \sqrt{2}\;\t \tilde{H}_{1}(x)
                 +  \f{i}{\sqrt{2}}\;\t\t\;\tb\bar{\s}^{m}
                           \P_{m}\tilde{H}_{1}(x)
                 +  \t\t\;F_{1}(x) \,\ , \nonumber \\
                 \label{Component Field Expansion prop 5}\SL
  \hat{H}_{2}(x,\t,\tb) &=& \LP( \BA{c}   \hat{H}_{2}^{1}(x,\t,\tb) \\
                                          \hat{H}_{2}^{2}(x,\t,\tb)\EA \RP) \nn
             &=&    H_{2}(x)
                 +  i\;\TSTB \;\P_{m}H_{2}(x)
                 +  \f{1}{4}\;\t\t\;\tb\tb\;\P^{m}\P_{m}H_{2}(x) \nn
             & & \mbox{}
                 +  \sqrt{2}\;\t \tilde{H}_{2}(x)
                 +  \f{i}{\sqrt{2}}\;\t\t\;\tb\bar{\s}^{m}\P_{m}
                           \tilde{H}_{2}(x)
                 +  \t\t\;F_{2}(x) \,\ , \nonumber \\
                 \label{Component Field Expansion prop 6}
\EEA
onde as componentes s\~ao
\BEA
 \BA{lllll}
    H_{1}(x)     = \LP( \BA{c}  H_{1}^{0}(x) \\ H_{1}^{-}(x) \EA \RP)
         & \hspace{0.4cm} &
   \tilde{H}_{1}(x) =
        \LP( \BA{c}  \psi^{1}_{H_{1}}(x) \\ \psi_{H_{1}}^{2}(x) \EA \RP)
         & \hspace{0.4cm} &
   F_{1}(x) = \LP( \BA{c}  f_{1}^{1}(x) \\ f_{1}^{2}(x) \EA \RP) \,\ ,
 \EA \nonumber \\
    \label{Component Field Expansion prop 7}
\EEA
e de maneira semelhante para o segundo dubleto
\BEA
  \BA{lllll}
    H_{2}(x)     = \LP( \BA{c}  H_{2}^{+}(x) \\ H_{2}^{0}(x) \EA \RP)
         & \hspace{0.4cm} &
    \tilde{H}_{2}(x) =
            \LP( \BA{c}  \psi^{1}_{H_{2}}(x) \\ \psi_{H_{2}}^{2}(x) \EA \RP)
         & \hspace{0.4cm} &
   F_{2}(x) = \LP( \BA{c}  f_{2}^{1}(x) \\ f_{2}^{2}(x) \EA \RP) \,\ .
 \EA \nonumber \\
     \label{Component Field Expansion prop 8}
\EEA
Repare que {\em todos} os campos $F$ s\~ao campos auxiliares, que depois
ser\~ao removidos usando as equa\c c\~oes de movimento.

Por quest\~ao de conveni\^encia, n\'os iremos trabalhar no gauge 
de Wess-Zumino. Neste gauge a expans\~ao em componentes dos supercampos de
gauge de SU(3), SU(2) e U(1) $\hat{V}_{c}={\bf T}^{a}\hat{V}^{a}_{c}$, 
$\hat{V}={\bf T}^{i}\hat{V}^{i}$ com ${\bf T}^{a}= {\bf \lambda}^{a}/2$, 
${\bf T}^{i}= {\bf \sigma}^{i}/2$ e $\hat{V}'={\bf Y} \hat{v}$, s\~ao escritas 
da seguinte maneira
\BEA
\hspace{-1.3cm}
 \hat{V}^{a}_{c}(x,\t,\tb)
     &=& - \;\TSTB\;G^{a}_{m}(x)
         + i\;\t\t\;\tb\bar{\tilde{g}}^{a}(x)
         - i\;\tb\tb\;\t\tilde{g}^{a}(x)
         + \frac{1}{2}\;\t\t\,\tb\tb\;D^{a}(x) \,\ , \nonumber \\
         \label{Component GField Expansion prop 9}
\EEA
e
\BEA
\hspace{-1.3cm}
 \hat{V}^{i}(x,\t,\tb)
     &=& - \;\TSTB\;V^{i}_{m}(x)
         + i\;\t\t\;\tb\bar{\lambda}^{i}_{A}(x)
         - i\;\tb\tb\;\t\lambda^{i}_{A}(x)
         + \frac{1}{2}\;\t\t\,\tb\tb\;D^{i}(x) \,\ , \nonumber \\
         \label{Component Field Expansion prop 9}
\EEA
e
\BEA
\hspace{-1.3cm}
 \hat{v}(x,\t,\tb)
     &=& - \;\TSTB\;V_{m}(x)
         + i\;\t\t\;\tb\bar{\lambda}_{B}(x)
         - i\;\tb\tb\;\t\lambda_{B}(x)
         + \frac{1}{2}\;\t\t\,\tb\tb\;D(x) \,\ , \nonumber \\
         \label{Component Field Expansion prop 10}
\EEA
Onde $\la^{i}_{A}(x)$ e $\la_{B}(x)$ s\~ao os campos dos gauginos, que s\~ao 
espinores de  Weyl de duas componentes. Eles s\~ao os parceiros supersim\'etricos dos 
b\'osons de gauge, e os campos $D$ tamb\'em s\~ao campos auxiliares, como os
campos F.

\subsection{Defini\c c\~oes dos Estados F\'\i sicos.}

As defini\c c\~oes de nossos estados f\'\i sicos em termos dos estados de 
intera\c c\~ao s\~ao
\BEA
   A_{m}(x) &=&   \cos\theta_{\mbox{w}}\, V_{m}(x)
              + \sin\theta_{\mbox{w}}\,V^{3}_{m}(x) \,\ , 
              \nonumber \\
   Z_{m}(x)  &=&   - \sin\theta_{\mbox{w}}\,V_{m}(x)
                  + \cos\theta_{\mbox{w}} \,V^{3}_{m}(x) \,\ , 
                  \nonumber \\ 
   W^{\pm}_{m}(x) &=& \f{V^{1}_{m}(x) \mp i V^{2}_{m}(x)}{\sqrt{2}} \,\ ,
              \label{svtl10}
\EEA
e os correspondentes estados para os gauginos spin-$1/2$ s\~ao introduzidos de 
maneira an\'aloga
\BEA
   \la_{ \gamma}(x)   &=&     \cos\theta_{\mbox{w}} \,\la_{B}(x)
                   + \sin\theta_{\mbox{w}} \,\la^{3}_{A}(x) \,\ ,
                   \nonumber \\
    \la_{Z}(x)   &=&   - \sin\theta_{\mbox{w}} \,\la_{B}(x)
                   + \cos\theta_{\mbox{w}} \,\la^{3}_{A}(x) \,\ ,
                   \nonumber \\
   \la^{\pm}(x) &=& \f{\la^{1}_{A}(x) \mp i \la^{2}_{A}(x)}{\sqrt{2}} \,\ .
                    \label{svtl11}
\EEA
O \^angulo de mistura $ \theta_{W}$ e os acoplamentos de 
gauge est\~ao relacionados da mesma maneira como no modelo padr\~ao, ou seja, 
pelas seguintes rela\c c\~oes 
\BEA
\sin \theta_{W}&=& \frac{g^{ \prime}}{ \sqrt{g^{2}+g^{ \prime 2}}} 
\nonumber \\
\cos \theta_{W}&=& \frac{g}{ \sqrt{g^{2}+g^{ \prime 2}}} 
\nonumber \\
e&=&g \sin \theta_{W}=g^{ \prime} \cos \theta_{W} \,\ .
\label{operador2}
\EEA  

Definiremos o o estado do fotino~($\tilde{A}$), do
Zino~($\tilde{Z}$) e dos dois Higgsinos neutros 
($\tilde{H}_{1}^{n}$, $\tilde{H}_{2}^{n}$)
em termos dos espinores de Weyl de duas componentes definidos acima 
por
\BEA
  \tilde{A}(x) &=& \LP( \BA{r}  -i \la_{\gamma}(x) \\
                              i \bar{\la}_{\gamma}(x)   \EA   \RP) \,\ ,
                              \label{svtl12}\SL
  \tilde{Z}(x) &=& \LP( \BA{r}   -i \la_{Z}(x) \\
                               i \bar{\la}_{Z}(x)   \EA   \RP) \,\ ,
                               \label{svtl13}\SL
  \tilde{H}_{1}^{n} &=& \LP( \BA{r}  \psi_{H_{1}}^{1} \\
                                 \bar{\psi}_{H{1}}^{1}   \EA   \RP) \,\ ,
                                 \label{svtl14}\SL
  \tilde{H}_{2}^{n} &=& \LP( \BA{r}  \psi_{H_{2}}^{2} \\
                                 \bar{\psi}_{H{2}}^{2}   \EA   \RP) \,\ .
                                 \label{svtl15}
\EEA
Repare que todos os espinores dos b\'osons neutros s\~ao do tipo de Majorana.

Para os espinores de Dirac, temos os Winos~($\tilde{W}$) e
os Higgsinos carregados~($\tilde{H}$) e os seus respectivos 
conjugados de carga s\~ao definidos de uma maneira an\'aloga
\BEA
 & &  \tilde{W}(x) \;=\; \LP( \BA{r}   -i  \la^{+}(x) \\
                               i  \bar{\la}^{-}(x)   \EA   \RP) \,\ , \SL
                               \label{svtl16} 
 & &  \tilde{H}(x)  \;=\; \LP( \BA{r}  \psi_{H_{2}}^{1} \\
                              \bar{\psi}_{H_{1}}^{2}   \EA   \RP) \,\ , \SL
                              \label{svtl17}
 & & \tilde{W}^{c}(x) \;=\; \LP( \BA{r}   -i  \la^{-}(x) \\
                               i  \bar{\la}^{+}(x)   \EA   \RP) \,\ , \SL
       \label{svtl18}
 & & \tilde{H}^{c}(x)  \;=\; \LP( \BA{r}  \psi_{H_{1}}^{2} \\
                              \bar{\psi}_{H_{2}}^{1}   \EA   \RP) \,\ .
         \label{svtl19}
\EEA
Aqui nos b\'osons carregados, os dois \'ultimos espinores s\~ao espinores conjugado de 
carga como nas Eqs.(\ref{Dirac Spinor}) e (\ref{Charge conjugation}).

Os espinores de quatro componentes dos l\'eptons carregados tamb\'em s\~ao espinores de 
Dirac, e t\^em a seguinte forma
\BEA
\Psi(e) \;=\; \LP( \BA{r} e_{L}(x) \\
                    e_{R}(x) \EA \RP) \,\ , 
                        \label{eletron spinor}
\EEA
j\'a para o neutrino
\BEA
\Psi(\nu) \;=\; \LP( \BA{r} \nu_{L}(x) \\
                    0 \EA \RP) \,\ . 
                        \label{neutrino spinor}
\EEA

Introduzindo os seguintes operadores
\BEA
    {\bf T}^{\pm} &=& {\bf T}^{1} \pm i {\bf T}^{2} \,\ , \nonumber \\
    {\bf Q} &=& {\bf T}^{3} +\f{{\bf Y}}{2} \,\ 
\label{operador1}
\EEA
onde o operador ${\bf Q}$ \'e o operador carga, com
autovalores em unidades da carga elementar $e$. J\'a 
$T^{i}= \frac{ \sigma^{i}}{2}$ conforme j\'a hav\'\i amos definido antes.

\subsection{O Termo 
$\bar{{\bf \Phi}} \exp[2(g \frac{\sigma^i}{2}\hat{V}^i 
+ \frac{g^\prime}{2}Y \hat{v})] {\bf \Phi}$.}

A expans\~ao deste termo em componentes \'e a seguinte
\BEA
 e^{2g\hat{V}+g'\hat{V}'} &=&
        \LP(    1
             +  2gT^{i}\hat{V}^{i}
             +  2g^{2}T^{i}T^{j}\hat{V}^{i}\hat{V}^{j}  \RP)
     \nmb
        \times
        \LP(    1
             +  g'Y\hat{v}'
             +  \HA g'^{2}Y^{2}\hat{v}'^{2}   \RP) \nn
     &=&      1
           +  g'Y\hat{v}'
           +  2gT^{i}\hat{V}^{i}
           +  \f{g'^{2}}{2}Y^{2}\hat{v}'^{2}
           +  2g^{2}T^{i}T^{j}\hat{V}^{i}\hat{V}^{j}
     \nmb
           +  2gg'YT^{i}\hat{V}^{i}\hat{v} \,\ ,
            \EEA
usamos o fato que $V^{n}=0$ para $n \geq 3$ no gauge de Wess-Zumino 
analogamente ao que foi feito 
na se\c c\~ao 6.2, a expans\~ao em componentes deste termo, \'e a seguinte
\BEA
&\int& d^4\theta
     \bar{{\bf \Phi}} \exp[2(g \frac{\sigma^i}{2}\hat{V}^i +
\frac{g^\prime}{2}Y \hat{v})] {\bf\Phi} \nn
&=& 
\vert {\bf F} \vert^2 - \vert \partial_m {\bf A}\vert ^2 
        - i{\bf \psi} \sigma^m \partial_m \bar{\bf \psi} \nn
      &+& \frac{g}{2}\left[
        (\bar{\bf\psi}\bar\sigma^m\sigma^i{\bf\psi}
        -i{\bf A}\sigma^i\partial_n\bar{{\bf A}} 
          +i\bar{{\bf A}}\sigma^i\partial^m {\bf A})V^i_m 
          \right. \nn
     &-& \left. i\sqrt{2}( \bar{\bf\psi}\sigma^i{\bf A}\bar{\lambda}^i_A 
                  - \bar{{\bf A}}\sigma^i{\bf\psi}\lambda^i_A) 
     + \bar{{\bf A}}\sigma^i{\bf A} D^i \right] \nn
     &+& \frac{g^\prime}{2}Y \left[
        (\bar{\bf\psi}\bar{\sigma}^m{\bf\psi} 
        -iA\partial^m\bar{A} 
            +i\bar{{\bf A}}\partial^m{\bf A})V_m \right. \nonumber \\ 
       &-& \left. i\sqrt{2}( \bar{\bf\psi}{\bf A}\bar{\lambda}_{B} 
                    -\bar{{\bf A}}{\bf\psi}\lambda_B)
     +\bar{{\bf A}}{\bf A} D \right] \nn
     &+& \frac{1}{4}[(g^2V_m^i \sigma^i V^{jm} \sigma^j +Y^2g^{\prime 2}V^m V_m)
                   \bar{{\bf A}}{\bf A} 
                  +2Ygg^\prime V^i_m V^m(\bar{{\bf A}}\sigma^i{\bf A})] 
                  \,\ , \nonumber \\ \label{svtl20}
\EEA
com $i=1,2,3$ e $m=0,1,2,3$.

\subsubsection{Reescrevendo Termos de Intera\c c\~ao Contendo Gauginos.}

Antes de continuarmos a nossa discuss\~ao, um c\'alculo geral e \'util 
ser\'a feito. Considere os seguintes termos da terceira e da quinta linha
\BEA
  \sqrt{2}i\;\bar{A}\LP[\;gT^{i}\lambda^{i}_{A} +\HA g'Y\lambda_{B}\;\RP]\psi
         - \sqrt{2}i\;\bar{\psi}\LP[\;gT^{i}\bar{\lambda}^{i}_{A}
                +\HA g'Y\bar{\lambda}_{B}\;\RP] A \,\ ,
         \label{RMFL prop 3}
\EEA
onde novamente $T^{i}=\frac{ \sigma^{i}}{2}$.

O primeiro termo da Eq.(\ref{RMFL prop 3}),
no parentese quadrado pode, em analogia com a derivada covariante, ser 
escrito usando os operadores da Eqs.(\ref{operador1}) e (\ref{operador2}) 
da seguinte maneira
\BEA
  \lefteqn{gT^{i}\lambda^{i}_{A} +\HA g'Y\lambda_{B}}\hspace{1.5cm} \nn
    &=&
    \f{g}{\sqrt{2}}\LP(T^{+}\la^{+}+T^{-}\la^{-}\RP)
          + e Q\,\la_{\gamma}
         +\f{g}{\CWA} \LP[T_{3}-Q\SWAS\RP]\la_{Z} \,\ , \nonumber \\
         \label{ig1}
\EEA
j\'a para o segundo termo de maneira an\'aloga teremos
\BEA
  \lefteqn{gT^{i}\bar{ \lambda}^{i}_{A} +\HA g'Y
  \bar{ \lambda}_{B}}\hspace{1.5cm} \nn
    &=&
    \f{g}{\sqrt{2}}\LP(T^{-}\bar{ \la}^{+}+T^{+}\bar{ \la}^{-}\RP)
          + e Q\,\bar{ \la}_{\gamma}
         +\f{g}{\CWA} \LP[T_{3}-Q\SWAS\RP]\bar{ \la}_{Z} \,\ , \nonumber \\
         \label{ig2}
\EEA

Usando Eq.(\ref{ig1}) e Eq.(\ref{ig2}), obtemos para a Eq.(\ref{RMFL prop 3})
\BEA
   \lefteqn{\sqrt{2}i\;\bar{A}\LP[\;gT^{i}\lambda^{i}_{A}
              +\HA g'Y\lambda_{B}\;\RP]\psi
         - \sqrt{2}i\;\bar{\psi}\LP[\;gT^{i}\bar{\lambda}^{i}_{A}
                 +\HA g'Y\bar{\lambda}_{B}\;\RP] A}\hspace{1.5cm} \nn
     &=&   ig \LP(\bar{A} T^+ \psi\;\la^+ -\bar{\la}^+\;\bar{\psi} T^- A \RP)
         + ig \LP(\bar{A} T^- \psi\;\la^- -\bar{\la}^-\;\bar{\psi} T^+ A \RP)
     \nmb
         +\sqrt{2}ie \LP(\bar{A} Q \psi\;\la_{\gamma}-
\bar{\la}_{\gamma}\;\bar{\psi} Q A \RP)
     \nmb
         +\f{\sqrt{2}ig}{\CWA}\LP(\bar{A} \LP[T_{3}-Q\SWAS\RP]
          \psi\;\la_{Z}-\bar{\la}_{Z}\;\bar{\psi} \LP[T_{3}-Q\SWAS\RP] A\RP)\NN
     &=&
           ig \LP(\bar{A} T^+ \psi\;\la^+ -\bar{\la}^+\;\bar{\psi} T^- A \RP)
         + ig \LP(\bar{A} T^- \psi\;\la^- -\bar{\la}^-\;\bar{\psi} T^+ A \RP)
     \nmb
         +\sqrt{2}ie Q_i \LP(\bar{A}
           \psi^{i}\;\la_{\gamma}-
\bar{\la}_{\gamma}\;\bar{\psi}^{i}  A \RP)
     \nmb
        + \f{\sqrt{2}ig}{\CWA}\LP[{\cal T}_{3i}-Q_{i}
              \SWAS\RP]\LP(\bar{A}  \psi^{i}\;\la_{Z}-
              \bar{\la}_{Z}\;\bar{\psi}^{i} A\RP) \,\ .
            \label{RMFL prop 4}
\EEA
Sendo que ${\cal T}_{3i}$ e $Q_{i}$ s\~ao os auto-valores de $T_{3}$ e $Q$ e $i=1,2$.
respectivamente.

\subsection{Termo dos L\'eptons.}

Para os $\hat{L}$ usando Eq.(\ref{svtl20}) e o fato que  
\BEA
\sigma^{i}_{ab} \sigma^{i}_{cd}=2 \delta_{ad} \delta_{bc}- 
\delta_{ab} \delta_{cd} \,\ ,
\EEA
acharemos
\BEA
&\int& d^4\theta
     \hat{\bar{L}} \exp[2(g \frac{\sigma^i}{2}\hat{V}^i +
\frac{g'}{2}Y_{L} \hat{v})] \hat{L} \nn
&=& 
\vert {\bf F_{L}} \vert^2 - \vert \partial_m {\bf \tilde{L}}\vert ^2 
        - i{\bf L} \sigma^m \partial_m \bar{\bf L} \nn
      &+& \frac{g}{2}\left[
        (\bar{L}\bar\sigma^m\sigma^i L
        -i \tilde{L}\sigma^i\partial^m\bar{\tilde{L}} 
          +i\bar{\tilde{L}}\sigma^i\partial^m \tilde{L})V^i_m 
          \right. \nn 
     &-& \left.i\sqrt{2}( \bar{L}\sigma^i\tilde{L}\bar{\lambda}^i_{A} 
                  - \bar{\tilde{L}}\sigma^iL\lambda^i_{A})
     + \bar{\tilde{L}}\sigma^i\tilde{L} D^i \right] \nn
     &+& \frac{g^\prime}{2}Y_{L} \left[
        (\bar{L}\bar{\sigma}^mL
        -i\tilde{L}\partial^m\bar{\tilde{L}} 
            +i\bar{\tilde{L}}\partial^m\tilde{L})V_m 
            \right. \nn 
       &-& \left. i\sqrt{2}( \bar{L}\tilde{L}\bar{\lambda}_{B} 
                    -\bar{\tilde{L}}L\lambda_{B})
     + \bar{\tilde{L}}\tilde{L} D \right] \nn
     &+& \frac{1}{4}[(g^2V_m^i V^{im} +Y_{L}^2g^{\prime 2}V^m V_m)
                   \bar{\tilde{L}}\tilde{L} 
                  +2Y_{L}gg^\prime V^i_m V^m(\bar{\tilde{L}}\sigma^i
\tilde{L})] \,\ .
\label{lint}
\EEA
e para os $\hat{R}$ de maneira an\'aloga, encontraremos
\BEA
&\int& d^4\theta
     \hat{\bar{R}} \exp[2( \frac{g'}{2}Y_{R} \hat{v})] \hat{R} \nn
&=& 
\vert {\bf F_{R}} \vert^2 - \vert \partial_\mu {\bf \tilde{R}}\vert ^2 
        - i{\bf R} \sigma^m \partial_m \bar{\bf R} \nn
     &+& \frac{g^\prime}{2}Y_{R} \left[
        (\bar{R}\bar{\sigma}^mR
        -i\tilde{R}\partial^m\bar{\tilde{R}} 
            +i\bar{\tilde{R}}\partial^m\tilde{R})V_m 
            \right. \nn 
       &-&\left. i\sqrt{2}( \bar{R}\tilde{R}\bar{\lambda}_{B} 
                    -\bar{\tilde{R}}R\lambda_{B})
     +\bar{\tilde{R}}\tilde{R} D \right] \nn
     &+& \frac{1}{4}[Y_{R}^2g^{\prime 2}V^m V_m
                   \bar{\tilde{R}}\tilde{R}] \,\ .
\label{rint}
\EEA

\subsection{Termos dos Higgs.}

Para os $\hat{H_{1}}$ obteremos
\BEA
&\int& d^4\theta
     \hat{\bar{H}}_{1} \exp[2(g \frac{\sigma^i}{2}\hat{V}^i +
\frac{g'}{2}Y_{H_{1}} \hat{v})] \hat{H}_{1} \nn
&=& 
\vert {\bf F_{1}} \vert^2 - \vert \partial_m {\bf H_{1}}\vert ^2 
        - i{\bf \tilde{H}_{1}} \sigma^m \partial_m 
\bar{\bf \tilde{H}_{1}} \nn
      &+& \frac{g}{2}\left[
        (\bar{ \tilde{H}_{1}}\bar\sigma^m\sigma^i \tilde{H}_{1}
        -iH_{1}\sigma^i\partial^m\bar{H}_{1} 
          +i\bar{H_{1}}\sigma^i\partial^m H_{1})V^i_m 
          \right. \nn 
     &-& \left. i\sqrt{2}( \bar{\tilde{H}_{1}}\sigma^iH_{1}\bar{\lambda}^i_{A} 
                  - \bar{H_{1}}\sigma^i\tilde{H}_{1}\lambda^i_{A})
     + \bar{H_{1}}\sigma^iH_{1} D^i \right] \nn
     &+& \frac{g^\prime}{2}Y_{H_{1}} \left[
        (\bar{\tilde{H}_{1}}\bar{\sigma}^m\tilde{H}_{1}
        -iH_{1}\partial^m\bar{H}_{1} 
            +i\bar{H_{1}}\partial^mH_{1})V_m \right. \nn 
       &-& \left. i\sqrt{2}( \bar{\tilde{H}_{1}}H_{1}\bar{\lambda}_{B} 
                    -\bar{H_{1}}\tilde{H}_{1}\lambda_{B})
     +\bar{H_{1}}H_{1} D \right] \nn
     &+& \frac{1}{4}[(g^2V_m^i V^{im} +Y_{H_{1}}^2g^{\prime 2}V^m V_m)
                   \bar{H_{1}}H_{1} 
                  +2Y_{H_{1}}gg^\prime V^i_m V^m(\bar{H_{1}}\sigma^i
H_{1})] \,\ , \nonumber \\
\label{h1int}
\EEA
j\'a para os $\hat{H_{2}}$ teremos
\BEA
&\int& d^4\theta
     \hat{\bar{ H}}_{2} \exp[2(g \frac{\sigma^i}{2}\hat{V}^i +
\frac{g'}{2}Y_{H_{2}} \hat{v})] \hat{H}_{2} \nn
&=& 
\vert {\bf F_{2}} \vert^2 - \vert \partial_m {\bf H_{2}}\vert ^2 
        - i{\bf \tilde{H}_{2}} \sigma^m \partial_m 
\bar{\bf \tilde{H}_{2}} \nn
      &+& \frac{g}{2}\left[
        (\bar{\tilde{H}_{2}}\bar\sigma^m\sigma^i \tilde{H}_{2}
        -iH_{2}\sigma^i\partial^m\bar{H}_{2} 
          +i\bar{H_{2}}\sigma^i\partial^m H_{2})V^i_m
          \right. \nn 
     &-& \left. i\sqrt{2}( \bar{\tilde{H}_{2}}\sigma^iH_{2}\bar{\lambda}^i_{A} 
                  - \bar{H_{2}}\sigma^i\tilde{H}_{2}\lambda^i_{A})
     + \bar{H_{2}}\sigma^iH_{2} D^i \right] \nn
     &+& \frac{g^\prime}{2}Y_{H_{2}} \left[
        (\bar{\tilde{H}_{2}}\bar{\sigma}^m\tilde{H}_{2}
        -iH_{2}\partial^m\bar{H}_{2} 
            +i\bar{H_{2}}\partial^mH_{2})V_m
            \right. \nn 
       &-&\left. i\sqrt{2}( \bar{\tilde{H}_{2}}H_{2}\bar{\lambda}_{B} 
                    -\bar{H_{2}}\tilde{H}_{2}\lambda_{B})
     + \bar{H_{2}}H_{2} D \right] \nn
     &+& \frac{1}{4}[(g^2V_m^i V^{im} +Y_{H_{2}}^2g^{\prime 2}V^m V_m)
                   \bar{H_{2}}H_{2} 
                  +2Y_{H_{2}}gg^\prime V^i_m V^m(\bar{H_{2}}\sigma^i
H_{2})] \,\ . \nonumber \\
\label{h2int}
\EEA

\section{As Componentes do Superpotencial.}

A expans\~ao em componentes de $W$, de acordo com a Eq.(\ref{caos}), \'e a seguinte
\BEA
 \int d^{4}\t\; W
   &=&   \int d^{4}\t\; \LP\{   \mu\; \e_{ij}\hat{H}_{1}^{i}\hat{H}_{2}^{j} +
          f\;\e_{ij}\hat{H}^{i}_{1}\hat{L}^{j}\hat{R} +h.c. \RP\}  \nn
   &=&
      \mu \e_{ij}\;
           \LP[\,H^{i}_{1}F_{2}^{j}
              + F_{1}^{i}H_{2}^{j}
              - \tilde{H}_{1}^{i}\tilde{H}_{2}^{j}\, \RP] \nn
  & & \mbox{}
       + f\e_{ij}\;
           \LP[\;F_{1}^{i}\tilde{L}^{j}\tilde{R}
              +  H_{1}^{i}F_{L}^{j}\tilde{R}
              +  H_{1}^{i}\tilde{L}^{j} F_{R} \RP.
   \nmb  \hspace{1.8cm} \LP.
              -  \tilde{H}_{1}^{i}L^{j}\tilde{R}
              -  H_{1}^{i}L^{j}R
           -  R\tilde{H}_{1}^{i}\tilde{L}^{j} \,\RP]+ h.c. \nn
&=& \L_{F}+\L_{llH}+\L_{H1}+\L_{\tilde{l}l\tilde{H}} \,\ . \hspace{0.4cm}
                        \label{Komp higgs property 1}
\EEA

Onde
\BEA
\L_{llH}&=&-f\e_{ij}(H_{1}^{i}L^{j}R+\bar{H}_{1}\bar{L}^{j}\bar{R}) \,\ , 
\nonumber \\
\L_{\tilde{l}l\tilde{H}}&=&-f\e_{ij}(\tilde{H}_{1}^{i}L^{j}\tilde{R}+
R\tilde{H}_{1}^{i}\tilde{L}^{j}+\tilde{\bar{H}}_{1}\bar{L}^{j}
\tilde{\bar{R}}+\bar{R}\tilde{\bar{H}}_{1}^{i}\tilde{\bar{L}}^{j}) \,\ ,
\nonumber \\
\L_{H1}&=&-\mu \e_{ij} \LP[ \tilde{H}_{1}^{i}\tilde{H}_{2}^{j}+ 
\bar{\tilde{H}}_{1}^{i}\bar{\tilde{H}}_{2}^{j} \RP] \,\ ,
\EEA
a parte dos termos $F$ analisaremos mais adiante.

Usando Eqs.(\ref{Component Field Expansion prop 3}), 
(\ref{Component Field Expansion prop 4}) e 
(\ref{Component Field Expansion prop 7}) encontraremos
\BEA
\L_{llH}&=&- f \epsilon_{ij} \left( 
H^{i}_{1}L^{j}R+h.c. \right) \nonumber \\
&=&-f(e_{L}e^{*}_{R}H^{0}_{1}+e^{*}_{L} e_{R}
\bar{H}^{0}_{1}- \nu_{L}e^{*}_{R} H^{-}_{1}- \nu^{*}_{L}
e_{R} \bar{H}^{-}_{1}) \,\ ,
\label{eeH} 
\EEA
usando Eqs.(\ref{eletron spinor}) e (\ref{neutrino spinor}) podemos escrever 
a equa\c c\~ao acima em termos dos espinores de quatro componentes da seguinte 
maneira
\BEA
\L_{llH}= f \bar{\Psi}(e)L \Psi(e)H^{0}_{1}+f \bar{\Psi}(e)R \Psi(e)
\bar{H}^{0}_{1}-f \bar{\Psi}(e)R \Psi(\nu)H^{-}_{1}-f \bar{\Psi}(\nu)L \Psi(e)
\bar{H}^{-}_{1} \,\ . \nonumber \\
\label{eeHint}
\EEA

De maneira an\'aloga poderemos escrever $\L_{\tilde{l}lH}$ da seguinte maneira
\BEA
\L_{\tilde{l}l\tilde{H}}&=&- f \epsilon_{ij} \left( 
\tilde{H}^{i}_{1}L^{j}\tilde{R}+\tilde{H}^{i}_{1} 
\tilde{L}^{j}R+h.c. \right) \nonumber \\
&=&- f \left( 
\tilde{H}^{1}_{1}L^{2}\tilde{R}-\tilde{H}^{2}_{1}L^{1}\tilde{R}+ 
\tilde{H}^{1}_{1}\tilde{L}^{2}R-\tilde{H}^{2}_{1}\tilde{L}^{1}R+h.c. 
\right) \nonumber \\
&=&- f \left[
(\psi^{1}_{H_{1}}e_{L}-\psi^{2}_{H_{1}}\nu)\tilde{e}^{*}_{R}+
(\bar{\psi}^{1}_{H_{1}}e^{*}_{L}-\bar{\psi}^{2}_{H_{1}}\nu^{*})\tilde{e}_{R}+
\psi^{1}_{H_{1}}e^{*}_{R}\tilde{e}_{L}-\psi^{2}_{H_{1}}e^{*}_{R}\tilde{\nu} 
\right . \nonumber \\
&+& \left. \bar{\psi}^{1}_{H_{1}}e_{R}\tilde{e}^{*}_{L}-
\bar{\psi}^{2}_{H_{1}}e_{R}\tilde{\nu}^{*} \right] \nonumber \\
&=&f \left[
\bar{\tilde{H}}^{n}_{1}R\Psi(e)\tilde{e}^{*}_{R}-\bar{\tilde{H}}R\Psi(\nu)
\tilde{e}_{R}^{*}+\bar{\Psi}(e)L
\tilde{H}^{n}_{1}\tilde{e}_{R}
-\bar{\Psi}(\nu)L\tilde{H}\tilde{e}_{R} \right. \nonumber \\
&+& \left. \bar{\Psi}(e)R
\tilde{H}^{n}_{1}\tilde{e}_{L} -
\bar{\Psi}(e)R\tilde{H}^{c}\tilde{\nu}+
\bar{\tilde{H}}^{n}_{1}L\Psi(e)\tilde{e}^{*}_{L}-\bar{\tilde{H}}^{c}L
\Psi(e)\tilde{\nu}^{*} \right] \,\ .
\EEA

Analisando a parte contendo os dois Higgs, acharemos
\BEA
\L_{H1}&=&-\mu\e^{ij}\LP[\;
   \tilde{H}^{i}_{1}\tilde{H}^{j}_{2}
   +\bar{\tilde{H}}^{i}_{1}\bar{\tilde{H}}^{j}_{2}\;\RP]
           \hspace{1.5cm} \nn
  &=&-\mu \LP[ 
  \tilde{H}^{1}_{1}\tilde{H}^{2}_{2}-\tilde{H}^{2}_{1}\tilde{H}^{1}_{2}+
  \bar{\tilde{H}}^{1}_{1}\bar{\tilde{H}}^{2}_{2}-
  \bar{\tilde{H}}^{2}_{1}\bar{\tilde{H}}^{1}_{2} \RP] \hspace{1.5cm} \nn
  &=&
    \mu\LP[ \psi^{2}_{H_{1}}\psi^{1}_{H_{2}}
         +  \bar{\psi}^{2}_{H_{1}}\bar{\psi}^{1}_{H_{2}}
         -  \psi^{1}_{H_{1}}\psi^{2}_{H_{2}}
         -  \bar{\psi}^{1}_{H_{1}}\bar{\psi}^{2}_{H_{2}} \RP] \nn
   &=&
        \mu\,\bar{\tilde{H}}\tilde{H}
      - \f{\mu}{2}\,\bar{\tilde{H}}^{n}_{1}\tilde{H}^{n}_{2}
      - \f{\mu}{2}\,\bar{\tilde{H}}^{n}_{2}\tilde{H}^{n}_{1} \,\ ,
      \label{higgsino massa}
\EEA

\section{Expans\~ao em Componentes de $\L_{GMT}$.}
    \label{SECT: Component Field Expansion}

Nesta parte vamos abrir em componentes $\L_{GMT}$, da 
Eq.(\ref{The Soft SUSY-Breaking Term prop 3}), e usando as 
defini\c c\~oes dos estados f\'\i sicos Eq.(\ref{svtl11}) e os espinores de 
quatro componentes Eq.(\ref{svtl16}) podemos escrever
\BEA
\lefteqn{-\HA M\LP(\la^{1}_{A}\la^{1}_{A}+
\bar{\la}^{1}_{A}\bar{\la}^{1}_{A}\RP)
   -\HA M\LP(\la^{2}_{A}\la^{2}_{A}+
\bar{\la}^{2}_{A}\bar{\la}^{2}_{A}\RP)}\hspace{2.5cm}\nn
     &=&  - M \LP(\la^{-}\la^{+}+\bar{\la}^{-}\bar{\la}^{+}\RP) \nn
     &=&   M_{\tilde{W}}\,\bar{\tilde{W}}\tilde{W} \,\ ,
\EEA
onde $M_{\tilde{W}} \equiv M$. Semelhantemente para as outras componentes e usando 
as Eqs.(\ref{svtl11}), (\ref{svtl12}) e (\ref{svtl13}) para os termos 
que faltam teremos
\BEA
\lefteqn{-\HA M\LP(\la^{3}_{A}\la^{3}_{A}+
\bar{\la}^{3}_{A}\bar{\la}^{3}_{A}\RP)
   -\HA M'\LP(\la_{B}\la_{B}+
\bar{\la}_{B}\bar{\la}_{B}\RP)}\hspace{1.5cm}\nn
   &=& -\HA\LP( M \SWAS+ M' \CWAS \RP)
          \LP(\la_{\gamma}\la_{\gamma}+
\bar{\la}_{\gamma}\bar{\la}_{\gamma}\RP)
   \nmb
       -\HA\LP( M \CWAS+ M' \SWAS \RP)
          \LP(\la_{Z}\la_{Z}+\bar{\la}_{Z}\bar{\la}_{Z}\RP)
  \nmb
       -\HA\LP( M -  M' \RP) \sin{2\t_{\mbox{w}}}\;
          \LP(\la_{\gamma}\la_{Z}+\bar{\la}_{\gamma}\bar{\la}_{Z}\RP) \nn
   &=& \HA\LP( M \SWAS+ M' \CWAS \RP)
           \;\bar{\!\!\tilde{A}}\tilde{A}
       +\HA\LP( M \CWAS+ M' \SWAS \RP)
           \bar{\tilde{Z}}\tilde{Z}
   \nmb
       +\HA\LP( M -  M' \RP) \sin{2\t_{\mbox{w}}}\;
           \;\,\bar{\!\!\tilde{A}}\tilde{Z} \nn
   &=&  \HA\,  M_{\tilde{A}}\;\,\bar{\!\!\tilde{A}}\tilde{A}
         + \HA\,M_{\tilde{Z}}\,\bar{\tilde{Z}}\tilde{Z}
        +\HA\LP( M_{\tilde{Z}} -  M_{\tilde{A}} \RP)
           \tan{2\t_{\mbox{w}}}\;\,\bar{\!\!\tilde{A}}\tilde{Z} \,\ ,
           \label{Component Field Expansion of L sub Soft prop 6}
\EEA
onde introduz\'\i mos a seguinte nota\c c\~ao simplificadora
\BEA
    M_{\tilde{A}} &=&  M'\CWAS+ M \SWAS \,\ , \nonumber \\
    M_{\tilde{Z}} &=&   M'\SWAS+ M \CWAS \,\ .
       \label{Component Field Expansion of L sub Soft prop 8}
\EEA
Assim $\L_{GMT}$ tem a seguinte expans\~ao em termos de componentes
\BEA
  \L_{GMT} &=&
          M_{\tilde{W}}\,\bar{\tilde{W}}\tilde{W}
          +  \HA\,  M_{\tilde{A}}\;\,\bar{\!\!\tilde{A}}\tilde{A}
          + \HA\,M_{\tilde{Z}}\,\bar{\tilde{Z}}\tilde{Z}
           +  \HA\LP( M_{\tilde{Z}} -  M_{\tilde{A}} \RP)\tan{2\t_{\mbox{w}}}
          \;\,\bar{\!\!\tilde{A}}\tilde{Z} \,\ . \nonumber \\ 
              \label{Component Field Expansion of L sub Soft prop 9}
\EEA

\section{Expans\~ao em Componentes de $\L_{Gauge}$.}

A parte cin\'etica dos b\'osons de gauge, que \'e dada por
$\L_{Gauge}$, cont\'em termos provenientes do grupo SU(2) e U(1), e toma 
a seguinte forma, conforme visto na se\c c\~ao 6.3, em componentes
\BEA
  \L_{Gauge}
     &=&  \f{1}{4} \int  d^{4}\t\;
         \LP[\,
                W_{i\,\a} W_{\a}^{i}
             +  W_{\,\a} W_{\a} \,
         \RP]   + h.c. \NN
     &=&
      - i\;\bar{\lambda}^{i}_{A}\bar{\s}^{m}\LP(
        \P_{m}\lambda^{i}_{A} -g\e_{ijk}V^{j}_{m}\la^{k}\RP)
      - i\;\bar{\lambda}_{B}\bar{\s}^{m}\P_{m}\lambda_{B} \nn
  & & \mbox{}
      - \f{1}{4}\;\LP(\;V^{i\;mn}V^{i}_{mn}
                       + V^{mn}V_{mn}\;\RP)
      + \HA\;\LP(\;D^{i}D^{i}+DD\;\RP) + h.c. \,\ . 
      \nonumber \\
      \label{avanco}
\EEA
Desta lalgrangeana obtemos a parte cin\'etica do b\'oson de gauge, a terceira 
linha dada por
\BEA
\L_{Vkin}=- \f{1}{4}\;\LP(\;V^{i\;mn}V^{i}_{mn}
                       + V^{mn}V_{mn}\;\RP)
\EEA
que \'e a mesma do {\bf SM}.

\subsection{Termo Cin\'etico do Gaugino.}

Da primeira linha da Eq.(\ref{avanco}) a parte cin\'etica do gaugino \'e dada por
\BEA
   \lefteqn{-i\bar{\la}^{i}_{A}\bar{\s}^{m} D_{m}\la^{i}_{A}
             -i\bar{\la}_{B}\bar{\s}^{m} \partial_{m}\la_{B}}\hspace{1.5cm} \nn
          &=& -i\bar{\la}^{i}_{A}\bar{\s}^{m}\P_{m}\la^{i}_{A}
             -i\bar{\la}_{B}\bar{\s}^{m} \P_{m}\la_{B}
             +ig\, \e_{ijk}\bar{\la}^{i}_{A}\bar{\s}^{m}V^{j}_{m}\la^{i}_{A} \nn
             &=&\L_{\tilde{V}kin}+\L_{\tilde{V}\tilde{V}V} \,\ .
               \label{RMFL prop 5}
\EEA

Podemos escrever $\L_{\tilde{V}kin}$, usando Eq.(\ref{svtl11}), da seguinte maneira
\BEA
\L_{\tilde{V}kin}&=&-i \bar{\la}^{i}_{A}\bar{\s}^{m}\P_{m}\la^{i}_{A}-i 
\bar{\la}_{B}\bar{\s}^{m}\P_{m}\la_{B} \nonumber \\
&=&i \la^{i}_{A}\s^{m}\P_{m}\bar{\la}^{i}_{A}+i 
\la_{B}\s^{m}\P_{m}\bar{\la}_{B} \nonumber \\
&=&i(\la^{+}\s^{m}\P_{m}\bar{\la}^{+}+\la^{-}\s^{m}\P_{m}\bar{\la}^{-}+
\la_{Z}\s^{m}\P_{m}\bar{\la}_{Z}+\la_{\gamma}\s^{m}\P_{m}\bar{\la}_{\gamma}) 
\,\ . \nonumber \\
\EEA
Para o \'ultimo termo da Eq.(\ref{RMFL prop 5}) temos
\BEA
\L_{\tilde{V}\tilde{V}V}&=&ig\, \e_{ijk}\bar{\la}^{i}_{A}\bar{\s}^{m}V^{j}_{m}\la^{i}_{A} \nn
&=&ig(\bar{\la}^{1}_{A}\bar{\s}^{m}V^{2}_{m}-\bar{\la}^{2}_{A}\bar{\s}^{m}V^{1}_{m})\la^{3}_{A}+
ig(\bar{\la}^{2}_{A}\bar{\s}^{m}\la^{1}_{A}-\bar{\la}^{1}_{A}\bar{\s}^{m}\la^{2}_{A})V^{3}_{m} 
\nonumber \\
&+&ig\bar{\la}^{3}_{A}\bar{\s}^{m}(\la^{2}_{A}V^{1}_{m}-\la^{1}_{A}V^{2}_{m}) \,\ .
                 \label{hdskjhsdjkhkjs} 
		 \nonumber \\
\EEA

Vamos analisar cada termo da Eq.(\ref{hdskjhsdjkhkjs}) em separado. Os 
resultados em termos dos campos f\'\i sicos, usando as Eqs.(\ref{svtl10}) 
e (\ref{svtl11}), s\~ao
\BEA
i(\la^{1}_{A}V^{2}_{m}-\la^{2}_{A}V^{1}_{m})&=&\la^{+}W^{-}_{m}-\la^{-}W^{+}_{m} \nonumber \\
i(\bar{\la}^{1}_{A}V^{2}_{m}-\bar{\la}^{2}_{A}V^{1}_{m})&=&
\bar{\la}^{+}W^{+}_{m}-\bar{\la}^{-}W^{-}_{m} \nonumber \\
i(\bar{\la}^{1}_{A}\la^{2}_{A}-\bar{\la}^{2}_{A}\la^{1}_{A})&=&
\bar{\la}^{-}\la^{-}-\bar{\la}^{+}\la^{+} \,\ .
\EEA

Com isto podemos escrever nossa lagrangiana da seguinte maneira
\BEA
\L_{\tilde{V}\tilde{V}V}&=&g(\bar{\la}^{+}W^{+}_{m}-\bar{\la}^{-}W^{-}_{m})\bar{\s}^{m}
(\la_{\gamma}\sin \theta_{W}+\la_{Z}\CWA) \nonumber \\
&-&
g(\bar{\la}^{-}\la^{-}-\bar{\la}^{+}\la^{+})\bar{\s}^{m}(A_{m}\sin \theta_{W}+Z_{m}\CWA) 
\nonumber \\ 
&-&
g(\bar{\la}_{\gamma}\sin \theta_{W}+\bar{\la}_{Z}\CWA)\bar{\s}^{m}(
\la^{+}W^{-}_{m}-\la^{-}W^{+}_{m}) \nonumber \\
      &=&  g\CWA \LP[
                 \LP(\bar{\la}_{Z}\bar{\s}^{m}\la^{-}
      -\bar{\la}^{+}\bar{\s}^{m}\la_{Z}\RP)W^{+}_{m}
                -\LP(\bar{\la}_{Z}\bar{\s}^{m}\la^{+}
      +\bar{\la}^{-}\bar{\s}^{m}\la_{Z}\RP)W^{-}_{m}
                \RP.
     \nmb \hspace{2cm} \LP.
                - \LP(\bar{\la}^{+}\bar{\s}^{m}\la^{+}
      -\bar{\la}^{-}\bar{\s}^{m}\la^{-}\RP)Z_{m}\RP]
     \nmb
          + e \LP[
                  \LP(\bar{\la}_{A}\bar{\s}^{m}\la^{-}
       -\bar{\la}^{+}\bar{\s}^{m}\la_{A}\RP)W^{+}_{m}
               +   \LP(\bar{\la}_{A}\bar{\s}^{m}\la^{+}
       -\bar{\la}^{-}\bar{\s}^{m}\la_{A}\RP)W^{-}_{m}\RP.
     \nmb \hspace{0.8cm} \LP.
                -  \LP(\bar{\la}^{+}\bar{\s}^{m}\la^{+}
       -\bar{\la}^{-}\bar{\s}^{m}\la^{-}\RP)A_{m} \RP] \,\ ,
\EEA
que colocando na nota\c c\~ao de quatro componentes Eqs.(\ref{svtl12}), 
(\ref{svtl13}) e (\ref{svtl16}), obteremos finalmente
\BEA
\L_{\tilde{V}\tilde{V}V}&=&-e(\bar{\tilde{A}}\gamma^{m}\tilde{W}W^{-}_{m}-
\bar{\tilde{W}}\gamma^{m}\tilde{A}W^{+}_{m}-
\bar{\tilde{W}}\gamma^{m}\tilde{W}A_{m}) \nonumber \\
&-&g \cos \theta_{W}(
\bar{\tilde{Z}}\gamma^{m}\tilde{W}W^{-}_{m}-
\bar{\tilde{W}}\gamma^{m}\tilde{Z}W^{+}_{m}-
\bar{\tilde{W}}\gamma^{m}\tilde{W}Z_{m}) \,\ .
                  \label{ingve er en proff}
\EEA

\section{Intera\c c\~ao L\'epton L\'epton B\'oson de Gauge}

Estas intera\c c\~oes resultam dos seguintes termos
\BEA
\L_{llV}&\ni& \int d^{4} \theta \LP\{ 
\hat{\bar{L}} \exp \LP[ 2 \LP( g \frac{\sigma^{i}}{2} \hat{V}^{i}+ 
\frac{g'}{2}Y_{L} \hat{v} \RP) \RP] \hat{L}+
\hat{\bar{R}} \exp \LP[ 2 \LP(  \frac{g'}{2}Y_{R} \hat{v} \RP) \RP] 
\hat{R} \RP\} \nonumber \\
&=& \frac{g}{2}(\bar{L} \bar{\sigma}^{m} \sigma^{i}L)V^{i}_{m}+
\frac{g'}{2}Y_{L} \bar{L} \bar{\sigma}^{m}LV_{m}+
\frac{g'}{2}Y_{R} \bar{R} \bar{\sigma}^{m}RV_{m} \nonumber \\
&=& \L^{carregada}_{llV}+\L^{neutra}_{llV}
\,\ .
\EEA

A parte carregada \'e escrita usando Eq.(\ref{svtl10}) da seguinte maneira
\BEA
\L^{carregada}_{llV}&=&
\frac{g}{2}\left[ \bar{L} \bar{\sigma}^{m} 
(\sigma^{1}V^{1}_{m}+\sigma^{2}V^{2}_{m})L \right] \nonumber \\
&=&\frac{g}{\sqrt{2}} \bar{L} \bar{\sigma}^{m}
\LP( \BA{cc} 
O & W^{+}_{m} \\
W^{-}_{m} & 0          \EA \RP) L \,\ .
\EEA
Abrindo em componentes Eq.(\ref{Component Field Expansion prop 1}) teremos
\BEA
\L^{carregada}_{llV}=\frac{g}{\sqrt{2}}(
\nu^{*} \bar{\sigma}^{m}e_{L}W^{+}_{m}+
e^{*}_{L} \bar{\sigma}^{m} \nu W^{-}_{m}) \,\ ,
\EEA
usando os espinores de quatro componentes Eqs.(\ref{eletron spinor}) e 
(\ref{neutrino spinor}), a lagrangiana acima 
pode ser escrita da seguinte maneira
\BEA
\L^{carregada}_{llV}=\frac{-g}{\sqrt{2}}( 
\bar{\Psi}(\nu) \gamma^{m}R\Psi(e)W^{+}_{m}+
\bar{\Psi}(e) \gamma^{m}R\Psi(\nu)W^{-}_{m}) \,\ .
\EEA

Agora vamos analisar a parte neutra analogamente ao que foi 
feito na parte carregada obteremos
\BEA
\L^{neutra}_{llV}&=&
\frac{g}{2}\left[ \bar{L} ( \bar{\sigma}^{m} 
\sigma^{3}V^{3}_{m})L \right]+
\frac{g'}{2}Y_{L} \bar{L} \bar{\sigma}^{m}LV_{m}+
\frac{g'}{2}Y_{R} \bar{R} \bar{\sigma}^{m}RV_{m} \nonumber \\
&=&
eQ_{e}\bar{L} \bar{\sigma}^{m}LA_{m}+eQ_{e}\bar{R} 
\bar{\sigma}^{m}RA_{m} \nonumber \\
&+& \frac{g}{\CWA}\LP( T^{3}\cos^{2} \theta_{W}-\frac{Y_{L}}{2}
\sin^{2} \theta_{W} 
\RP)\bar{L} \bar{\sigma}^{m}LZ_{m} \nonumber \\
&-&\frac{g}{\CWA}Q_{e}\sin^{2} \theta_{W} \bar{R} \bar{\sigma}^{m}
RZ_{m} \nonumber \\
&=&
-eQ_{e}\bar{\Psi}(e) \gamma^{m}\Psi(e)A_{m}-
\frac{g}{\CWA}T^{3}_{\nu}\bar{\Psi}(\nu) \gamma^{m}
\Psi(\nu)Z_{m} \nonumber \\
&-&\frac{g}{\CWA}[(T^{3}_{e}-Q_{e}\sin^{2} \theta_{W})
\bar{\Psi}(e) \gamma^{m}R\Psi(e)+Q_{e}\sin^{2} \theta_{W} \bar{\Psi}(e) \gamma^{m}
L\Psi(e)]Z_{m} \,\ . \nonumber \\
\EEA

Dessa maneira obtemos
\BEA
\L_{llV}&=&\frac{-g}{\sqrt{2}}( 
\bar{\Psi}(\nu) \gamma^{m}R\Psi(e)W^{+}_{m}+
\bar{\Psi}(e) \gamma^{m}R\Psi(\nu)W^{-}_{m}) \nonumber \\
&-&
eQ_{e}\bar{\Psi}(e) \gamma^{m}\Psi(e)A_{m}-
\frac{g}{\CWA}T^{3}_{\nu}\bar{\Psi}(\nu) \gamma^{m}
\Psi(\nu)Z_{m} \nonumber \\
&-&\frac{g}{\CWA}[(T^{3}_{e}-Q_{e}\sin^{2} \theta_{W})
\bar{\Psi}(e) \gamma^{m}R\Psi(e)+Q_{e}\sin^{2} \theta_{W} \bar{\Psi}(e) 
\gamma^{m}L\Psi(e)]Z_{m} \,\ . \nonumber \\
\EEA

\section{Intera\c c\~ao Higgsino Higgsino B\'oson Vetorial}

A intera\c c\~ao entre estas part\'\i culas vem do seguinte termo
\BEA
\L_{\tilde{H}\tilde{H}V}&\ni& \int d^{4} \theta \LP\{ 
\hat{\bar{H}}_{1} \exp \LP[2 \LP( g \frac{\sigma^{i}}{2} \hat{V}^{i}+ 
\frac{g'}{2}Y_{H_{1}} \hat{v} \RP) \RP] \hat{H}_{1}+
\hat{\bar{H}}_{2} \exp \LP[ 2 \LP( g \frac{\sigma^{i}}{2} \hat{V}^{i}+ 
\frac{g'}{2}Y_{H_{2}} \hat{v} \RP) \RP] \hat{H}_{2} \RP\} \nonumber \\
&=& \frac{g}{2} \bar{\tilde{H}}_{1} 
\bar{\sigma}^{m} \sigma^{i} \tilde{H}_{1}V^{i}_{m}+ 
\frac{g'}{2}Y_{H_{1}} \bar{\tilde{H}}_{1} \bar{\sigma}^{m} 
\tilde{H}_{1}V_{m} \nonumber \\
&+& \frac{g}{2} \bar{\tilde{H}}_{2} 
\bar{\sigma}^{m} \sigma^{i} \tilde{H}_{2}V^{i}_{m}+ 
\frac{g'}{2}Y_{H_{2}} \bar{\tilde{H}}_{2} \bar{\sigma}^{m} 
\tilde{H}_{2}V_{m} \nonumber \\
&=& \L^{1}_{\tilde{H}\tilde{H}V}+\L^{2}_{\tilde{H}\tilde{H}V}
\,\ .
\EEA

Vamos analisar primeiro o termo com $\tilde{H}_{1}$ e usando Eqs.(\ref{svtl10}) 
e (\ref{operador2}), teremos
\BEA
\L^{1}_{\tilde{H}\tilde{H}V}&=& \frac{g}{2} \bar{\tilde{H}}_{1} 
\bar{\sigma}^{m} \sigma^{i} \tilde{H}_{1}V^{i}_{m}+ 
\frac{g'}{2}Y_{H_{1}} \bar{\tilde{H}}_{1} \bar{\sigma}^{m} 
\tilde{H}_{1}V_{m} \nonumber \\
&=& \frac{g}{\sqrt{2}} \bar{\tilde{H}}_{1} \bar{\sigma}^{m}
\LP( \BA{cc} 
O & W^{+}_{m} \\
W^{-}_{m} & 0          \EA \RP) \tilde{H}_{1}+eQ_{H_{1}} 
\bar{\tilde{H}}_{1} \bar{\sigma}^{m} \tilde{H}_{1}A_{m} 
\nonumber \\
&-&\frac{g}{\CWA}(Q_{H_{1}}\SWA-T^{3}) 
\bar{\sigma}^{m} \tilde{H}_{1}Z_{m} \,\ ,
\EEA
que abrindo em componentes com Eq.(\ref{Component Field Expansion prop 7}) 
pode ser colocada da seguinte maneira
\BEA
\L^{1}_{\tilde{H}\tilde{H}V}
&=&\f{g}{2\CWA}\; \bar{\psi}^{1}_{H_{1}}
            \bar{\s}^{m}\psi^{1}_{H_{1}}\;Z_{m} 
            -e \bar{\psi}^{2}_{H1}\bar{\s}^{m}\psi^{2}_{H1}A_{m}
            \nonumber \\
&+& \f{g}{\sqrt{2}}\; \bar{\psi}^{1}_{H_{1}}\bar{\s}^{m}
            \psi^{2}_{H_{1}}\;W^{+}_{m}
    +\f{g}{\sqrt{2}}\; \bar{\psi}^{2}_{H_{1}}\bar{\s}^{m}
             \psi^{1}_{H_{1}}\;W^{-}_{m} \nonumber \\
&-& \f{g}{2\CWA}\LP(1-2\SWAS \RP)\bar{\psi}^{2}_{H_{1}}
             \bar{\s}^{m}\psi^{2}_{H_{1}}\;Z_{m} \,\ ,
             \EEA
usando os espinores de quatro componentes, Eqs.(\ref{svtl14}) e (\ref{svtl17}) 
teremos
\BEA
\L^{1}_{\tilde{H}\tilde{H}V}
&=&- \f{g}{4\CWA}\;\bar{\tilde{H}}^{n}_{1}\g^{m} \g_{5} \tilde{H}^{n}_{1}\;Z_{m}
-e\bar{\tilde{H}}\g^{m}L\tilde{H}A_{m} 
\nonumber \\
&+&\f{g}{\sqrt{2}}\;\bar{\tilde{H}}\g^{m}
                 L\tilde{H}^{n}_{1} \;W^{+}_{m}
      -\f{g}{\sqrt{2}}\;\bar{\tilde{H}}^{n}_{1}\g^{m}
                  L\tilde{H} \;W^{-}_{m} 
                  \nonumber \\
&+&\f{g}{2\CWA}\LP(1-2\SWAS\RP)\bar{\tilde{H}}
                  \g^{m} L\tilde{H} \; Z_{m} \,\ .
\EEA

J\'a para o termo com $\tilde{H}_{2}$ de maneira an\'aloga ao feito 
acima obteremos
\BEA
\L^{2}_{\tilde{H}\tilde{H}V}&=& \frac{g}{2} \bar{\tilde{H}}_{2} 
\bar{\sigma}^{m} \sigma^{i} \tilde{H}_{2}V^{i}_{m}+ 
\frac{g'}{2}Y_{H_{2}} \bar{\tilde{H}}_{2} \bar{\sigma}^{m} 
\tilde{H}_{2}V_{m} \nonumber \\
&=& \frac{g}{\sqrt{2}} \bar{\tilde{H}}_{2} \bar{\sigma}^{m}
\LP( \BA{cc} 
O & W^{+}_{m} \\
W^{-}_{m} & 0          \EA \RP) \tilde{H}_{2}+eQ_{H_{2}} 
\bar{\tilde{H}}_{2} \bar{\sigma}^{m} \tilde{H}_{2}A_{m} 
\nonumber \\
&-&\frac{g}{\CWA}(Q_{H_{2}}\SWA-T^{3}) 
\bar{\sigma}^{m} \tilde{H}_{2}Z_{m} \,\ ,
\EEA
que em componentes, Eq.(\ref{Component Field Expansion prop 8}), torna-se
\BEA
\L^{2}_{\tilde{H}\tilde{H}V}
&=&\f{-g}{2\CWA} \bar{\psi}^{2}_{H_{2}}
            \bar{\s}^{m}\psi^{2}_{H_{2}}Z_{m} 
            +e \bar{\psi}^{1}_{H2}\bar{\s}^{m}\psi^{1}_{H2}A_{m}
            \nonumber \\
&+& \f{g}{\sqrt{2}}\; \bar{\psi}^{1}_{H_{2}}\bar{\s}^{m}
            \psi^{2}_{H_{2}}\;W^{+}_{m}
    +\f{g}{\sqrt{2}}\; \bar{\psi}^{2}_{H_{2}}\bar{\s}^{m}
             \psi^{1}_{H_{2}}\;W^{-}_{m} \nonumber \\
&+& \f{g}{2\CWA}\LP(1-2\SWAS \RP)\bar{\psi}^{1}_{H_{2}}
             \bar{\s}^{m}\psi^{1}_{H_{2}}\;Z_{m} \,\ ,
             \EEA
podemos escrever a lagrangiana acima em termos de espinores de 
quatro componentes Eqs.(\ref{svtl15}) e (\ref{svtl17}) da seguinte maneira             
\BEA
\L^{2}_{\tilde{H}\tilde{H}V}
&=& \f{g}{4\CWA}\;\bar{\tilde{H}}^{n}_{2}\g^{m} \g_{5} \tilde{H}^{n}_{2}\;Z_{m}
-e\bar{\tilde{H}}\g^{m}R\tilde{H}A_{m} 
\nonumber \\
&-&\f{g}{\sqrt{2}}\;\bar{\tilde{H}}\g^{m}
                 R\tilde{H}^{n}_{2} \;W^{+}_{m}
      -\f{g}{\sqrt{2}}\;\bar{\tilde{H}}^{n}_{2}\g^{m}
                  R\tilde{H} \;W^{-}_{m} 
                  \nonumber \\
&-&\f{g}{2\CWA}\LP(1-2\SWAS\RP)\bar{\tilde{H}}
                  \g^{m} R\tilde{H} \; Z_{m} \,\ .
\EEA

Juntando os dois termos obteremos
\BEA
\L_{\tilde{H}\tilde{H}V}
&=&- \f{g}{4\CWA}\;\bar{\tilde{H}}^{n}_{1}\g^{m} \g_{5} \tilde{H}^{n}_{1}\;Z_{m}
-e\bar{\tilde{H}}\g^{m}L\tilde{H}A_{m} 
\nonumber \\
&+&\f{g}{\sqrt{2}}\;\bar{\tilde{H}}\g^{m}
                 L\tilde{H}^{n}_{1} \;W^{+}_{m}
      -\f{g}{\sqrt{2}}\;\bar{\tilde{H}}^{n}_{1}\g^{m}
                  L\tilde{H} \;W^{-}_{m} 
                  \nonumber \\
&+&\f{g}{2\CWA}\LP(1-2\SWAS\RP)\bar{\tilde{H}}
                  \g^{m} L\tilde{H} \; Z_{m} \nonumber \\
&+&
\f{g}{4\CWA}\;\bar{\tilde{H}}^{n}_{2}\g^{m} \g_{5} \tilde{H}^{n}_{2}\;Z_{m}
-e\bar{\tilde{H}}\g^{m}R\tilde{H}A_{m} 
\nonumber \\
&-&\f{g}{\sqrt{2}}\;\bar{\tilde{H}}\g^{m}
                 R\tilde{H}^{n}_{2} \;W^{+}_{m}
      -\f{g}{\sqrt{2}}\;\bar{\tilde{H}}^{n}_{2}\g^{m}
                  R\tilde{H} \;W^{-}_{m} 
                  \nonumber \\
&-&\f{g}{2\CWA}\LP(1-2\SWAS\RP)\bar{\tilde{H}}
                  \g^{m} R\tilde{H} \; Z_{m} \,\ .
\EEA

\section{Intera\c c\~ao Sl\'epton L\'epton Gaugino.}

A intera\c c\~ao $\tilde{l}l\tilde{V}$ vem de dois termos, que escreveremos 
da seguinte maneira
\BEA
L_{\tilde{l}l\tilde{V}}
&\ni& \int d^{4} \theta \LP\{ 
\hat{\bar{L}} \exp \LP[ 2 \LP( g \frac{\sigma^{i}}{2} \hat{V}^{i}+ 
\frac{g'}{2}Y_{L} \hat{v} \RP) \RP] \hat{L}+
\hat{\bar{R}} \exp \LP[ 2 \LP(  \frac{g'}{2}Y_{R} \hat{v} \RP) \RP] 
\hat{R} \RP\} \nonumber \\
&=&L^{1}_{\tilde{l}l\tilde{V}}+
L^{2}_{\tilde{l}l\tilde{V}}
\EEA
O primeiro termo \'e dado por
\BEA
L^{1}_{\tilde{l}l\tilde{V}}&=&
\sqrt{2}i\;\bar{\tilde{L}}
\LP(gT^{i}\lambda^{i}_{A}+\frac{Y_{L}}{2} g'\lambda_{B}\RP)L
      - \sqrt{2}i\;\bar{L}\LP(gT^{i}\bar{\lambda}^{i}_{A}
      +\frac{Y_{L}}{2} g'\bar{\lambda}_{B}\RP)\tilde{L}\hspace{1cm} 
      \nonumber \\
      &=&
         ig\LP( \bar{\tilde{L}}T^{+}\;L\la^{+}
               - \bar{\la}^{+}\bar{L}T^{-}\;\tilde{L}\RP)
       + ig\LP( \bar{\tilde{L}}T^{-}\;L\la^{-}
               - \bar{\la}^{-}\bar{L}T^{+}\;\tilde{L}\RP)
     \nmb
       + \sqrt{2}ieQ_{i}\LP( \bar{\tilde{L}}^{i}\;L^{i}\la_{\gamma}
               - \bar{\la}_{\gamma}\bar{L}^{i}\;\tilde{L}^{i} \RP)
     \nmb
       +\f{\sqrt{2}ig}{\CWA} \LP({\cal T}^{3}_{i} -Q_{i}\SWAS\RP)
          \LP(  \bar{\tilde{L}}^{i}\;L^{i}\la_{Z}
               - \bar{\la}_{Z}\bar{L}^{i}\;\tilde{L}^{i} \RP)\NN
     &=&
        ig(\tilde{\nu}^{*}e_{L}\la^{+}-\bar{\la}^{+}e^{*}_{L}\tilde{\nu})+
        ig(\tilde{e}^{*}_{L}\nu\la^{-}-\bar{\la}^{-}\nu^{*}\tilde{e}_{L}) \nonumber \\
     &-&i\sqrt{2}e(\tilde{e}^{*}_{L}e_{L}\la_{\gamma}-\bar{\la}_{\gamma}e^{*}_{L}\tilde{e}_{L}) 
     \nonumber \\
     &+&\frac{ig}{\sqrt{2} \cos \theta_{W}}(\tilde{\nu}^{*}\nu\la_{Z}-\bar{\la}_{Z}\nu^{*}
     \tilde{\nu}) \nonumber \\
     &-&\frac{ig}{\sqrt{2} \cos \theta_{W}}(1-2 \sin^{2} \theta_{W}) \,\ 
     (\tilde{e}^{*}_{L}e_{L}\la_{Z}-\bar{\la}_{Z}e^{*}_{L}\tilde{e}_{L}) \nonumber \\
     &=&g(\bar{\tilde{W}}^{c}R\Psi(e)\tilde{\nu}^{*}+\bar{\Psi}(e)L\tilde{W}^{c}\tilde{\nu})+g
     (\bar{\tilde{W}}R\Psi(\nu)\tilde{e}^{*}_{L}+\bar{\Psi}(\nu)L\tilde{W}\tilde{e}_{L}) 
     \nonumber \\
     &-&\sqrt{2}e(\bar{\tilde{A}}R\Psi(e)\tilde{e}^{*}_{L}+\bar{\Psi}(e)L\tilde{A}\tilde{e}_{L}) 
     \nonumber \\
     &+&\frac{g}{\sqrt{2} \cos \theta_{W}}(\bar{\tilde{Z}}R\Psi(\nu)\tilde{\nu}^{*}+
     \bar{\Psi}(\nu)L\tilde{Z}\tilde{\nu}) \nonumber \\
     &-&\frac{g}{\sqrt{2} \cos \theta_{W}}(1-2 \sin^{2} \theta_{W}) \,\ 
     (\bar{\tilde{Z}}R\Psi(e)\tilde{e}^{*}_{L}-\bar{\Psi}(e)L\tilde{Z}\tilde{e}_{L}) \,\ .
           \label{RIT prop 2}
\EEA

O termo correspondente ao l\'epton de m\~ao direita \'e escrito da 
seguinte maneira
\BEA
L^{2}_{\tilde{l}l\tilde{V}}&=&
\sqrt{2}i\;\bar{\tilde{R}} \left( g'\frac{Y_{R}}{2}\lambda_{B} \right) R
      - \sqrt{2}i\;\bar{R} \left( g'\frac{Y_{R}}{2}\bar{\lambda}_{B} \right) \tilde{R}\hspace{1.5cm} \NN
   &=&
      \sqrt{2}ig'\;\bar{\tilde{R}}
            \LP(\la_{\gamma}\CWA-\la_{Z}\SWA\RP)R
   \nmb
      - \sqrt{2}ig'\;\bar{R}
             \LP(\bar{\la}_{\gamma}\CWA-\bar{\la}_{Z}\SWA\RP)\tilde{R}
	     \nonumber \\
   &=&
   ig'\sqrt{2}(e^{*}_{R}\la_{\gamma}\tilde{e}_{R} \CWA-
   e^{*}_{R}\la_{Z}\tilde{e}_{R} \SWA)-ig'\sqrt{2}
   (e_{R}\bar{ \la}_{ \gamma} \tilde{e}^{*}_{R} \CWA-
   e_{R}\bar{\la}_{Z}\tilde{e}^{*}_{R}\SWA) 
   \nonumber \\
   &=&\sqrt{2}e(\bar{\Psi}(e)R\tilde{A}\tilde{e}_{R}+
   \bar{\tilde{A}}L\Psi(e)\tilde{e}^{*}_{R}) \nonumber \\
   &-&g \frac{\sin^{2} \theta_{W}}{\CWA}
   (\bar{\Psi}(e)R\tilde{Z}\tilde{e}_{R}+
   \bar{\tilde{Z}}L\Psi(e)\tilde{e}^{*}_{R}) \,\ .
       \label{RIT prop 3}
\EEA

Logo a intera\c c\~ao sl\'epton l\'epton gaugino \'e dada por
\BEA
L_{\tilde{l}l\tilde{V}}
&=&g(\bar{\tilde{W}}^{c}R\Psi(e)\tilde{\nu}^{*}+\bar{\Psi}(e)L\tilde{W}^{c}\tilde{\nu})+g
     (\bar{\tilde{W}}R\Psi(\nu)\tilde{e}^{*}_{L}+\bar{\Psi}(\nu)L\tilde{W}\tilde{e}_{L}) 
     \nonumber \\
     &-&\sqrt{2}e(\bar{\tilde{A}}R\Psi(e)\tilde{e}^{*}_{L}+\bar{\Psi}(e)L\tilde{A}\tilde{e}_{L}) 
     \nonumber \\
     &+&\frac{g}{\sqrt{2} \cos \theta_{W}}(\bar{\tilde{Z}}R\Psi(\nu)\tilde{\nu}^{*}+
     \bar{\Psi}(\nu)L\tilde{Z}\tilde{\nu}) \nonumber \\
     &-&\frac{g}{\sqrt{2} \cos \theta_{W}}(1-2 \sin^{2} \theta_{W}) \,\ 
     (\bar{\tilde{Z}}R\Psi(e)\tilde{e}^{*}_{L}-\bar{\Psi}(e)L\tilde{Z}\tilde{e}_{L}) \nonumber \\
&+&
\sqrt{2}e(\bar{\Psi}(e)R\tilde{A}\tilde{e}_{R}+
   \bar{\tilde{A}}L\Psi(e)\tilde{e}^{*}_{R}) \nonumber \\
   &-&g \frac{\sin^{2} \theta_{W}}{\CWA}
   (\bar{\Psi}(e)R\tilde{Z}\tilde{e}_{R}+
   \bar{\tilde{Z}}L\Psi(e)\tilde{e}^{*}_{R}) \,\ .
\EEA

\section{Intera\c c\~ao Sl\'epton Sl\'epton B\'oson de Gauge}

Neste caso as intera\c c\~oes s\~ao as seguintes
\BEA
\L_{\tilde{l}\tilde{l}V}&\ni& \int d^{4} \theta \LP\{ 
\hat{\bar{L}} \exp \LP[ 2 \LP( g \frac{\sigma^{i}}{2} \hat{V}^{i}+ 
\frac{g'}{2}Y_{L} \hat{v} \RP) \RP] \hat{L}+
\hat{\bar{R}} \exp \LP[ 2 \LP( \frac{g'}{2}Y_{R} \hat{v} \RP) \RP] 
\hat{R} \RP\} \nonumber \\
&=& \frac{ig}{2}\bar{\tilde{L}} \sigma^{i} \partial^{m}\tilde{L}V^{i}_{m}-
\frac{ig}{2}\tilde{L}\sigma^{i} \partial^{m}\bar{\tilde{L}}V^{i}_{m}-
\frac{ig'}{2}Y_{L} \tilde{L} \partial^{m}\bar{\tilde{L}}V_{m} \nonumber \\
&+&\frac{ig'}{2}Y_{L}\bar{\tilde{L}}\partial^{m}\tilde{L}V_{m}+
\frac{ig'}{2}Y_{R} \bar{\tilde{R}} \partial^{m}
\tilde{R}V_{m}-\frac{ig'}{2}Y_{R}\tilde{R}\partial^{m}\bar{\tilde{R}}
V_{m} \nonumber \\
&=& \L^{carregada}_{\tilde{l}\tilde{l}V}+\L^{neutra}_{\tilde{l}\tilde{l}V}
\,\ .
\EEA

Primeiro vamos analisar a parte carregada que pode ser escrita da seguinte maneira
\BEA
\L^{carregada}_{\tilde{l}\tilde{l}V}&=&
\frac{ig}{2}\left[ \bar{\tilde{L}}(\sigma^{1}V^{1}_{m}+\sigma^{2}V^{2}_{m})
\partial^{m}\tilde{L}- \tilde{L}(\sigma^{1}V^{1}_{m}+\sigma^{2}V^{2}_{m})
\partial^{m}\bar{\tilde{L}} \right] \nonumber \\
&=&\frac{ig}{\sqrt{2}} \bar{\tilde{L}} \LP( \BA{cc} 
O & W^{+}_{m} \\
W^{-}_{m} & 0          \EA \RP) 
\stackrel{\leftrightarrow}{\partial}^{m} \tilde{L} \nonumber \\
&=&
\frac{ig}{\sqrt{2}}( 
\tilde{\nu}^{*} \stackrel{\leftrightarrow}{\partial}^{m}\tilde{e}_{L}
W^{+}_{m}+
\tilde{e}^{*}_{L} \stackrel{\leftrightarrow}{\partial}^{m} \tilde{\nu} 
W^{-}_{m}) \,\ ,
\EEA
com o seguinte operador da teoria qu\^antica de campos
\BEA
\bar{\Phi}\stackrel{\leftrightarrow}{\partial}\Phi=
\bar{\Phi}\partial\Phi-\Phi\partial\bar{\Phi} \,\ .
\label{ucraniano}
\EEA

Com rela\c c\~ao a parte neutra encontraremos
\BEA
\L^{neutra}_{\tilde{l}\tilde{l}V}&=&
\frac{ig}{2}\left[ \bar{\tilde{L}} \sigma^{3}V^{3}_{m} \partial^{m}\tilde{L}- 
\tilde{L}\sigma^{3}V^{3}_{m}\partial^{m}\bar{\tilde{L}}\right]+
\frac{ig'}{2}Y_{L} \bar{\tilde{L}} \partial^{m}\tilde{L}V_{m}-
\frac{ig'}{2}Y_{L}\tilde{L}\partial^{m}\bar{\tilde{L}}V_{m} \nonumber \\
&+&
\frac{ig'}{2}Y_{R} \bar{\tilde{R}} \partial^{m}\tilde{R}V_{m}-
\frac{ig'}{2}Y_{R} \tilde{R}\partial^{m}\bar{\tilde{R}}V_{m}
\nonumber \\
&=&ieQ_{e}(\tilde{e}^{*}_{L}\stackrel{\leftrightarrow}{\partial}^{m}\tilde{e}_{L}+
\tilde{e}^{*}_{R}\stackrel{\leftrightarrow}{\partial}^{m}\tilde{e}_{R})A_{m}+
\frac{ig}{\CWA}[(T^{3}_{e}-Q_{e} \sin^{2} \theta_{W})
\tilde{e}^{*}_{L}\stackrel{\leftrightarrow}{\partial}^{m}\tilde{e}_{L}
\nonumber \\
&-&Q_{e} 
\sin^{2} \theta_{W} \tilde{e}^{*}_{R} \stackrel{\leftrightarrow}{\partial}^{m} 
\tilde{e}_{R}]Z_{m} \,\ . \nonumber \\
\end{eqnarray}

Portanto a intera\c c\~ao completa \'e a seguinte
\BEA
\L_{\tilde{l}\tilde{l}V}&=&
\frac{ig}{\sqrt{2}}( 
\tilde{\nu}^{*} \stackrel{\leftrightarrow}{\partial}^{m}\tilde{e}_{L}
W^{+}_{m}+
\tilde{e}^{*}_{L} \stackrel{\leftrightarrow}{\partial}^{m} \tilde{\nu} 
W^{-}_{m}) \nonumber \\
&+&
ieQ_{e}(\tilde{e}^{*}_{L}\stackrel{\leftrightarrow}{\partial}^{m}\tilde{e}_{L}+
\tilde{e}^{*}_{R}\stackrel{\leftrightarrow}{\partial}^{m}\tilde{e}_{R})A_{m}+
\frac{ig}{\CWA}[(T^{3}_{e}-Q_{e} \sin^{2} \theta_{W})
\tilde{e}^{*}_{L}\stackrel{\leftrightarrow}{\partial}^{m}\tilde{e}_{L}
\nonumber \\
&-&Q_{e} 
\sin^{2} \theta_{W} \tilde{e}^{*}_{R} \stackrel{\leftrightarrow}{\partial}^{m} 
\tilde{e}_{R}]Z_{m} \,\ . \nonumber \\
\end{eqnarray}

\section{Intera\c c\~ao B\'oson de Gauge B\'oson Higgs B\'osons Higgs}

A lagrangiana de intera\c c\~ao envolvendo estas part\'\i culas \'e 
a seguinte
\BEA
\L_{VHH}&\ni& \int d^{4} \theta \LP\{ 
\hat{\bar{H}}_{1} \exp \LP[2 \LP( g \frac{\sigma^{i}}{2} \hat{V}^{i}+ 
\frac{g'}{2}Y_{H_{1}} \hat{v} \RP) \RP] \hat{H}_{1}+
\hat{\bar{H}}_{2} \exp \LP[ 2 \LP( g \frac{\sigma^{i}}{2} \hat{V}^{i}+ 
\frac{g'}{2}Y_{H_{2}} \hat{v} \RP) \RP] \hat{H}_{2} \RP\} \nonumber \\
&=& \frac{-ig}{2}H_{1}\sigma^i\partial^m\bar{H}_{1}V^i_m+
\frac{ig}{2}\bar{H}_{1}\sigma^i\partial^mH_{1}V^i_m \nonumber \\
&-& \frac{ig'}{2}Y_{H1}H_{1}\partial^m\bar{H}_{1}V_m+
\frac{ig'}{2}Y_{H1}\bar{H}_{1}\partial^mH_{1}V_m \nonumber \\
&-& \frac{ig}{2}H_{2}\sigma^i\partial^m\bar{H}_{2}V^i_m+
\frac{ig}{2}\bar{H}_{2}\sigma^i\partial^mH_{2}V^i_m \nonumber \\
&-& \frac{ig'}{2}Y_{H2}H_{2}\partial^m\bar{H}_{2}V_m+
\frac{ig'}{2}Y_{H2}\bar{H}_{2}\partial^mH_{2}V_m \nonumber \\
&=& \L^{carregada}_{VHH}+\L^{neutra}_{VHH}
\,\ .
\EEA

Podemos escrever a parte carregada da seguinte maneira
\BEA
\L^{carregada}_{VHH}&=& \frac{-ig}{2} \left[ H_{1}(
\sigma^1V^1_m+\sigma^2V^2_m)\partial^m\bar{H}_{1} \right] +
\frac{ig}{2} \left[ \bar{H}_{1}(
\sigma^1V^1_m+\sigma^2V^2_m)\partial^{m}H_{1} \right] \nonumber \\
&-& \frac{ig}{2} \left[ H_{2}(
\sigma^1V^1_m+\sigma^2V^2_m)\partial^m\bar{H}_{2} \right] +
\frac{ig}{2} \left[ \bar{H}_{2}(
\sigma^1V^1_m+\sigma^2V^2_m)\partial^{m}H_{2} \right] \nonumber \\
&=&\frac{ig}{\sqrt{2}} [\bar{H}_{1}\stackrel{\leftrightarrow}{\partial}^{m}
H_{1}+\bar{H}_{2}\stackrel{\leftrightarrow}{\partial}^{m}H_{2}]
\LP( \BA{cc} 
O & W^{+}_{m} \\
W^{-}_{m} & 0          \EA \RP) \nonumber \\
&=& \frac{-ig}{\sqrt{2}}\left[
H_{1}^{1*}\stackrel{\leftrightarrow}{\partial}^{m}H_{1}^{2}W^{+}_{m}+
H_{1}^{2*}\stackrel{\leftrightarrow}{\partial}^{m}H_{1}^{1}W^{-}_{m}
\right. \nonumber \\
&+&\left. 
H_{2}^{1*}\stackrel{\leftrightarrow}{\partial}^{m}H_{2}^{2}W^{+}_{m}+
H_{2}^{2*}\stackrel{\leftrightarrow}{\partial}^{m}H_{2}^{1}W^{-}_{m}
\right] \,\ .
\EEA

J\'a para a parte neutra encontraremos
\BEA
\L^{neutra}_{VHH}&=&
\frac{-ig}{2}H_{1}\sigma^3V^3_m\partial^m\bar{H}_{1}+
\frac{ig}{2}\bar{H}_{1}\sigma^3V^3_m\partial^mH_{1} \nonumber \\
&-& \frac{ig'}{2}Y_{H1}H_{1}\partial^m\bar{H}_{1}V_m+
\frac{ig'}{2}Y_{H1}\bar{H}_{1}\partial^mH_{1}V_m \nonumber \\
&-&
\frac{ig}{2}H_{2}\sigma^3V^3_m\partial^m\bar{H}_{2}+
\frac{ig}{2}\bar{H}_{2}\sigma^3V^3_m\partial^mH_{2} \nonumber \\
&-& \frac{-ig'}{2}Y_{H2}H_{2}\partial^m\bar{H}_{2}V_m+
\frac{ig'}{2}Y_{H2}\bar{H}_{2}\partial^mH_{2}V_m \nonumber \\
&=&
\frac{ig}{2}[\bar{H}_{1}\stackrel{\leftrightarrow}{\partial}^{m}H_{1}+
\bar{H}_{2}\stackrel{\leftrightarrow}{\partial}^{m}H_{2}]
\sigma^3V^3_m \nonumber \\
&+& \frac{ig'}{2}Y_{H1}\bar{H}_{1}\stackrel{\leftrightarrow}{\partial}^{m}
H_{1}V_m+
\frac{ig'}{2}Y_{H2}\bar{H}_{2}\stackrel{\leftrightarrow}{\partial}^{m}
H_{2}V_m \nonumber \\
&=&-ie(H_{1}^{2*}
\stackrel{\leftrightarrow}{\partial}^{m}H_{1}^{2}-
H_{2}^{1*}\stackrel{\leftrightarrow}{\partial}^{m}H_{2}^{1}
)A_{m} \nonumber \\
&+&\frac{ig}{2\CWA}\left[(H_{2}^{2*}
\stackrel{\leftrightarrow}{\partial}^{m}H_{2}^{2}-
H_{1}^{1*}\stackrel{\leftrightarrow}{\partial}^{m}H_{1}^{1}
) \right. \nonumber \\
&+& \left.(2\sin^{2} \theta_{W}-1)(H_{1}^{2*}
\stackrel{\leftrightarrow}{\partial}^{m}H_{1}^{2}-
H_{2}^{1*}\stackrel{\leftrightarrow}{\partial}^{m}H_{2}^{1}
) \right]Z_{m} \,\ .
\EEA

Logo a lagrangiana completa de intera\c c\~ao \'e
\BEA
\L_{VHH}&=&
\frac{-ig}{\sqrt{2}}\left[
H_{1}^{1*}\stackrel{\leftrightarrow}{\partial}^{m}H_{1}^{2}W^{+}_{m}+
H_{1}^{2*}\stackrel{\leftrightarrow}{\partial}^{m}H_{1}^{1}W^{-}_{m}
\right. \nonumber \\
&+&\left. 
H_{2}^{1*}\stackrel{\leftrightarrow}{\partial}^{m}H_{2}^{2}W^{+}_{m}+
H_{2}^{2*}\stackrel{\leftrightarrow}{\partial}^{m}H_{2}^{1}W^{-}_{m}
\right] \nonumber \\
&-&
ie(H_{1}^{2*}
\stackrel{\leftrightarrow}{\partial}^{m}H_{1}^{2}-
H_{2}^{1*}\stackrel{\leftrightarrow}{\partial}^{m}H_{2}^{1}
)A_{m} \nonumber \\
&+&\frac{ig}{2\CWA}\left[(H_{2}^{2*}
\stackrel{\leftrightarrow}{\partial}^{m}H_{2}^{2}-
H_{1}^{1*}\stackrel{\leftrightarrow}{\partial}^{m}H_{1}^{1}
) \right. \nonumber \\
&+& \left.(2\sin^{2} \theta_{W}-1)(H_{1}^{2*}
\stackrel{\leftrightarrow}{\partial}^{m}H_{1}^{2}-
H_{2}^{1*}\stackrel{\leftrightarrow}{\partial}^{m}H_{2}^{1}
) \right]Z_{m} \,\ .
\EEA
Onde $\stackrel{\leftrightarrow}{\partial}$ est\'a dada pela 
Eq.(\ref{ucraniano}).

\section{Intera\c c\~ao B\'oson Higgs Higgsino Gaugino}

As intera\c c\~oes deste setor prov\^em dos seguintes termos
\BEA
\L_{H\tilde{H}\tilde{V}}&\ni& \int d^{4} \theta \LP\{ 
\hat{\bar{H}}_{1} \exp \LP[2 \LP( g \frac{\sigma^{i}}{2} \hat{V}^{i}+ 
\frac{g'}{2}Y_{H_{1}} \hat{v} \RP) \RP] \hat{H}_{1}+
\hat{\bar{H}}_{2} \exp \LP[ 2 \LP( g \frac{\sigma^{i}}{2} \hat{V}^{i}+ 
\frac{g'}{2}Y_{H_{2}} \hat{v} \RP) \RP] \hat{H}_{2} \RP\} \nonumber \\
&=& \frac{-ig}{\sqrt{2}}(\bar{\tilde{H}}_{1} \sigma^{i}H_{1}
\bar{\lambda}^i_A-\bar{H}_{1}\sigma^i\tilde{H}_{1}
\lambda^i_A)- \frac{ig'}{\sqrt{2}}Y_{H1}(\bar{\tilde{H}}_{1}H_{1}
\bar{\lambda}_B-\bar{H}_1\tilde{H}_1\lambda_B)
\nonumber \\
&-& \frac{ig}{\sqrt{2}}(\bar{\tilde{H}}_{2} \sigma^{i}H_{2}
\bar{\lambda}^i_A-\bar{H}_{2}\sigma^i\tilde{H}_{2}
\lambda^i_A)- \frac{ig'}{\sqrt{2}}Y_{H2}(\bar{\tilde{H}}_{2}H_{2}
\bar{\lambda}_B-\bar{H}_2\tilde{H}_2\lambda_B)
\nonumber \\
&=& \L^{carregada}_{H\tilde{H}\tilde{V}}+\L^{neutra}_{H\tilde{H}\tilde{V}}
\,\ .
\EEA

Onde a parte carregada \'e a seguinte
\BEA
\L^{carregada}_{H\tilde{H}\tilde{V}}&=& \frac{-ig}{2}\left[ 
\bar{\tilde{H}}_{1}(\sigma^1\bar{\la}^1_A+\sigma^2\bar{\la}^2_A)H_{1}-
\bar{H}_{1}(\sigma^1\la^1_A+\sigma^2\la^2_A)\tilde{H}_{1} \right]
\nonumber \\
&-& \frac{ig}{2}\left[ 
\bar{\tilde{H}}_{2}(\sigma^1\bar{\la}^1_A+\sigma^2\bar{\la}^2_A)H_{2}-
\bar{H}_{2}(\sigma^1\la^1_A+\sigma^2\la^2_A)\tilde{H}_{2} \right] 
\nonumber \\
&=&g \left[
H_{1}^{1*}\lambda^{+}\psi^{2}_{H1}+H_{1}^{2*}\lambda^{-}
\psi^{1}_{H1} \right. \nonumber \\
&+& \left. H_{2}^{2*}\lambda^{-}\psi^{1}_{H2}+H_{2}^{1*}
\lambda^{+}\psi^{2}_{H2}+h.c. \right] \nonumber \\
&=&ig \left[
H_{1}^{1*}\bar{\tilde{H}}R\tilde{W}+
H_{1}^{2*}\bar{\tilde{W}}R\tilde{H}^{n}_{1} \right. \nonumber \\
&+& \left. H_{2}^{2*}\bar{\tilde{W}}R\tilde{H}+
H_{2}^{1*}\bar{\tilde{H}}^{n}_{2}R\tilde{W}+
h.c. \right] \,\ .
\EEA

Agora vamos analisar a parte neutra que \'e
\BEA
\L^{neutra}_{H\tilde{H}\tilde{V}}&=&
\frac{-ig}{\sqrt{2}}(\bar{\tilde{H}}_{1}\sigma^3\bar{\la}^3_A
H_{1}-\bar{H}_{1}\sigma^3\la^3_A\tilde{H}_{1})-
\frac{ig}{\sqrt{2}}(\bar{\tilde{H}}_{2}\sigma^3\bar{\la}^3_A
H_{2}-\bar{H}_{2}\sigma^3\la^3_A\tilde{H}_{2}) \nonumber \\
&-& \frac{ig'}{2}Y_{H1}(\bar{\tilde{H}}_{1}\bar{\la}_{B}H_{1}-
\bar{H}_{1}\la_B\tilde{H}_{1})-
\frac{ig'}{2}Y_{H2}(\bar{\tilde{H}}_{2}\bar{\la}_{B}H_{2}-
\bar{H}_{2}\la_B\tilde{H}_{2}) \nonumber \\
&=&
\frac{1}{\sqrt{2}}\left[
e(H_{2}^{1*}\bar{\tilde{A}}R\tilde{H}-H_{1}^{2*}\bar{\tilde{H}}R
\tilde{A}) \right. \nonumber \\
&+&\frac{g}{2\CWA}\left(H_{1}^{1*}\bar{\tilde{H}}^{n}_{1}R\tilde{Z}
-H_{2}^{2*}\bar{\tilde{Z}}R\tilde{H}^{n}_{2}+(1-2\sin^{2} \theta_{W})(
H_{2}^{1*}\bar{\tilde{Z}}R\tilde{H}-
H_{1}^{2*}\bar{\tilde{H}}R\tilde{Z})+h.c. \right] \,\ . \nonumber \\
\EEA

Ou seja obtemos
\BEA
\L^{neutra}_{H\tilde{H}\tilde{V}}&=&
g \left[
H_{1}^{1*}\bar{\tilde{H}}R\tilde{W}+
H_{1}^{2*}\bar{\tilde{W}}R\tilde{H}^{n}_{1} \right. \nonumber \\
&+& \left. H_{2}^{2*}\bar{\tilde{W}}R\tilde{H}+
H_{2}^{1*}\bar{\tilde{H}}^{n}_{2}R\tilde{W}+
h.c. \right] \nonumber \\
&+&
\frac{1}{\sqrt{2}}\left[
e(H_{2}^{1*}\bar{\tilde{A}}R\tilde{H}-H_{1}^{2*}\bar{\tilde{H}}R
\tilde{A}) \right. \nonumber \\
&+&\frac{g}{2\CWA}\left(H_{1}^{1*}\bar{\tilde{H}}^{n}_{1}R\tilde{Z}
-H_{2}^{2*}\bar{\tilde{Z}}R\tilde{H}^{n}_{2}+(1-2\sin^{2} \theta_{W})(
H_{2}^{1*}\bar{\tilde{Z}}R\tilde{H}-
H_{1}^{2*}\bar{\tilde{H}}R\tilde{Z})+h.c. \right] \,\ . \nonumber \\
\EEA

\section{Os Campos Auxiliares.}

Nesta se\c c\~ao iremos mostrar como eliminar os campos auxiliares da 
Tab.\ref{Table:  Auxiliary Component fields}. Se n\'os apanhamos todos os 
termos $F$ e D das Eqs.(\ref{Component Field Expansion prop 1}), 
(\ref{Component Field Expansion prop 2}), 
(\ref{Component Field Expansion prop 3}), 
(\ref{Component Field Expansion prop 4}), 
(\ref{Component Field Expansion prop 5}), 
(\ref{Component Field Expansion prop 6}), 
(\ref{Component Field Expansion prop 9}), 
(\ref{Component Field Expansion prop 10}), (\ref{lint}),(\ref{rint}),(\ref{h1int}), 
(\ref{h2int}) e (\ref{Komp higgs property 1}) teremos

\BEA
  \L_{Aux} &=& \L_{Aux-F} + \L_{Aux-D} \,\ ,
     \label{Elimination of the Auxiliary Fields prop 1}
\EEA
com
\BEA
    \L_{Aux-F}  &=&
          \bar{F}_{L}F_{L}
       +  \bar{F}_{R}F_{R}
       +  \bar{F}_{1}F_{1}
       +  \bar{F}_{2}F_{2} \nn
   & & \mbox{}
      + \mu\;\e_{ij}\LP[\,
            H_{1}^{i}F_{2}^{j}
         +   \bar{H}_{1}^{i}\bar{F}_{2}^{j}
         +   F_{1}^{i}H_{2}^{j}
         +   \bar{F}_{1}^{i}\bar{H}_{2}^{j}\,\RP] \nn
    & & \mbox{}
    +   f\;\e_{ij}\LP[\,
            F_{1}^{i}\tilde{L}^{j}\tilde{R}
         +  \bar{F}_{1}^{i}\tilde{\bar{L}}^{j}\tilde{\bar{R}}
         +  H_{1}^{i}F_{L}^{j}\tilde{R}
         +  \bar{H}_{1}^{i}\bar{F}_{L}^{j}\tilde{\bar{R}}\RP. \nn
     & & \mbox{}  \LP. \hspace{1.2cm}
         +  H_{1}^{i}\tilde{L}^{j}F_{R}
         +  \bar{H}_{1}^{i}\tilde{\bar{L}}^{j}F_{\bar{R}} \, \RP] \,\ ,
         \label{Elimination of the Auxiliary Fields prop 2}
\EEA
e
\BEA
   \L_{Aux-D} &=&
       \HA\;\LP(\;D^{i}D^{i}+DD\;\RP) \nn
  & & \mbox{}
      +\tilde{\bar{L}}\LP(gT^{i}D^{i}-\HA g'D\RP)\tilde{L}
      +\tilde{\bar{R}}g'D\tilde{R}  \nn
  & & \mbox{}
      + \bar{H}_{1}\LP(gT^{i}D^{i}-\HA g'D\RP)H_{1}
      + \bar{H}_{2}\LP(gT^{i}D^{i}+\HA g'D\RP)H_{2} \,\ . \nonumber \\
      \label{Elimination of the Auxiliary Fields prop 3}
\EEA

Aplicando a equa\c c\~ao simplificada vista na se\c c\~ao 6.1, obtemos as 
seguintes rela\c c\~oes para os campos auxiliares
\BEA
    \bar{F}_{L}^{j}   &=& -f\;\e^{ij}H_{1}^{i}\tilde{R} \,\ ,
       \nonumber \\
    \bar{F}_{R}      &=& -f\;\e^{ij}H_{1}^{i}\tilde{L}^{j} \,\ , \nonumber \\
    \bar{F}_{1}^{i}   &=& -\mu\;\e^{ij}H_{2}^{j}
                             - f\;\e^{ij}\tilde{L}^{j}\tilde{R} \,\ ,\nonumber \\
    \bar{F}_{2}^{j}   &=& -\mu\;\e^{ij}H_{1}^{i} \,\ ,
        \label{aux prop 4}
\EEA
e
\BEA
  D^{i} &=& -g\LP[\,
            \tilde{ \bar{L}}T^{i}\tilde{L}
          + \bar{H}_{1}T^{i}H_{1}
          + \bar{H}_{2}T^{i}H_{2} \RP] \,\ ,
            \nonumber \\
  D &=&   \f{g'}{2}\,\tilde{ \bar{L}}\tilde{L}
         - g'\, \tilde{ \bar{R}}\tilde{R}
         + \f{g'}{2}\, \bar{H}_{1}H_{1}
         - \f{g'}{2}\, \bar{H}_{2}H_{2} \,\ .
            \label{aux prop 6}
\EEA

\subsection{Elimina\c c\~ao do Campo Auxiliar $F$.}

Com as Eqs.\r{aux prop 4} temos
\BEA
    \L_{Aux-F}  &=&
         \LP(-f\e^{ij}\,H_{1}^{i}\tilde{R}\RP)
              \LP(-f\e^{kj}\,\bar{H}_{1}^{k}\tilde{\bar{R}}\RP)
       +  \LP(-f\e^{ij}\,H_{1}^{i}\tilde{L}^{j}\RP)
              \LP(-f\e^{kl}\,\bar{H}_{1}^{k}\tilde{\bar{L}}^{l}\RP)
    \nmb
       +  \LP(-\mu\e^{ij}\,H_{2}^{j}-f\e^{ij}\tilde{L}^{j}\tilde{R}\RP)
              \LP(-\mu\e^{ik}\,\bar{H}_{2}^{k}
             -f\e^{ik}\tilde{\bar{L}}^{k}\tilde{\bar{R}}\RP)
    \nmb
       +  \LP(-\mu\e^{ij}\,H_{1}^{i}\RP)
              \LP(-\mu\e^{kj}\,\bar{H}_{1}^{k}\RP)
     \nmb
         + \mu\;\e^{ij}\, H_{1}^{i}\LP(-\mu\e^{kj}\,\bar{H}_{1}^{k}\RP)
         + \mu\;\e^{ij}\,  \bar{H}_{1}^{i}\LP(-\mu\e^{kj}\,H_{1}^{k}\RP)
     \nmb
         + \mu\;\e^{ij}\,  \LP(-\mu\e^{ik}\,\bar{H}_{2}^{k}
         -f\e^{ik}\tilde{\bar{L}}^{k}\tilde{\bar{R}}\RP) H_{2}^{j}
     \nmb
         + \mu\;\e^{ij}\,  \LP(-\mu\e^{ik}\,
          H_{2}^{k}-f\e^{ik}\tilde{L}^{k}\tilde{R}\RP)\bar{H}_{2}^{j}\,
     \nmb
         + f\;\e^{ij}\,\LP(-\mu\e^{ik}\,\bar{H}_{2}^{k}
   -f\e^{ik}\tilde{\bar{L}}^{k}\tilde{\bar{R}}\RP)\tilde{L}^{j}\tilde{R}
     \nmb
               +  f\;\e^{ij}\,\LP(-\mu\e^{ik}\,H_{2}^{k}
              -f\e^{ik}\tilde{L}^{k}\tilde{R}\RP)
                     \tilde{\bar{L}}^{j}\tilde{\bar{R}}
     \nmb
         +  f\;\e^{ij}\,H_{1}^{i}\LP(-f\e^{kj}
                   \,\bar{H}_{1}^{k}\tilde{\bar{R}}\RP)\tilde{R}
         +  f\;\e^{ij}\,\bar{H}_{1}^{i}\LP(-f\e^{kj}\,H_{1}^{k}\tilde{R}\RP)
                 \tilde{\bar{R}}
     \nmb
         +  f\;\e^{ij}\,H_{1}^{i}\tilde{L}^{j}\LP(-f\e^{kl}\,\bar{H}_{1}^{k}
                 \tilde{\bar{L}}^{l}\RP)
         +  f\;\e^{ij}\,\bar{H}_{1}^{i}\tilde{\bar{L}}^{j}\LP(-f\e^{kl}
                          \,H_{1}^{k}\tilde{L}^{l}\RP)  \NN
   &=&
       -\mu^{2}\, \bar{H}_{1}H_{1}
        -\mu^{2}\, \bar{H}_{2}H_{2}
        -\mu f \LP[\, \bar{H}_{2}\tilde{L}\,\tilde{R}
                  +\tilde{\bar{L}}H_{2}\,\tilde{\bar{R}}\,\RP]\nn
   & & \mbox{}
     - f^{2}\LP[ \,
       \tilde{\bar{L}}\tilde{L}\,\tilde{\bar{R}}\tilde{R}
     + \bar{H}_{1}H_{1}\LP( \tilde{\bar{L}}\tilde{L}
                                + \tilde{\bar{R}}\tilde{R} \RP)
     - \bar{H}_{1}\tilde{L}\LP(\bar{H}_{1}
                   \tilde{L}\RP)^{\dagger}\,\RP] \,\ .
         \label{aux prop 6aaaaaa}
\EEA
Na \'ultima passagem usamos as seguintes rela\c c\~oes:
\BEA
  \e^{ij}\e^{kj} &=& \d^{ik} \,\ ,\nonumber \SL
  \e^{ij}\e^{kl} &=& \d^{ik}\d^{jl}-\d^{il}\d^{jk} \,\ . \nonumber
\EEA

\subsection{Campos Auxiliares D.}

Vamos reescrever $\L_{Aux-D}$, introduzindo as seguintes abrevia\c c\~oes 
tempor\'arias
\BEA
   A &=& \tilde{\bar{L}}T^{i}\tilde{L} \,\ ,\nn
   B &=& \bar{H}_1 T^i H_1 \,\ ,\nn
   C &=& \bar{H}_2 T^i H_2 \,\ , \nn
   D &=& \frac{Y_{L}}{2}\tilde{\bar{L}}\tilde{L} \,\ , \nn
   E &=& \frac{Y_{R}}{2}\tilde{\bar{R}}\tilde{R} \,\ ,\nn
   F &=& \frac{Y_{H1}}{2}\bar{H}_1 H_1 \,\ ,\nn
   G &=& \frac{Y_{H2}}{2}\bar{H}_2 H_2 \,\ . \nonumber
\EEA
Onde \'\i ndece de $SU(2)$ ``i"  foi suprimido por conveni\^encia.

Com estas abrevia\c c\~oes Eqs.\r{aux prop 6} adquirem a 
seguinte forma
\BEA
   D^i &=& -g \LP[A+B+C \RP] \,\ ,\nn
   D  &=&  g'\LP[D+E+F+G \RP] \,\ .  \nonumber
\EEA

Para $\L_{Aux-D}$ isto implica
\BEA
  \L_{Aux-D}
     &=& - \f{g^2}{2} \LP(A+B+C\RP)\LP(A+B+C\RP)
    \nmb
        - \f{g'^2}{2}
               \LP( D+E+F+G \RP)
                \LP( D+E+F+G \RP) \,\ ,
                 \nonumber
\EEA
ou em termos de nossos campos
\BEA
  \L_{Aux-D}
        &=& -\f{g^2}{2}
         \LP(\,\tilde{\bar{L}}T^{i}
          \tilde{\bar{L}}+H_{1}T^{i}H_{1}+\bar{H}_{2}T^{i}H_{2}\,\RP)
         \LP(\,\tilde{\bar{L}}T^{i}\tilde{L}
             +\bar{H}_{1}T^{i}H_{1}+\bar{H}_{2}T^{i}H_{2}\,\RP)
    \nmb
        -\f{g'^2}{2}
          \LP(\,\tilde{\bar{L}}\f{Y_{L}}{2} \tilde{L}+
\tilde{\bar{R}}\f{Y_{R}}{2}\tilde{R}
+\bar{H}_{1}\f{Y_{H1}}{2}H_{1}+
\bar{H}_{2}\f{Y_{H2}}{2}H_{2}\,\RP)^2 \,\ .
                             \label{sabado}
\EEA

Lembrando que
\BEA
\sigma^{i}_{ab} \sigma^{i}_{cd}=2\delta_{ad}\delta_{bc}-\delta_{ab}\delta{cd} 
\,\ ,
\EEA
podemos mostrar que
\BEA
\LP(\,\tilde{\bar{L}}T^{i}
          \tilde{\bar{L}}+H_{1}T^{i}H_{1}+\bar{H}_{2}T^{i}H_{2}\,\RP)^2&=& \nn
&&\f{1}{4} \LP[ \bar{\tilde{L}}\tilde{L} \bar{\tilde{L}}\tilde{L}+
4 \bar{\tilde{L}}H_{1} \bar{H}_{1}\tilde{L}-
2 \bar{\tilde{L}}\tilde{L} \bar{H}_{1}H_{1} \RP. \nn
&+&
\LP. 4 \bar{\tilde{L}}H_{2} \bar{H}_{2}\tilde{L} - 
2 \bar{\tilde{L}}\tilde{L} \bar{H}_{2}H_{2} \RP. \nn
&+& \LP.
(\bar{H}_{1}H_{1}-\bar{H}_{2}H_{2})^{2}+
4 \vert \bar{H}_{2}H_{1} \vert^{2} \RP] \,\ , \nn
\LP(\,\tilde{\bar{L}}\f{Y_{L}}{2} \tilde{L}+
\tilde{\bar{R}}\f{Y_{R}}{2}\tilde{R}
+\bar{H}_{1}\f{Y_{H1}}{2}H_{1}+
\bar{H}_{2}\f{Y_{H2}}{2}H_{2}\,\RP)^2&=& \nn
&&\f{1}{4} \LP[ Y^{2}_{L} \bar{ \tilde{L}} \tilde{L} \bar{ \tilde{L}}
 \tilde{L}+  Y^{2}_{R} \bar{ \tilde{R}} \tilde{R} \bar{ \tilde{R}}
 \tilde{R} \RP. \nn
&+& \LP. Y^{2}_{H1} \bar{H}_{1}H_{1} \bar{H}_{1}H_{1}+ 
 Y^{2}_{H2} \bar{H}_{2}H_{2} \bar{H}_{2}H_{2} \RP. \nn
 &+& \LP. 2( Y_{L}Y_{R} \bar{ \tilde{L}} \tilde{L} 
\bar{ \tilde{R}} \tilde{R}+ Y_{L}Y_{H1} \bar{ \tilde{L}} \tilde{L} 
\bar{H}_{1}H_{1}) \RP. \nn
&+& \LP. 2( Y_{L}Y_{H2} \bar{ \tilde{L}} \tilde{L} 
\bar{H}_{2}H_{2}+  Y_{R}Y_{H1} \bar{ \tilde{R}} \tilde{R} 
\bar{H}_{1}H_{1}) \RP. \nn
&+& \LP. 2(  Y_{R}Y_{H2} \bar{ \tilde{R}} \tilde{R} 
\bar{H}_{2}H_{2}+  Y_{H1}Y_{H2} \bar{H}_{1} H_{1} \bar{H}_{2}H_{2}) \RP] 
\,\ . \nonumber \\      
\label{domingo}
\EEA

\section{Condi\c c\~ao de Quebra.}

A quebra de simetria de gauge est\'a no {\bf MSSM} diretamente relacionada 
a quebra de supersimetria. Agora iremos estudar sobre que circunst\^ancias 
ocorrem estas quebras.

Em teorias de supersimetria, temos dois tipos de potenciais o superpotencial 
e os potenciais escalares. O superpotencial j\'a foi discutido anteriormente
neste estudo, assim agora iremos estudar o potencial escalar, que tem sua 
analogia com o {\bf SM} como mostraremos a seguir.

As contribui\c c\~oes ao potencial escalar do {\bf MSSM}, $V_{MSSM}$, vem de 
tr\^es fontes, os termos $F$ e $D$ mais os termos soft. Dessa maneira escreveremos
\BEA
   V_{MSSM} &=& V_{D} + V_{F} + V_{Soft} \,\ ,
      \label{Radiative Breaking prop 1}
\EEA
onde
\BEA
     V_{D} &=& - \L_{Aux-D} \,\ , \nn
     V_{F} &=& -\L_{Aux-F} \,\ , \nn
    V_{Soft} &=& - \L_{SMT} \,\ .  
\label{Radiative Breaking prop 1aaaa}
\EEA

Agora iremos abandonar este potencial escalar geral, e iremos nos concentrar 
apenas no potencial de Higgs porque \'e este o potencial de interesse na discuss\~ao 
de quebra de simetria de gauge.

\subsection{Potencial Escalar de Higgs.}

Assim, para o setor puro de Higgs da teoria, o potencial de Higgs $V\equiv V_{Higgs}$ 
\'e escrito da seguinte maneira\footnote{Este potencial \'e 
um caso especial do geral de dois dubletos de Higgs.}
\BEA
    V &=&   \LP(M_{1}^{2}+\mu^{2}\RP) \bar{H}_{1}\!H_{1}
          + \LP(M_{2}^{2}+\mu^{2}\RP) \bar{H}_{2}\!H_{2}
          - M_{12}^{2}\,\e_{ij}\LP(H_{1}^{i}H_{2}^{j} +h.c.\RP)
      \nmb
          +\f{g^{2}}{2} \LP( \bar{H}_{1}T^{i}H_{1}
                 +\bar{H}_{2}T^{i}H_{2}\RP)
                        \LP( \bar{H}_{1}T^{i}H_{1}
                  +\bar{H}_{2}T^{i}H_{2}\RP)
      \nmb
          + \f{g'^{2}}{8} \LP( \bar{H}_{1}\!H_{1}
                   -  \bar{H}_{2}\!H_{2}   \RP)^{2} \,\ .
              \label{Radiative Breaking prop 1aaa}
\EEA
Este potencial pode ser escrito, usando a Eq.(\ref{domingo}), 
da seguinte maneira
\BEA
    V &=&   m_{1}^{2}\,\bar{H}_{1}H_{1}
          + m_{2}^{2}\,\bar{H}_{2}H_{2}
          - M_{12}^{2}\,\e_{ij}\LP(H_{1}^{i}H_{2}^{j} +h.c.\RP)
      \nmb
          +\f{1}{8}\LP(g^{2}+g'^{2}\RP)
                \LP( \bar{H}_{1}H_{1}-\bar{H}_{2}H_{2}\RP)^{2}
          + \f{g^{2}}{2} \LP|\bar{H}_{1}H_{2}\RP|^{2} \,\ .
            \label{Tree-level scalar potential}
\EEA
Onde
\BEA
m^{2}_{1}&=& M_{1}^{2}+\mu^{2} \,\ , \nn
m^{2}_{2}&=& M_{2}^{2}+\mu^{2} \,\ .
\label{stine}
\EEA

Sem perda de generalidade, podemos escolher as fases dos campos escalares dos 
Higgs de tal maneira que todos os par\^ametros de massa $m_{i}^{2}$
($i=1,2$) e $M^{2}_{12}$ sejam reais e que os valores esperados dos v\'acuos 
(v.e.v.) 
dos campos dos Higgs sejam positivos. Como no {\bf SM} o grupo de
simetria de gauge $SU(2)\otimes U(1)$ quebre para a seguinte simetria 
$U(1)_{EM}$. Isto significa que o eletromagnetismo n\~ao \'e quebrado e 
portanto as componentes carregadas dos dubletos de Higgs n\~ao podem adquirir 
(v.e.v.). Com base no escrito acima e usando Eq.(\ref{Component Field Expansion prop 7}) 
podemos escrever
\BEA
   \LP< H_{1}\RP> &=& \LP(\BA{c} v_{1}\\ 0 \EA \RP) \,\ ,
      \nonumber \\
   \LP< H_{2}\RP> &=& \LP(\BA{c} 0 \\ v_{2} \EA \RP) \,\ ,
      \label{Radiative Breaking prop 3}
\EEA
e ap\'os esta quebra o potencial torna-se
\BEA
    V   &=&   m_{1}^{2}\,v_{1}^{2}
          + m_{2}^{2}\,v_{2}^{2}
          - 2 M_{12}^{2}\,v_{1}v_{2}
          +\f{1}{8}\LP(g^{2}+g'^{2}\RP)\LP[v_{1}^{2}-v_{2}^{2}\RP]^{2} \,\ .
              \label{Radiative Breaking prop 4}
\EEA

Este termo \'e positivo, assim para que o potencial tenha um valor m\'\i nimo, na dire\c c\~ao 
$v_{1}=v_{2}$, temos que ter
\BEA
  {\cal B} \equiv m_{1}^{2}+m_{2}^{2}-2 M_{12}^{2} \geq 0 \,\ .
       \label{Radiative Breaking prop 5}
\EEA
Esta rela\c c\~ao \'e conhecida como {\it condi\c c\~ao de estabilidade}.

Do mecanismo de Higgs do {\bf SM} \'e bem conhecido o fato que quando o Higgs 
adquire um $(v.e.v)$ diferente de zero, quebra a simetria  
$SU(2) \otimes U(1)$ porque a origem \'e ``inst\'avel". Vamos reescrer 
Eq.~\r{Radiative Breaking prop 4} da seguinte maneira
\BEA
  V &=& \mbox{v}^{T}{\cal M}^{2}\mbox{v} +
          \f{1}{8}\LP(g^{2}+g'^{2}\RP)\LP[v_{1}^{2}-v_{2}^{2}\RP]^{2} \,\ ,
                 \label{Radiative Breaking prop 6}
\EEA
com as seguintes identifica\c c\~oes
\BEA
     \mbox{v}      &=&  \LP( \BA{c} v_{1}\\ -v_{2}\EA\RP) \,\ ,
                 \label{Radiative Breaking prop 7} \nn
     {\cal M}^{2}  &=&  \LP( \BA{cc}  m_{1}^{2} & M_{12}^{2}\\
                                      M_{12}^{2} & m_{2}^{2}   \EA  \RP) \,\ .
                                       \label{Radiative Breaking prop 8}
\nonumber
\EEA
Como ${\cal M}^{2}$ \'e uma matriz sim\'etrica, devemos ter
\BEA
     \la_{-}\ABS{\mbox{v}}^{2}\;\, \leq \;\,
    \mbox{v}^{T}{\cal M}^{2}\mbox{v}\;\,  \leq \;\,  \la_{+}\ABS{\mbox{v}}^{2} \,\ .
         \label{Radiative Breaking prop 9}
\EEA
Onde $\la_{\pm}$ s\~ao os auto-valores de ${\cal M}^{2}$ dados por
\BEA
   \la_{\pm} &=& \HA \LP(m_{1}^{2}+m_{2}^{2}
        \pm \sqrt{\LP( m_{1}^{2}+m_{2}^{2} \RP)^{2}
          -4\LP(m_{1}^{2}m_{2}^{2}-M_{12}^{4}\RP)}\RP) \,\ ,
         \label{Radiative Breaking prop 10}
\EEA

Como o \'ultimo termo da Eq.(\ref{Radiative Breaking prop 6}) \'e sempre positivo, 
a forma quadr\'atica $\mbox{v}^{T}{\cal M}^{2}\mbox{v}$ tem que estar em seu 
valor m\'\i nimo para que $V$ tamb\'em esteja no seu valor m\'\i nimo, isto \'e
\BEA
   \mbox{v}^{T}{\cal M}^{2}\mbox{v} &=& \la_{-}\ABS{\mbox{v}}^{2} \,\ . \nonumber
\EEA
Portanto para obtermos $V_{min}<0$ devemos ter $\la_{-}< 0$,
ou seja, obtemos a seguinte condi\c c\~ao
\BEA
   \det{{\cal M}^{2}} = m_{1}^{2}m_{2}^{2}-M_{12}^{4} < 0 \,\ .
      \label{Radiative Breaking prop 12}
\EEA
Assim se a Eq.(\ref{Radiative Breaking prop 12}),
e a condi\c c\~ao de estabilidade Eq.(\ref{Radiative Breaking prop 5}), 
s\~ao satisfeitas teremos a quebra da simetria de gauge 
$SU(2)\otimes U(1)$. \'E importante comentar que as condi\c c\~oes 
Eqs.(\ref{Radiative Breaking prop 12}) e (\ref{Radiative Breaking prop 5}) 
n\~ao podem ser simultaneamente satisfeitas se $m_{1}^{2}=m_{2}^{2}$. 
Al\'em disto, 
Eq.(\ref{stine}) mostra que a contribui\c c\~ao supersim\'etrica a 
$m_{1}^{2}$ e $m_{2}^{2}$ \'e a mesma; qualquer diferen\c ca entre estas 
duas quantidades \'e devida aos termos $M_{1}^{2}$ e $M_{2}^{2}$ que vem 
do termo de quebra da supersimetria. Em outras palavras; no {\bf MSSM} existe uma 
conex\~ao entre quebra da simetria de gauge e quebra de supersimetria. Ou 
seja primeiro precisamos quebrar supersimetria para depois quebrar a simetria 
de gauge.

Vamos supor que estas condi\c c\~oes s\~ao satisfeitas e mostraremos 
que isto implica a correta quebra de simetria do modelo.

Ap\'os quebrar a simetria de gauge, tr\^es dos oitos graus de liberdade 
contido nos dois dubletos de Higgs s\~ao ``comidos'' pelos modos longitudinais 
dos b\'osons de gauge $W^{\pm}$ e $Z^{0}$. Os cincos graus de liberdade 
restantes que permanecem formam um Higgs pseudoescalar neutro, dois escalares 
neutros e dois b\'osons de Higgs carregados. A obten\c c\~ao de todo o espectro 
de massa do modelo \'e o assunto da pr\'oxima se\c c\~ao.

\section{Determina\c c\~ao das Massas.}

Nesta se\c c\~ao iremos calcular as massas dos nossos estados 
f\'\i sicos deste modelo.

\subsection{As Massas dos B\'osons de Gauge}

O termo de massa dos b\'osons de gauge vem dos seguintes termos
\BEA
\int d^{4} \theta \hat{ \bar{H}}_{1}e^{2g\hat{V}+g'\hat{V}'}
                           \hat{H}_{1}
                         + \hat{ \bar{H}}_{2}e^{2g\hat{V}+g'\hat{V}'}
                           \hat{H}_{2} \,\ ,
\EEA
e as componentes respons\'aveis pela massa est\~ao escritas abaixo
\BEA
 \L^{massa}_{gauge}&=& \frac{1}{4}(g^2V_m^i V^{im} +Y_{H1}^2 
 g^{\prime 2}V^m V_m) \bar{H}_{1} H_{1}+
\frac{1}{4}(g^2V_m^i V^{im} +Y_{H2}^2 
 g^{\prime 2}V^m V_m) \bar{H}_{2} H_{2} \nonumber \\
 &+& \frac{Y_{H1}}{2} gg^\prime V^i_m V^m(\bar{H}_{1} \sigma^iH_{1})+ 
 \frac{Y_{H2}}{2} gg^\prime V^i_m V^m(\bar{H}_{2} \sigma^iH_{2}) \,\ ,  
\EEA
usando os valores da Tab.2 podemos escrever a lagrangiana acima da 
seguinte maneira
\BEA
 \L^{massa}_{gauge}&=& \frac{1}{4}[(g^2V_m^i V^{im} + g^{\prime 2}V^m V_m) 
 (\bar{H}_{1} H_{1}+\bar{H}_{2} H_{2}) \nonumber \\
 &+& \frac{gg^\prime}{2} V^i_m V^m [ 
(\bar{H}_{2} \sigma^iH_{2})-(\bar{H}_{1} \sigma^iH_{1})] \,\ .
\label{boson1}
\EEA
Usando a defini\c c\~ao dos b\'osons carregados Eqs.(\ref{svtl10}) e as matrizes de Pauli, 
Eq.(\ref{pauli}), e 
os valores esperados do v\'acuo de $H_{1}$ e $H_{2}$ podemos reescrever a 
Eq.(\ref{boson1}) da seguinte maneira
\BEA
 \L^{massa}_{gauge}&=& \frac{1}{2} g^2(v_1^2+v_2^2) W^{+m}W^{-}_{m}+ 
 \frac{(v_1^2+v_2^2)}{4}(g^2V_m^3 V^{3m}+g^{\prime 2}V^m V_m) \nonumber \\
 &-& \frac{gg^\prime}{2} (v_1^2+v_2^2) V_m^3 V^m \nonumber \\
 & \equiv& 
 \L^{massa}_{carregado}+\L^{massa}_{neutro} \,\ .
\label{boson2}
\EEA

Onde identificamos
\BEA
\L^{massa}_{carregado}= \frac{1}{2} g^2(v_1^2+v_2^2) W^{+m}W^{-}_{m} \,\ ,
\EEA
de onde conclu\'\i mos que a massa do b\'oson carregado \'e dada por
\BEA
M^{2}_{W}= \frac{1}{2} g^2(v_1^2+v_2^2) \,\ ,
\label{carregado massa}
\EEA
mas este valor \'e muito bem medido $v_1^2+v_2^2 \approx(174 GeV)^2$. Podemos portanto descrever os dois valores 
esperados do v\'acuo em termos de um \'unico par\^ametro definido da 
seguinte maneira
\BEA
    \tan\beta &=& \f{v_{2}}{v_{1}} \,\ .
      \label{Radiative Breaking prop 12f}
\EEA
como $v_{1},v_{2} \geq 0$ teremos
\BEA
       0 \;\;\leq\;\; \b \;\;\leq\;\;  \f{\pi}{2} \,\ .
\EEA
Da Eq.(\ref{Radiative Breaking prop 12f}) podemos escrever
\BEA
\sin \beta &=& \frac{gv_{2}}{\sqrt{2}M_{W}} \,\ , \nonumber \\
\cos \beta &=& \frac{gv_{1}}{\sqrt{2}M_{W}} 
\label{beta} \,\ .
\EEA
Usando Eq.(\ref{Radiative Breaking prop 12f}), podemos escrever a massa do 
b\'oson carregado da seguinte maneira
\BEA
M^{2}_{W}= \frac{1}{2} g^2v_1^2(1+\tan^{2} \beta) \,\ .
\EEA

J\'a a parte neutra \'e a seguinte
\BEA
\L^{massa}_{neutro}= \frac{(v_1^2+v_2^2)}{4} \left[ (g^2V_m^3 V^{3m}+
g^{\prime 2} V^m V_m)- 
2gg^\prime (v_1^2+v_2^2) V_m^3 V^m \right] \,\ ,
\EEA
que em forma matricial torna-se
\BEA
\L^{massa}_{neutro}= \frac{(v_1^2+v_2^2)}{4} 
\LP(\BA{cc} V^{3}_{m} & V_{m} \EA \RP)
      \LP(\BA{cc}  
      g^{2} & -gg^{\prime} \\
      -gg^{\prime} & g^{\prime 2}
             \EA \RP)
      \LP( \BA{c} V^{3m} \\ V^{m} \EA \RP) \,\ ,
      \label{boson3}
      \EEA
usando a defini\c c\~ao do \^angulo de Weinberg dada pela 
Eq.(\ref{operador2}) podemos 
escrever Eq.(\ref{boson3}) da seguinte maneira
\BEA
\L^{massa}_{neutro}= g^{2} \frac{(v_1^2+v_2^2)}{4} 
\LP(\BA{cc} V^{3}_{m} & V_{m} \EA \RP)
      \LP(\BA{cc}
      1 & - \tan \theta_{W} \\
      - \tan \theta_{W} & \tan^{2} \theta_{W}
      \EA \RP)
      \LP( \BA{c} V^{3m} \\ V^{m} \EA \RP) \,\ . 
      \nonumber \\
      \label{boson4}
      \EEA
      
Para obter os estados f\'\i sicos dos b\'osons vetoriais neutros e suas 
respectivas massas, temos que diagonalizar Eq.(\ref{boson4}) desde que os 
estados f\'\i sicos s\~ao ortogonais um ao outro. Fazendo a 
diagonaliza\c c\~ao acharemos que os b\'osons f\'\i sicos s\~ao dados por
\BEA
\LP( \BA{c} A_{m} \\ Z_{m} \EA \RP)= 
\LP(\BA{cc}
      \sin \theta_{W} & \cos \theta_{W} \\
      \cos \theta_{W} & - \sin \theta_{W}
      \EA \RP)
      \LP( \BA{c} V^{3}_{m} \\ V_{m} \EA \RP) \,\ ,
      \label{boson5}
      \EEA
que coincide com a defini\c c\~ao apresentada na Eq.(\ref{svtl10}) e 
a matriz de massa diagonalizada \'e dada por 
\BEA
\LP(\BA{cc}
      0 & 0 \\
      0 & \frac{1}{ \cos^{2} \theta_{W}}
      \EA \RP) \,\ ,
      \label{boson6}
      \EEA
usando Eq.(\ref{boson5}) e Eq.(\ref{boson6}) podemos escrever 
Eq.(\ref{boson4}) da seguinte maneira
\BEA
\L^{massa}_{neutro}= g^{2} \frac{(v_1^2+v_2^2)}{4}
\LP( \BA{cc} A_{m} & Z_{m} \EA \RP)
\LP(\BA{cc}
      0 & 0 \\
      0 & \frac{1}{ \cos^{2} \theta_{W}}
      \EA \RP)
      \LP( \BA{c} A_{m} \\ Z_{m} \EA \RP) \,\ .
      \EEA
Da lagrangiana acima percebemos que
\BEA
M^{2}_{ \gamma}&=&0 \,\ , \nonumber \\
M^{2}_{Z}&=& \frac{1}{2 \cos^{2} \theta_{W}} g^2(v_1^2+v_2^2)= 
\frac{M^{2}_{W}}{ \cos^{2} \theta_{W}} \,\ .
\label{z-mass}
\EEA
Como no {\bf SM} o f\'oton n\~ao adquire massa e a massa do $Z^{0}$ \'e 
relacionada a massa dos b\'osons carregados pela mesma rela\c c\~ao.

\subsection{Espectro do B\'oson de Higgs F\'\i sico.}

No {\bf SM} n\'os comecamos por expandir em torno do valor esperado 
do v\'acuo do Higgs e identificamos os novos estados como sendo os 
estados f\'\i sicos. Por\'em, fazendo a mesma coisa para o 
{\bf MSSM}, estes novos autoestados da intera\c c\~ao fraca n\~ao 
representam os autoestados da massa, como iremos ver.

Quando a condi\c c\~ao Eq.(\ref{Radiative Breaking prop 12}) \'e satisfeita, as
componentes neutras de $H_{1}$ e $H_{2}$ adquirem v.e.v. 
($v_{1}, v_{2} \neq 0$). Antes de obtermos as massas dos Higgs vamos obter 
algumas rela\c c\~oes \'uteis. Em $V_{\min}$, o potencial tem que satisfazer 
a seguintes rela\c c\~ao, que fornece os extremos do potencial
\BEA
\PD{V_{min}}{v_{1}}=\PD{V_{min}}{v_{2}}=0 \,\ , \nonumber
\EEA
bem como a condi\c c\~ao de m\'\i nimo, que \'e expressa por
\BEA
\f{\P^{2}V_{min}}{\P v_{1}\P v_{2}}>0 \,\ . \nonumber
\EEA
Estas equa\c c\~oes e Eq.(\ref{Radiative Breaking prop 4})
nos fornecem as seguintes rela\c c\~oes
\BEA
  m_{1}^{2}v_{1}-M_{12}^{2}v_{2}
            +\f{1}{4}\LP(g^{2}+g'^{2}\RP)\LP[v_{1}^{2}-v_{2}^{2}\RP]
             v_{1} &=& 0 \,\ ,
            \label{Radiative Breaking prop 12a}\SL
  m_{2}^{2}v_{2}-M_{12}^{2}v_{1}
            -\f{1}{4}\LP(g^{2}+g'^{2}\RP)\LP[v_{1}^{2}-v_{2}^{2}\RP]
              v_{2} &=& 0 \,\ ,
            \label{Radiative Breaking prop 12b} \SL
     -2M_{12}^{2}-\LP(g^{2}+g'^2\RP)\,v_1 v_2 & > & 0 \,\ .
         \label{Radiative Breaking prop 12c}
\EEA
Multiplicando as Eqs.(\ref{Radiative Breaking prop 12a}) e
(\ref{Radiative Breaking prop 12b}) por $v_{1}^{-1}$ e $v_{2}^{-1}$ 
respectivamente, e ent\~ao somar e subtrair as equa\c c\~oes 
obtidas, obteremos
\BEA
  m_{1}^{2}+m_{2}^{2} &=& M_{12}^{2}\LP(\tan{\b}+\cot{\b}\RP) \,\ ,
     \nonumber \\
  v_{1}^{2}-v_{2}^{2} &=& \f{-2}{g^2+g'^2}\LP[m_1^2-m_2^2
               -\LP(m_1^2+m_2^2\RP)\f{\tan{\b}
           -\cot{\b}}{\tan{\b}+\cot{\b}}\RP]\nn
      &=&  \f{-2}{g^2+g'^2}\LP[m_1^2-m_2^2
               +\LP(m_1^2+m_2^2\RP)\cos{2\b}\RP] \,\ ,
           \label{Radiative Breaking prop 12e}
\EEA
n\'o utilizamos Eq.(\ref{Radiative Breaking prop 12f}) para obter as equa\c c\~oes 
acima.

Com as Eqs.(\ref{Radiative Breaking prop 12a}), (\ref{Radiative Breaking prop 12b})
e (\ref{Radiative Breaking prop 12e}) o potencial m\'\i nimo pode ser escrito da 
seguinte maneira
\BEA
  V_{min} &=& \f{-1}{2(g^{2}+g^{'\,2})}
    \LP[\LP(m_{1}^{2}-m_{2}^{2}\RP)+
     \LP(m_1^2+m_2^2\RP)\cos{2\b}\,\RP]^{2} \,\ ,
      \label{Radiative Breaking prop 13}
\EEA
quero lembrar s\'o mais uma coisa os par\^ametros da equa\c c\~ao acima 
dependem do ponto de renormaliza\c c\~ao $Q$.

Agora j\'a estamos aptos para calcular as massas dos Higgs.
Os autoestados f\'\i sicos s\~ao obtidos diagonalizando a matriz de 
massa dos b\'osons de Higgs. Isto \'e feito mais facilmente na base 
real onde podemos escrever
\BEA
   H_1 &=& \LP(\BA{r} h_1+i h_2  \\
                      h_3+i h_4  \EA \RP)=  
\LP( \BA{r} H_{1}^{1} \\
            H_{1}^{2} \EA \RP) \,\ ,
                       \label{The Physical Higgs Boson Spectrum prop 3}\SL
   H_2 &=& \LP(\BA{r} h_5+i h_6  \\
                      h_7+i h_8  \EA \RP)= 
\LP( \BA{r} H_{2}^{1} \\
            H_{2}^{2} \EA \RP) \,\ .
                       \label{The Physical Higgs Boson Spectrum prop 4}
\EEA
Nesta base real o potencial de Higgs, ver 
Eq.(\ref{Tree-level scalar potential}), torna-se
\BEA
  V(h_i) &=&    m_1^2\sum_{i=1}^{4}h_i^2
              + m_2^2\sum_{j=5}^8h_j^2
              - 2 M_{12}^2\LP(h_1h_7+h_4h_6-h_3h_5-h_2h_8\RP)
         \nmb
              + \f{1}{8}\LP(g^2+g'^2\RP)
           \LP[\sum_{i=1}^4 h_i^2-\sum_{j=5}^{8}h_j^2\RP]^2
         \nmb
              + \f{g^2}{2}\LP(h_1h_5+h_2h_6+h_3h_7+h_4h_8\RP)^2
         \nmb
              +\f{g^2}{2}\LP(h_1h_6+h_3h_8-h_2h_5-h_4h_7\RP)^2  \,\ .
              \label{The Physical Higgs Boson Spectrum prop 4a}
\EEA
Deste potencial \'e evidente que a base dos campos de Higgs em que 
estamos trabalhando n\~ao pode ser a base f\'\i sica porque cont\'em 
termos de massa fora da diagonal. Assim temos que mudar para a base 
dos auto-estados da massa.

O estado f\'\i sico dos b\'osons de Higgs s\~ao obtidos diagonalizando 
a matriz de massa do Higgs que \'e dada por ~\cite{HAB79}
\BEA
   M^2_{ij} &=& \LP. \HA\;\f{\P^2 V}{\P h_i\; \P h_j}\RP|_{\;\mbox{min}} \,\ .
            \label{The Physical Higgs Boson Spectrum prop 5}
\EEA
O termo ``min" significa fazer $\LP<h_1\RP>=v_1$, $\LP<h_7\RP>=v_2$
e $\LP<h_i\RP> =0$ para todos os outros $i$.

Agora iremos analizar os diferentes setores dos Higgs deste modelo.

\subsubsection{Setor do Higgs Carregados; \'\i ndices 3, 4, 5 e 6.}

Usando as Eqs.(\ref{The Physical Higgs Boson Spectrum prop 4a}) e 
(\ref{The Physical Higgs Boson Spectrum prop 5})
a matriz de massa do b\'oson de Higgs \'e facilmente calculada.
Observe que a parte real e imagin\'ario do setor do Higgs carregado se 
desacoplam isto se deve ao fato de que
\BEA
  M^2_{56}=M^2_{54}=M^2_{36}=M^2_{34} = 0 \,\ .  \nonumber
\EEA
As componentes reais s\~ao dadas por
\BEA
  M^2_{55} &=& m_2^2-\f{1}{4}\LP(g^2+g'^2\RP)
         \LP(v_1^2-v_2^2\RP)+\HA g^2v_1^2\nn
           &=& \HA\LP(g^2+\f{2M_{12}^2}{v_1v_2}\RP)v_1^2 
           \,\ ,\nn
  M^2_{53} &=& M_{12}^2+\HA g^2 v_1v_2\nn
           &=& \HA\LP(g^2+\f{2M_{12}^2}{v_1v_2}\RP)v_1v_2 
           \,\ ,\nn
  M^2_{33} &=& m_1^2+\f{1}{4}\LP(g^2+g'^2\RP)
          \LP(v_1^2-v_2^2\RP)+\HA g^2v_2^2\nn
           &=& \HA\LP(g^2+\f{2M_{12}^2}{v_1v_2}\RP) v_2^2 
           \,\ , 
\EEA
j\'a para a parte imagin\'aria obteremos
\BEA
  M^2_{66} &=& M^2_{55} \,\ , \nn
  M^2_{44} &=& M^2_{33} \,\ , \nn
  M^2_{64} &=& - M^2_{53} \,\ .
      \label{The Physical Higgs Boson Spectrum prop 6}
\EEA
Nas express\~oes acima usamos Eqs.(\ref{Radiative Breaking prop 12a}) e 
(\ref{Radiative Breaking prop 12b})
para eliminar os par\^ametros de massa $m_1^2$ e $m_2^2$.
Portanto nesta base $(h_5,h_3)$ e $(-h_6,h_4)$, a matriz de massa do 
Higgs carregado \'e expressa por\footnote{O sinal da base $(-h_6,h_4)$
\'e escolhida para que as duas base tenham a mesma matriz de massa.}
\BEA
  M_{\pm}^2 &=& \HA \LP(g^2+\f{2M_{12}^2}{v_1v_2}\RP)
                   \LP(\BA{cc} v_1^2    &  v_1 v_2 \\
                               v_1v_2   &  v_2^2     \EA \RP) \,\ .
                   \label{The Physical Higgs Boson Spectrum prop 7}
\EEA
Para obter os estados f\'\i sicos dos b\'osons de Higgs carregados e suas 
respectivas massas temos que diagonalizar esta matriz.

Calculando os autovalores, a matriz de massa pode ser escrita da seguinte maneira 
onde usamos que~($tan\b=v_{2}/v_{1}$)
\BEA
  M^2_{\pm} &=&
     \LP( \BA{cc}   -\sin{\b}  & \cos{\b}\\
                     \cos{\b}  & \sin{\b}   \EA \RP)
     \LP( \BA{cc}     0  & 0\\
                     0   & m^2_{H^{\pm}}   \EA \RP)
     \LP( \BA{cc}   -\sin{\b}  & \cos{\b}\\
                     \cos{\b}  & \sin{\b}   \EA \RP) \,\ ,
                     \label{The Physical Higgs Boson Spectrum prop 8}
\EEA
com
\BEA
  m^2_{H^{\pm}} &=& \HA\LP(g^2+\f{2M_{12}^2}{v_1v_2}\RP)\LP(v_1^2+v_2^2\RP) 
  \nonumber \\
  &=&M^{2}_{W}+\f{M_{12}^2}{v_1v_2}\LP(v_1^2+v_2^2\RP) \,\ ,
     \label{Physical Higgs Boson}
\EEA
\'e a matriz de massa dos Higgs carregados. Repare que neste procedimento de 
diagonaliza\c c\~ao, dois estados sem massa e dois estados massivos apareceram. 
Os estados de massa zero s\~ao associados com b\'osons de Goldstone carregado, 
como veremos a seguir.

Agora vamos obter os estados f\'\i sicos
\BEA
\lefteqn{ \LP( \BA{cc} h_5 & h_3 \EA \RP)
M^2_{\pm} \LP( \BA{r} h_5 \\ h_3 \EA \RP)
 +  \LP( \BA{cc} -h_6 & h_4 \EA \RP) M^2_{\pm}
\LP( \BA{r} -h_6 \\ h_4 \EA \RP)} \hspace{1.5cm} \nn
  &=&
    \LP( \BA{cc} h_5+i h_6 & h_3-ih_4 \EA \RP)
M^2_{\pm} \LP( \BA{r} h_5-ih_6 \\ h_3+i h_4 \EA \RP)\nn
  &=&
     \LP( \BA{cc} H_2^1 & \bar{H}_1^{2} \EA \RP) M^2_{\pm}
      \LP( \BA{l} \bar{H}_2^{1} \\ H_1^2  \EA \RP) \nn
  &=&  \LP( \BA{r} -H_2^1 \sin{\b}+\bar{H}_1^{2}\cos{\b} \\
                    H_2^1\cos{\b}+\bar{H}_1^{2}\sin{\b} \EA \RP)^{T}
       \LP( \BA{cc}     0  & 0\\
                        0   & m^2_{H^{\pm}}   \EA \RP)
       \LP( \BA{r} -\bar{H}_2^{1} \sin{\b}+H_1^{2}\cos{\b} \\
                    \bar{H}_2^{1}\cos{\b}+H_1^{2}\sin{\b} \EA \RP)\nn
   &=& \LP( \BA{cc} G^{+}& H^{+}\EA\RP)
       \LP( \BA{cc}     0  & 0\\
                        0   & m^2_{H^{\pm}}   \EA \RP)
       \LP(\BA{r}  G^{-}\\H^{-}  \EA\RP) \,\ .      \nonumber
\EEA
Onde fazemos as seguintes identifica\c c\~oes
\BEA
   G^{-} &=& H_1^{2}\cos{\b}-\bar{H}_2^{1} \sin{\b} \,\ \mbox{(b\'oson de
Goldstone)} \,\ ,    \label{The Physical Higgs Boson Spectrum prop 9a}\SL
   H^{-} &=& H_1^{2}\sin{\b}+\bar{H}_2^{1}\cos{\b} \,\ \mbox{(Higgs
carregado)} \,\ ,    \label{The Physical Higgs Boson Spectrum prop 9b}
\EEA
e
\BEA
   G^{+} &=& \bar{G}^{-} \,\ ,\nn
   H^{+} &=& \bar{H}^{-} \,\ . \nonumber
\EEA

\subsubsection{Setor do Higgs Neutro; \'\i ndices 2 e 8.}

J\'a vimos que o setor do Higgs carregado se desacopla em uma parte 
real e outra imagin\'aria. Isto tamb\'em ocorre com o setor neutro, para ver 
isto \'e so verificar que
\BEA
  M^2_{12}=M^2_{72}=M^2_{18}=M^2_{78} = 0 \,\ .  \nonumber
\EEA
Isto se deve ao fato que nossa teoria \'e invariante por CP. Comecaremos 
a discuss\~ao com o setor imagin\'ario (CP \'\i mpar) para depois vermos 
o setor real (CP par).

Fazendo um procedimento an\'alogo ao do setor carregado podemos mostrar que
\BEA
  M^2_{88} &=& m_2^2-\f{1}{4}\LP(g^2+g'^2\RP)
         \LP(v_1^2-v_2^2\RP)\nn
           &=& \HA\LP(\f{M_{12}^2}{v_1v_2}\RP)v_1^2 \,\ ,\nn
  M^2_{28} &=& M_{12}^2\nn
           &=& \HA\LP(\f{M_{12}^2}{v_1v_2}\RP)v_1v_2 \,\ ,\nn
  M^2_{22} &=& m_1^2+\f{1}{4}\LP(g^2+g'^2\RP)
          \LP(v_1^2-v_2^2\RP)\nn
           &=& \HA\LP(\f{M_{12}^2}{v_1v_2}\RP) v_2^2 \,\ , \nonumber
\EEA
onde outra vez usamos Eqs.(\ref{Radiative Breaking prop 12a}) e 
(\ref{Radiative Breaking prop 12b})
para eliminar os par\^ametros de massa $m_1^2$ e $m_2^2$. As equa\c c\~oes 
acima em forma matricial pode ser escrita da seguinte maneira
\BEA
   \f{M_{12}^2}{v_1v_2}\LP( \BA{cc}   v_1^2    & v_1 v_2   \\
                                   v_1 v_2  & v_2^2    \EA \RP) \,\ , \nonumber
\EEA
na base $(h_8, h_2)$. Diagonalizando esta matriz, que \'e um procedimento 
id\^entico ao utilizado no setor carregado, os autoestados f\'\i sico s\~ao
\BEA
\lefteqn{\f{M_{12}^2}{v_1v_2}\LP( \BA{cc} h_8 & h_2\EA \RP)
          \LP( \BA{cc} v_1^2  & v_1v_2 \\
                       v_1v_2 & v_2^2  \EA \RP)
          \LP(\BA{r} h_8 \\ h_2 \EA \RP)}\hspace{1.5cm}\nn
    &=&  \LP( \BA{r} -h_8\sin{\b}+h_2\cos{\b} \\
                      h_8\cos{\b}+h_2\sin{\b} \EA \RP)^{T}
       \LP( \BA{cc}     0  & 0\\
                        0   & m^2_{H^{0}_3}   \EA \RP)
       \LP( \BA{r} -h_8\sin{\b}+h_2\cos{\b} \\
                    h_8\cos{\b}+h_2\sin{\b} \EA \RP)\nn
   &=& \LP( \BA{cc} \f{G^{0}}{\sqrt{2}} & \f{H^{0}_3}{\sqrt{2}} \EA\RP)
       \LP( \BA{cc}     0  & 0\\
                        0   & m^2_{H^{0}_3}  \EA \RP)
       \LP(\BA{r}  \f{G^{0}}{\sqrt{2}} \\
      \f{H^{0}_3}{\sqrt{2}}  \EA\RP) \,\ .  \nonumber
\EEA
Onde fazemos as seguintes identifica\c c\~oes
\BEA
   G^{0} &=&  \sqrt{2} \LP(\, h_2\cos{\b}-h_8\sin{\b}\, \RP) \nn
         &=&  \sqrt{2} \LP(\, \mbox{Im\,}
               H_1^1\cos{\b}-\mbox{Im\,}H_2^2\sin{\b}\,\RP) \,\ , 
               \,\ \mbox{(b\'oson de Goldstone)} \nonumber \\
         \label{The Physical Higgs Boson Spectrum prop 10a}\SL
   H^0_3 &=&  \sqrt{2} \LP(\, h_2\sin{\b}+h_8\cos{\b}\,\RP) \nn
         &=&     \sqrt{2} \LP(\, \mbox{Im\,}
            H_1^1\sin{\b}+\mbox{Im\,}H_2^2\cos{\b}\,\RP) \,\ , 
            \,\ \mbox{(Higgs neutro CP$=-1$)} \nonumber \\
         \label{The Physical Higgs Boson Spectrum prop 10b}
\EEA
os fatores $\sqrt{2}$ s\~ao colocados para obtermos os termos 
cin\'eticos usuais.

A matriz de massa do b\'oson de Higgs neutro \'e
\BEA
     m_{H^0_3}^2 &=& \f{M_{12}^2}{v_1v_2}\LP( v_1^2+v_2^2\RP)   \nn
                 &=& m_{H^{\pm}}^2-M_{\mbox{w}}^2 \,\ ,
                 \label{The Physical Higgs Boson Spectrum prop 10dgdb}
\EEA
na \'ultima passagem usamos a Eq.(\ref{Physical Higgs Boson}).

\subsubsection{Setor de Higgs neutro; \'\i ndices 1 e 7.}

Ap\'os termos estudado o setor imagin\'ario do Higgs neutro iremos agora 
estudar a parte real, de $CP=+1$

Neste caso podemos mostrar
\BEA
   M^{2}_{11} &=& \HA \LP(g^2+g'^2\RP)v_1^2+M_{12}^2\f{v_2}{v_1} \equiv A \,\ ,
       \label{index 17 prop 3a}\nn
   M^{2}_{17} &=& -\HA \LP(g^2+g'^2\RP)v_1 v_2-M_{12}^2 \equiv B \,\ ,
       \label{index 17 prop 3b} \nn
   M^{2}_{77} &=& \HA \LP(g^2+g'^2\RP)v_2^2+M_{12}^2\f{v_1}{v_2} \equiv C \,\ ,
       \label{index 17 prop 3c}    \nonumber
\EEA
repare que $A, C \geq 0$ e $B \leq 0$.Usamos as 
Eqs.(\ref{Radiative Breaking prop 12a}) e 
(\ref{Radiative Breaking prop 12b}) para eliminar 
$m_1^2$ e $m_2^2$. Que colocada na forma matricial torna-se
\BEA
   M_0^2 &=& \LP( \BA{cc} A & B \\
                          B & C   \EA \RP) \,\ , \nonumber
\EEA
na base $(h_1, h_7)$.

O procedimento de diagonaliza\c c\~ao deste setor \'e ligeiramente diferente 
do apresentados nos outros dois setores do Higgs. Os autovalores de $M^2_0$ s\~ao
\BEA
   m^{2}_{H^{0}_{1},\,H^{0}_{2}}  &=&
      \HA \LP[\;A+C\pm \sqrt{\LP( A-C \RP)^2 +4B^2}\;\RP] \nn
      &=& \HA \LP[\; m^2_{H^0_3}+m_{\mbox{z}}^2\pm
            \sqrt{\LP(m^2_{H^0_3}+m^2_{\mbox{z}}\RP)^2
                    -4m^2_{\mbox{z}}m^2_{H^0_3}\cos^2{2\b} }\;\RP] 
                    \,\ , \nonumber \\
         \label{index 17 prop 4}
\EEA
onde o sinal positivo (negativo) \'e associado a $m^2_{H^0_1}$
($m^2_{H^0_2}$). Os correspondentes autovetores s\~ao
\footnote{Onde ${\bf v}_1$ e ${\bf v}_2$ correspondem aos autovalores 
$m^2_{H^0_1}$ e $m^2_{H^0_2}$ respectivamente.}
\BEA
  {\bf v}_{1,2} &=& N_{1,\,2}\LP( \BA{c}   1 \\
      \f{-(A-C)\pm \sqrt{\LP( A-C \RP)^2 +4B^2}}{2B} \EA \RP) \,\ .
      \label{index 17 prop 5}
\EEA
$N_{1,\,2}$ s\~ao constantes de normaliza\c c\~ao.

Como ficar\'a claro mais para frente, \'e \'util introduzir o seguinte 
\^angulo  de mistura $\a$~( n\~ao confundir com a constante de 
estrutura fina) definida  por
\BEA
  \sin{2\a} &=& \f{2B}{\sqrt{\LP(A-C\RP)^2+4B^2}}\nn
             &=& -\sin{2\b}\;\LP( \f{m^2_{H^0_1}+m^2_{H^0_2}}{m^2_{H^0_1}
                    -m^2_{H^0_2}}\RP) \,\ , \nn
  \cos{2\a} &=& \f{A-C}{\sqrt{ \LP(A-C\RP)^2+4B^2}}\nn
               &=& -\cos{2\b}\;\LP( \f{m^2_{H^0_3}
             -m^2_{\mbox{z}}}{m^2_{H^0_1}-m^2_{H^0_2}}\RP) \,\ . \nonumber
\EEA
Das seguintes identidades matem\'aticas  $\sin{2\a}=2\sin\a\cos\a$ e
$\cos{2\a}= \cos^2\a-\sin^2\a$, podemos mostrar a seguinte equa\c c\~ao
\BEA
      x^2+2\cot\LP(2\a\RP) x -1 = 0 \,\ ,  \nonumber
\EEA
onde $x=\tan\a$. Em geral esta equa\c c\~ao tem duas solu\c c\~oes distintas. 
Por\'em, 
anteriormente, j\'a hav\'\i amos escolhido $v_1, v_2 \geq 0$ ou 
equivalentemente  \newline 
$0\leq \b \leq \f{\pi}{2}$, isto implica que 
$-\f{\pi}{2} \leq \a \leq 0$. Tendo em mente esta restri\c c\~ao podemos achar 
uma \'unica solu\c c\~ao para x, e o resultado \'e 
~(lembre-se que $B\leq 0$)
\BEA
   \tan\a &=& \f{-\LP(A-C\RP)+\sqrt{\,\LP(A-C\RP)^2+4B^2}}{2B} \,\ ,
             \label{index 17 prop 9a}
\EEA
e invertendo teremos
\BEA
     \cot\a &=& \f{\LP(A-C\RP)+\sqrt{\,\LP(A-C\RP)^2+4B^2}}{2B} \,\ .
               \label{index 17 prop 9b}
\EEA
Comparando Eqs(\ref{index 17 prop 9a}) e (\ref{index 17 prop 9b}) com
Eq.(\ref{index 17 prop 5}), vemos que a segunda componente de 
${\bf v}_{1}$~(${\bf v}_{2}$) pode ser identificado com $\tan\a$
($\cot\a$). O \^angulo de mistura, $\a$, foi definido para termos este 
resultado.

Desta maneira escolhemos $ N_1 = \cos{\a}$ e $N_2 = -\sin{\a}$ para obtermos 
autovetores ortonormais, e a matriz de massa do setor de Higgs real neutro 
adquire a seguinte forma
\BEA
  M^2_0 &=&
       \LP( \BA{cc} \cos{\a}  & -\sin{\a} \\
                    \sin{\a}  &  \cos{\a}  \EA \RP)
       \LP(\BA{cc}   m^2_{H^0_1}  &  0 \\
                       0          &  m^2_{H^0_2}  \EA \RP)
       \LP( \BA{cc} \cos{\a}  & -\sin{\a} \\
                    \sin{\a}  &  \cos{\a}  \EA \RP)^{-1} 
                    \,\ . \nonumber \\
\EEA

O correspondente termo de massa da lagrangiana agora torna-se
\BEA
\lefteqn{ \LP( \BA{cc} h_1 & h_7 \EA \RP) M^2_0
   \LP(\BA{r} h_1 \\ h_7 \EA \RP)}\hspace{0.3cm} \nn
     &=& \LP( \BA{r} h_1 \cos{\a}+ h_7\sin{\a} \\
                     -h_1\sin{\a}+h_7\cos{\a}   \EA \RP)^{T}
         \LP(\BA{cc}   m^2_{H^0_1}  &  0 \\
                       0          &  m^2_{H^0_2}  \EA \RP)
         \LP( \BA{r} h_1 \cos{\a}+ h_7\sin{\a} \\
                     -h_1\sin{\a}+h_7\cos{\a}  \EA \RP) \,\ .
\nonumber \\  \hspace{0.5cm}
\EEA
Para identificarmos o estado f\'\i sico do Higgs $H^0_1$ e $H^0_2$, temos 
que ser cuidadosos, a raz\~ao \'e que estes estados, como qualquer estado 
f\'\i sico, tem que ter valor esperado do v\'acuo igual a zero. Portanto 
fazemos as seguintes identifica\c c\~oes
\BEA
 \f{H^0_1}{\sqrt{2}} + v_1\cos{\a}+v_2\sin{\a}
           &=& h_1\cos{\a}+h_7\sin{\a} \,\ , \nn
 \f{H^0_2}{\sqrt{2}} - v_1\sin{\a}+v_2\cos{\a}
           &=& -h_1\sin{\a}+h_7\cos{\a} \,\ ,\nonumber
\EEA
ou equivalentemente
\BEA
  H^0_1 &=& \sqrt{2}\LP[ \,\LP(\mbox{Re\,}
          H_1^1-v_1\RP)\cos{\a}+\LP(\mbox{Re\,}H_2^2-v_2\RP)\sin{\a}\RP] 
          \,\ ,
      \label{index 17 prop 18}\SL
  H^0_2 &=& \sqrt{2}\LP[ \,-\LP(\mbox{Re\,}
          H_1^1-v_1\RP)\sin{\a}+\LP(\mbox{Re\,}H_2^2-v_2\RP)\cos{\a}\RP] 
          \,\ .
      \label{index 17 prop 19}
\EEA

\subsubsection{Conclus\~ao e Coment\'arios.}

Nas tr\^es subse\c c\~oes acima derivamos o conteudo f\'\i sico do b\'oson 
de Higgs do {\bf MSSM}. Este consiste dos b\'osons de Higgs carregados~($H^{\pm}$),
os b\'osons de Higgs neutros\footnote{Alguns autores usam a nota\c c\~ao 
$H^{0}$, $h^{0}$ e $A^{0}$ em vez da nossa $H^{0}_{1}$,  $H^{0}_{2}$
e $H^{0}_{3}$.}~($H^{0}_{i}$,     $i=1,2,3$) e finalmente os b\'osons de 
Goldstone carregados~($G^{\pm}$) e neutros~($G^{0}$).

Os novos campos em termos dos ``velhos" est\~ao dados nas 
Eqs.(\ref{The Physical Higgs Boson Spectrum prop 9a}), 
(\ref{The Physical Higgs Boson Spectrum prop 9b}), 
(\ref{The Physical Higgs Boson Spectrum prop 10a}), 
(\ref{The Physical Higgs Boson Spectrum prop 10b}), 
(\ref{index 17 prop 18}) e (\ref{index 17 prop 19}).
Por\'em, para obter a lagrangiana em termos dos campos f\'\i sicos, temos que 
inverter as rela\c c\~oes acima. O resultado desta invers\~ao s\~ao
\BEA
   H_1  &=&
     \LP( \BA{l}
       v_1+\f{1}{\sqrt{2}}\LP[\,
           H_1^0\cos{\a}-H_2^0\sin{\a}+iH_3^0\sin{\b}+iG^0\cos{\b}\,\RP] \\
       H^-\sin{\b}+G^-\cos{\b}  \EA \RP) 
       \,\ , \nonumber \\
       \label{estado H1} \SL
   H_2  &=&
     \LP( \BA{l}
       H^+\cos{\b}-G^+\sin{\b} \\
       v_2+\f{1}{\sqrt{2}}\LP[\,H_1^0\sin{\a}
           +H^0_2\cos{\a}+i H^0_3\cos{\b}-iG^0\sin{\b}\,\RP] \EA\RP) 
           \,\ . \nonumber \\
              \label{estado H2}
\EEA
Inserindo estas express\~oes na lagrangiana do {\bf MSSM} as intera\c c\~oes (e as 
regras de Feynman) dos b\'osons de Higgs f\'\i sicos s\~ao obtidas.

Das f\'ormulas de massas dos Higgs obtidas acima, 
Eqs.(\ref{The Physical Higgs Boson Spectrum prop 7}), 
(\ref{The Physical Higgs Boson Spectrum prop 10dgdb}) e 
(\ref{index 17 prop 4}), \'e interessante notar que no limite 
$m_{H^{0}_{3}}\rightarrow \infty$ ($\tan \b$ fixo),
$H^{\pm}$, $H^{0}_{1}$ e ($H^{0}_{3}$) se desacoplam da teoria, e assim o 
setor de Higgs da teoria cont\'em apenas $H^{0}_{2}$. Neste limite, \'e 
poss\'\i vel mostrar que $H^{0}_{2}$ \'e id\^entico ao Higgs do Modelo Padr\~ao 
(m\'\i nimo).

\'E importante comentar que todas as massa dos Higgs aqui obtidas foram 
calculadas apenas \`a nivel de \'arvore, e satisfazem as seguintes 
rela\c c\~oes
\BEA
   m_{H^{\pm}} \;\;\geq \;\; M_{\mbox{w}} \,\ ,\nn
   m_{H^{0}_{2}} \;\;\leq \;\; m_{\mbox{z}} \;\;\leq \;\; m_{H^{0}_{1}} 
   \,\ ,\nn
   m_{H^{0}_{3}} \;\; \geq \;\; m_{H^{0}_{2}}\,\ . \nonumber
\EEA
Desde que $ m_{H^{0}_{2}} \leq m_{\mbox{z}}$ (n\'\i vel de \'arvore)
foi acreditado, devido ao quadro de intera\c c\~ao do $H^{0}_{2}$, 
que $H^{0}_{2}$ poderia ser produzido e esperansosamente detectado no LEP. 
Nenhum Higgs foi visto e isto pode ser visto como um problema. Atualmente 
muitos f\'\i sicos acreditam que os Higgs do {\bf MSSM} podem obter grandes 
corre\c c\~oes 
radiativas, t\~ao grandes como  ${\cal O}\LP(100\RP)$ GeV. Desta maneira isto 
coloca a massa do $H^{0}_{2}$ acima da massa do b\'oson Z (e por outro lado 
fora do alcance de descoberta do LEP~1). Estas grandes corre\c c\~oes 
radiativas tamb\'em tem implica\c c\~oes~\cite{ELL91 PL B257 p 91 ref 1},
devido ao insucesso das buscas dos Higgs no LEP~1, que\footnote{\'E usual 
deixar que $\tan \b$ varie no alcance $1\leq \tan \b \leq 50$.}
\BEA
   \tan \b & \geq & 1 \,\ ,
      \label{constraint tan-beta}
\EEA
no contexto do {\bf MSSM}.

\subsection{As Massas dos L\'eptons.}

Vamos analisar as massas dos l\'eptons. O termo 
$f\e^{ij}\,\bar{R}L^i H_1^j +h.c.$, da parte do superpotencial de 
Yukawa (e seu hermitiano conjugado) origina a massa dos l\'eptons.

A parte $f\e^{ij}\,\bar{R}L^i H_1^j +h.c.$, cont\'em os seguintes 
termos, ver Eq.(\ref{eeH}), ap\'os a quebra
\BEA
 \L_{llH}
    = -f v_1 \LP( l_{R}l_{L}+l^{*}_{L}l^{*}_{R}\RP) 
    \,\ . \nonumber
\EEA
Portanto, podemos fazer a seguinte identifica\c c\~ao
\BEA
   m_{f} &=& f v_1 \,\ , 
\label{massa do lepton}
\EEA
e como no {\bf SM}, notamos que as massas dos l\'eptons s\~ao 
indeterminadas pela teoria.

Uma \'ultima observa\c c\~ao, os acoplamentos de Yukawa podem ser 
escritos, usando Eq.(\ref{beta}), da seguinte maneira
\BEA
f= \frac{m_{f}}{v_{1}}=\frac{gm_{f}}{\sqrt{2}M_{W} \cos \beta} 
\,\ .
\label{constante de Yukawa}
\EEA

\subsection{As Massas dos Sl\'eptons.}

Nesta teoria existem dois sl\'eptons (representados por $\tilde{l}_{L}$ e 
$\tilde{l}_{R}$), que s\~ao os parceiros supersim\'etricos das partes de
helicidade left e right dos f\'ermions $l$. Antes da quebra de supersimetria, 
eles s\~ao degenerados em massa com $l$.

As contribui\c c\~oes para as massas dos sl\'eptons, vem dos termos $F$, 
dos termos $D$ e do termo soft.

Os termos que contribuem para a massa dos sl\'eptons da $\L_{soft}$, 
Eq.(\ref{burro}), 
s\~ao
\BEA
\L^{sleptons}_{soft}&=&-M^{2}_{L} \bar{ \tilde{L}}^{i} \tilde{L}^{i}+ 
M^{2}_{R} \bar{ \tilde{R}} \tilde{R}- M^{2}_{LR} \epsilon_{ij} 
(H^{i}_{1} \tilde{L}^{j} \tilde{R}+ \bar{H}_{1}^{i} \bar{ \tilde{L}}^{j} 
\bar{ \tilde{R}}) \nn
&=&-m^{2}_{ \tilde{ \nu}} \tilde{ \nu}^{*} \tilde{ \nu}- 
m^{2}_{ \tilde{l}} \tilde{l}^{*}_{L} \tilde{l}_{L}-
m^{2}_{R} \tilde{l}^{*}_{R} \tilde{l}_{R}-
A_{f}m_{f}( \tilde{l}_{R}^{*} \tilde{l}_{L}+ \tilde{l}_{L}^{*} 
\tilde{l}_{R}) \,\ , \nonumber \\
\EEA
usamos a Eq.(\ref{massa do lepton}) e fizemos a seguinte identifica\c c\~ao
\BEA
A_{f}= \frac{M^{2}_{LR}}{f} \,\ .
\EEA

A parte dos termos $F$ dada pela 
Eq.(\ref{aux prop 6aaaaaa}) que contribuem 
para a massa dos sl\'eptons s\~ao\footnote{Note que das 
Eqs.~\r{Radiative Breaking prop 12f} e \r{massa do lepton} que
$fv_{2} = fv_{1} \frac{v_{2}}{v_{1}}=m_{f} \tan \b$.}
\BEA
\L^{slepton}_{F}&=&- \mu f ( \bar{H}_{2} \tilde{L} \tilde{R}+ 
\bar{ \tilde{L}}H_{2} \bar{ \tilde{R}})-f^{2} \bar{H}_{1}H_{1} 
( \bar{ \tilde{L}} \tilde{L}+ \bar{ \tilde{R}} \tilde{R}) \nn
&=&- \mu m_{f} \tan \beta ( \tilde{l}^{*}_{R} \tilde{l}_{L}+ 
\tilde{l}^{*}_{L} \tilde{l}_{R})-m^{2}_{f}( \tilde{l}^{*}_{L} 
\tilde{l}_{L}+ \tilde{ \nu}^{*} \tilde{ \nu}+ \tilde{l}^{*}_{R} 
\tilde{l}_{R}) \,\ .
\EEA

No caso dos termos $D$ temos as seguintes contribui\c c\~oes
\BEA
\L^{sleptons}_{D}&=&- \f{g^{2}}{4}[( \bar{ \tilde{L}} \sigma^{i} \tilde{L})( 
\bar{H}_{1} \sigma^{i}H_{1})+( \bar{ \tilde{L}} \sigma^{i} \tilde{L})( 
\bar{H}_{2} \sigma^{i}H_{2})] \nn
&-&\f{g^{'}}{2}[Y_{L}Y_{H1}\bar{\tilde{L}}\tilde{L} 
\bar{H}_{1}H_{1}+ Y_{L}Y_{H2}\bar{\tilde{L}}\tilde{L} 
\bar{H}_{2}H_{2}+ Y_{R}Y_{H1}\bar{\tilde{R}}\tilde{R} 
\bar{H}_{1}H_{1}+ Y_{R}Y_{H2}\bar{\tilde{R}}\tilde{R}  
\bar{H}_{2}H_{2}] \nn
&=&- \f{g^{2}}{4}(v^{2}_{1}-v^{2}_{2}) \bar{ \tilde{L}} \sigma^{3} \tilde{L} 
-\f{g^{'}}{2}[Y_{L}(v^{2}_{2}-v^{2}_{1}) \bar{\tilde{L}}\tilde{L}+ 
Y_{R}(v^{2}_{2}-v^{2}_{1}) \bar{\tilde{R}}\tilde{R}] \,\ , 
\EEA
usando Eqs.(\ref{operador2}), (\ref{operador1}), (\ref{z-mass}) e 
(\ref{beta}) podemos escrever
\BEA
{\cal L}^{slepton}_{D}=-M^{2}_{Z} \cos(2 \beta) \left[
(T_{3f}- \sin^{2} \theta_{W} Q_{f}) 
( \tilde{ \nu}^{*} \tilde{ \nu}+ \tilde{l}^{*}_{L} \tilde{l}_{L})+Q_{f} 
\sin^{2} \theta_{W} \tilde{l}^{*}_{R} \tilde{l}_{R} \right] 
\,\ . \nonumber \\
\EEA

Juntando todas estas pe\c cas discutida acima teremos
\begin{eqnarray}
  \L_{slepton}^{massa} &=&
    -\mu m_{f} \tan \beta \;\tilde{l}_{L}^{*}\tilde{l}_{R}
    -\mu m_{f} \tan \beta \;\tilde{l}_{R}^{*}\tilde{l}_{L}
    -m^{2}_{f}\LP(\tilde{l}_{L}^{*}\tilde{l}_{L}
                      +\tilde{l}_{R}^{*}\tilde{l}_{R} \RP)
    -m^{2}_{ \nu} \tilde{ \nu}^{*} \tilde{ \nu}
  \nmb
    -m_{ \tilde{l}}^{2}\;\tilde{l}_{L}^{*}\tilde{l}_{L}
    -m_{R}^{2}\;\tilde{l}_{R}^{\dagger}\tilde{l}_{R} 
    -m_{ \tilde{ \nu}}^{2} \tilde{ \nu}^{*} \tilde{ \nu} 
    -A_{f}m_{f}(\tilde{l}_{R}^{*} \tilde{l}_{L}+ 
    \tilde{l}_{L}^{*} \tilde{l}_{R}) \nonumber \\
&-& M^{2}_{Z} \cos(2 \beta) \left[ 
(T_{3e}- \sin^{2} \theta_{W} Q_{e}) \tilde{l}_{L}^{*} \tilde{l}+ 
(T_{3 \nu}- \sin^{2} \theta_{W} Q_{ \nu}) \tilde{ \nu}_{L}^{*} \tilde{\nu}+ 
Q_{e} \sin^{2} \theta_{W} \tilde{l}^{*}_{R} \tilde{l}_{R} \right] \nonumber \\
&=&- 
\left[ m^{2}_{ \tilde{l}}-M^{2}_{Z} \cos(2 \beta) 
\left( \f{1}{2}- \sin^{2} \theta_{W} \right)+m^{2}_{f} \right] 
\tilde{l}^{*}_{L} \tilde{l}_{L} \nonumber \\
&-& m_{f}(A_{f}+ \mu \tan \beta) 
\left( \tilde{l}_{R}^{*}\tilde{l}_{L}+\tilde{l}_{L}^{*}\tilde{l}_{R} \right)- \left( 
m^{2}_{R}-M^{2}_{Z} \cos(2 \beta) \sin^{2} \theta_{W}+m^{2}_{f} \right) 
\tilde{l}^{*}_{R} \tilde{l}_{R} \nonumber \\
&-&
\left[ m^{2}_{ \tilde{ \nu}}+\f{M^{2}_{Z}}{2} \cos(2 \beta) \right] 
\tilde{ \nu}^{*} \tilde{ \nu} \,\ .
\end{eqnarray}
Vamos chamar
\BEA
\tilde{m}_{\tilde{f}_{L}}^{2}&=&m^{2}_{ \tilde{l}}-
M_{Z}^{2} \cos(2 \beta) \left( \f{1}{2}- \sin^{2} \t_{W} \right) \,\ , \nn
\tilde{m}_{\tilde{f}_{R}}^{2}&=&m^{2}_{R}-
M_{Z}^{2} \cos( 2 \beta) \sin^{2} \t_W \,\ , \nn
m^{2}_{\tilde{f}_{LR}}&=&m_{f}(\mu \tan\beta+A_{f}) \,\ , \nn 
\tilde{m}^{2}_{\tilde{ \nu}}&=&m^{2}_{ \tilde{ \nu}}-\f{M^{2}_{Z}}{2} \cos(2 \beta) 
\,\ ,
\EEA
isto leva, lembre que $m^{2}_{ \tilde{l}}=m^{2}_{ \tilde{ \nu}}$, a 
seguinte rela\c c\~ao importante
\BEA
\tilde{m}_{\tilde{f}_{L}}^{2}- \tilde{m}^{2}_{\tilde{ \nu}}=
M^{2}_{Z}\cos(2 \beta) \cos^{2} \theta_{W} \,\ ,
\EEA
que \'e v\'alida se a mistura entre os escalares pode ser desprezada; este 
\'e sempre o caso para o el\'etron e o m\'uon.

Voltando a lagrangiana das massas dos sl\'eptons temos agora
\BEA     
\L^{slepton}_{massa}
&=&
   -\LP(\BA{cc} \tilde{l}_{L}^{*} & \tilde{l}_{R}^{*} \EA \RP)
      \LP(\BA{cc}  
      \tilde{m}_{\tilde{f}_{L}}^{2} +m^{2}_{f} & m^{2}_{\tilde{f}_{LR}} \\
       m^{2}_{\tilde{f}_{LR}}       & \tilde{m}_{\tilde{f}_{R}}^{2} +m^{2}_{f}
             \EA \RP)
      \LP( \BA{c} \tilde{l}_{L} \\ \tilde{l}_{R} \EA \RP) + 
      \tilde{m}^{2}_{ \tilde{ \nu}} \tilde{ \nu}^{*} \tilde{ \nu} 
      \,\ .  \nonumber
\EEA

Diagonalizando a matriz de massa dos sl\'eptons carregados, obtemos que os auto-estados 
da massa s\~ao 
\BEA
\tilde{l}_{1} &=& \tilde{l}_{L}\cos{\t_{f}} +\tilde{l}_{R}\sin{\t_{f}} \,\ , \nn
\tilde{l}_{2} &=&  \tilde{l}_{L}\sin{\t_{f}} -\tilde{l}_{R}\cos{\t_{f}} \,\ ,
\nonumber 
\EEA
com o \^angulo de mistura $\t_{f}$ definido por
\BEA
\tan 2\t_{f}= 
\f{2m^{2}_{\tilde{f}_{LR}}}
{\tilde{m}_{\tilde{f}_{L}}-\tilde{m}_{\tilde{f}_{R}}^{2}} \,\ ,
\nonumber
\EEA
e as massas respectivamente s\~ao dadas por
\BEA
   M_{\tilde{l}_{1},\tilde{l}_{2}}^2
      &=& f^{2}v_{1}^{2}+\HA\LP[\LP(\tilde{m}_{\tilde{f}_{L}}^{2}+
\tilde{m}^{2}_{\tilde{f}_{R}}\RP)
       \pm \sqrt{\LP( \tilde{m}_{\tilde{f}_{L}}^{2}-
\tilde{m}^{2}_{\tilde{f}_{R}} \RP)^{2}
+4m^{4}_{\tilde{f}_{LR}}
}\;\RP]\nn
  &=&     m^2_f+
\HA\LP[\LP(\tilde{m}_{\tilde{f}_{L}}^{2}+
\tilde{m}^{2}_{\tilde{f}_{R}}\RP)
       \pm \sqrt{\LP( \tilde{m}_{\tilde{f}_{L}}^{2}-
\tilde{m}^{2}_{\tilde{f}_{R}} \RP)^{2}
+4m^{4}_{\tilde{f}_{LR}}
}\;\RP] \,\ .
\EEA

Iremos assumir m\'axima mistura, isto \'e $\theta_{f}= \pi/4$ ou 
\BEA
\tilde{m}_{\tilde{f}_{L}}^{2} = \tilde{m}_{\tilde{f}_{R}}^{2}= \tilde{m}^{2} \,\ .
\EEA
Uma motiva\c c\~ao para esta escolha vem da QED supersim\'etrica onde esta
escolha \'e feita para que paridade n\~ao seja violada. Portanto
 \BEA
  \tilde{l}_{1} &=& \f{\tilde{l}_{L} +\tilde{l}_{R}}{\sqrt{2}} \,\ ,\SL
  \tilde{l}_{2} &=& \f{\tilde{l}_{L} -\tilde{l}_{R}}{\sqrt{2}} \,\ ,
\EEA
e
\BEA
  M_{\tilde{l}_{1},\tilde{l}_{2}}^2
      &=& \tilde{m}^2+m^2_f \pm m_{\tilde{f}_{LR}}^{2} \,\ .
\EEA

\subsection{Gaugino e Higgsino}

O termo de mistura do gaugino e higgsino, prov\'em de termos da 
Eq.(\ref{RMFL prop 4}), que neste caso ser\'a
\BEA
\L^{mistura}_{\tilde{H} \tilde{V}}&=&
\sqrt{2}i\;\bar{H}_{1}
      \LP(gT^{i}\lambda^{i}_{A}-\HA g'\lambda_{B}\RP) \tilde{H}_{1}
      - \sqrt{2}i\;\bar{\tilde{H}}_{1}\LP(gT^{i}\bar{\lambda}^{i}_{A}
             -\HA g'\bar{\lambda}_{B}\RP) H_{1}\nn
&+&\sqrt{2}i\;\bar{H}_{2}
        \LP(gT^{i}\lambda^{i}_{A}+\HA g'\lambda_{B}\RP) \tilde{H}_{2}
      - \sqrt{2}i\;\bar{\tilde{H}}_{2}\LP(gT^{i}\bar{\lambda}^{i}_{A}
             +\HA g'\bar{\lambda}_{B}\RP) H_{2} \,\ , 
             \nonumber \\
\EEA
que podemos escrever, usando as defini\c c\~oes dos estados Eqs.(\ref{slvt11}) 
e os operadores das Eqs.(\ref{operador2}) 
e (\ref{operador1}), da seguinte maneira
\BEA
\L^{mistura}_{\tilde{H} \tilde{V}}&=&
ig\LP( \bar{H}_{1} T^{+} \tilde{H}_{1}\la^{+}
         - \bar{\la}^{+}\bar{\tilde{H}}_{1}T^{-}H_{1}\RP)+
ig\LP( \bar{H}_{2} T^{+} \tilde{H}_{2}\la^{+}
         - \bar{\la}^{+}\bar{\tilde{H}}_{2}T^{-}H_{2}\RP) 
         \nonumber \\
&+&ig\LP( \bar{H}_{1} T^{-} \tilde{H}_{1}\la^{-}
         - \bar{\la}^{-}\bar{\tilde{H}}_{1}T^{+}H_{1}\RP)+
ig\LP( \bar{H}_{2} T^{-} \tilde{H}_{2}\la^{-}
         - \bar{\la}^{-}\bar{\tilde{H}}_{2}T^{+}H_{2}\RP) 
         \nonumber \\
&+&i \sqrt{2}eQ_{i}( \bar{H}^{i}_{1} \tilde{H}^{i}_{1}\la_{\gamma}- 
\bar{\la}_{\gamma} \bar{\tilde{H}}^{i}_{1}H^{i}_{1}) +
i \sqrt{2}eQ_{i}( \bar{H}^{i}_{2} \tilde{H}^{i}_{2}\la_{\gamma}- 
\bar{\la}_{\gamma} \bar{\tilde{H}}^{i}_{2}H^{i}_{2}) 
\nonumber \\
&+&i \sqrt{2} \frac{g}{\cos \theta_{W}}
(T^{3}_{i}-Q_{i} \sin^{2} \theta_{W}) 
( \bar{H}^{i}_{1} \tilde{H}^{i}_{1}\la_{Z}- 
\bar{\la}_{Z} \bar{\tilde{H}}^{i}_{1}H^{i}_{1}) \nonumber \\
&+&
i \sqrt{2} \frac{g}{\cos \theta_{W}}
(T^{3}_{i}-Q_{i} \sin^{2} \theta_{W}) 
( \bar{H}^{i}_{2} \tilde{H}^{i}_{2}\la_{Z}- 
\bar{\la}_{Z} \bar{\tilde{H}}^{i}_{2}H^{i}_{2}) \,\ .
\EEA

Usando a Eq.(\ref{Component Field Expansion prop 7}) e 
(\ref{Component Field Expansion prop 8}) podemos 
escrever a equa\c c\~ao acima da seguinte maneira
\BEA
\L^{mistura}_{\tilde{H} \tilde{V}}&=&
ig\LP( \bar{H}^{0}_{1}\;\psi^{2}_{H_{1}}\la^{+}
         - \bar{\la}^{+}\bar{\psi}^{2}_{H_{1}}\;H^{0}_{1}\RP)+
ig\LP( \bar{H}^{+}_{2}\;\psi^{2}_{H_{2}}\la^{+}
         - \bar{\la}^{+}\bar{\psi}^{2}_{H_{2}}\;H^{+}_{2}\RP) 
\nonumber \\
&+&ig\LP( \bar{H}^{-}_{1}\;\psi^{1}_{H_{1}}\la^{-}
         - \bar{\la}^{-}\bar{\psi}^{1}_{H_{1}}\;H^{-}_{1}\RP)+
ig\LP( \bar{H}^{0}_{2}\;\psi^{1}_{H_{2}}\la^{-}
         - \bar{\la}^{-}\bar{\psi}^{1}_{H_{2}}\;H^{0}_{2}\RP) 
\nonumber \\
&-&\sqrt{2}i e
         \LP( \bar{H}^{-}_{1}\;\psi^{2}_{H_{1}}\la_{\gamma}
           - \bar{\la}_{\gamma}\bar{\psi}^{2}_{H_{1}}\;H^{-}_{1}\RP)+
\sqrt{2}i e
         \LP( \bar{H}^{+}_{2}\;\psi^{1}_{H_{2}}\la_{\gamma}
           - \bar{\la}_{\gamma}\bar{\psi}^{1}_{H_{2}}\;H^{+}_{2}\RP) 
\nonumber \\
&+&\f{ig}{\sqrt{2}\CWA}
         \LP( \bar{H}^{0}_{1}\;\psi^{1}_{H_{1}}\la_{Z}
                - \bar{\la}_{Z}\bar{\psi}^{1}_{H_{1}}\;H^{0}_{1}\RP)-
\f{ig}{\sqrt{2}\CWA}
         \LP( \bar{H}^{0}_{2}\;\psi^{2}_{H_{2}}\la_{Z}
                - \bar{\la}_{Z}\bar{\psi}^{2}_{H_{2}}\;H^{0}_{2}\RP) 
\nonumber \\
&+&\f{ig}{\sqrt{2}\CWA} \LP(1-2\SWAS\RP)
         \LP( \bar{H}^{+}_{2}\;\psi^{1}_{H_{2}}\la_{Z}
               - \bar{\la}_{Z}\bar{\psi}^{1}_{H_{2}}\;H^{+}_{2}\RP) 
               \nonumber \\
               &-&
\f{ig}{\sqrt{2}\CWA} \LP(1-2\SWAS\RP)
         \LP( \bar{H}^{-}_{1}\;\psi^{2}_{H_{1}}\la_{Z}
               - \bar{\la}_{Z}\bar{\psi}^{2}_{H_{1}}\;H^{-}_{1}\RP) \,\ .
\EEA

Finalmente obtemos usando os espinores de quatro componentes
\BEA
\L^{mistura}_{\tilde{H} \tilde{V}}&=&
+g \LP(\bar{\tilde{W}}L\tilde{H}\;H^{0}_{1}
       + \bar{\tilde{H}}^{n}_{1}L\tilde{W}\;H^{-}_{1}\RP)
      - \sqrt{2} e \;\bar{\tilde{A}}L\tilde{H}\;H^{-}_{1}
   \nmb
      +\f{g}{\sqrt{2}\CWA}\;\bar{\tilde{Z}}L\tilde{H}^{n}_{1}\;H^{0}_{1}
      -\f{g}{\sqrt{2}\CWA}\LP(1-2\SWAS\RP)
                \bar{\tilde{Z}}L\tilde{H}\;H^{-}_{1}
   \nmb
+g \LP(\bar{\tilde{W}}L\tilde{H}^{n}_{2}\;H^{+}_{2}
          + \bar{\tilde{H}}L\tilde{W}\;H^{0}_{2}\RP)
      + \sqrt{2} e \;\bar{\tilde{H}}L\tilde{A}\;H^{+}_{2}
   \nmb
      +\f{g}{\sqrt{2}\CWA}\LP(1-2\SWAS\RP)
            \bar{\tilde{H}}L\tilde{Z}\;H^{+}_{2}
      -\f{g}{\sqrt{2}\CWA} \bar{\tilde{H}}^{n}_{2}L\tilde{Z}\;H^{0}_{2}
   \nmb
      + h.c. \,\ .
\EEA

\subsubsection{Termo de Massa do Chargino.}

Os charginos $\tilde{\chi}^+_i$ ($i=1,2$), surgem devido a mistura 
dos Winos, $\tilde{W}^{\pm}$, e Higgsinos carregados, $\tilde{H}^{\pm}$. Os 
charginos  s\~ao espinores de Dirac de quatro componentes. Em princ\'\i pio 
existem 
duas misturas independentes, i.e. ($\tilde{W}^-,\tilde{H}^-$) e
($\tilde{W}^+,\tilde{H}^+$), ent\~ao necessitamos de duas matrizes 
unit\'arias para diagonalizar a matriz de massa.

O termo de massa do wino, ver 
Eq.(\ref{Component Field Expansion of L sub Soft prop 9}), 
\'e o seguinte
\BEA
\L^{carregado}_{\tilde{V}}&=&-M_{\tilde{W}}\,\bar{\tilde{W}}\tilde{W} 
\nonumber \\
&=&- M \LP(\la^{-}\la^{+}+\bar{\la}^{-}\bar{\la}^{+}\RP) \,\ ,
\EEA
j\'a no caso do higgsino carregado, Eq.(\ref{higgsino massa}), 
\'e
\BEA
\L^{carregado}_{\tilde{H}}&=&\mu \,\bar{\tilde{H}}\tilde{H} 
\nonumber \\
&=&- \mu \LP(\psi^{1}_{H_{1}} \psi_{H_{1}}^{2}
+\bar{\psi^{1}_{H_{1}}}\bar{\psi_{H_{1}}^{2}}\RP) \,\ .
\EEA
J\'a os termos de mistura que contribuem para a massa do chargino, ver 
p\'agina anterior, s\~ao
\BEA
g \LP(\bar{\tilde{W}}L\tilde{H}\;H^{0}_{1}
       + \bar{H}^{0}_{1}\bar{\tilde{H}}R\tilde{W}\RP)+
g \LP(\bar{\tilde{H}}L\tilde{W}H^{0}_{2}+ 
\bar{H}^{0}_{2}\bar{\tilde{W}}R\tilde{H}\RP)    
&=&
ig\LP( \bar{H}^{0}_{1}\;\psi^{2}_{H_{1}}\la^{+}
- \bar{\la}^{+}\bar{\psi}^{2}_{H_{1}}\;H^{0}_{1}\RP) 
\nonumber \\
&+& 
ig\LP( \bar{H}^{0}_{0}\;\psi^{1}_{H_{2}}\la^{-}
- \bar{\la}^{-}\bar{\psi}^{1}_{H_{2}}\;H^{0}_{2}\RP) 
\,\ , \nonumber \\
\EEA
Juntando todas estas pe\c cas obtemos
\BEA
\L_{\tilde{\chi}^{\pm}}^{mass}
    &=&
ig\LP( \bar{H}^{0}_{1}\;\psi^{2}_{H_{1}}\la^{+}
- \bar{\la}^{+}\bar{\psi}^{2}_{H_{1}}\;H^{0}_{1}\RP)
+ 
ig\LP( \bar{H}^{0}_{0}\;\psi^{1}_{H_{2}}\la^{-}
- \bar{\la}^{-}\bar{\psi}^{1}_{H_{2}}\;H^{0}_{2}\RP) \nonumber \\
&-& \mu \LP(\psi^{1}_{H_{1}} \psi_{H_{1}}^{2}
+\bar{\psi^{1}_{H_{1}}}\bar{\psi_{H_{1}}^{2}}\RP)- 
M \LP(\la^{-}\la^{+}+\bar{\la}^{-}\bar{\la}^{+}\RP) \,\ ,
\EEA
tomandos o (v.e.v.) dos Higgs teremos
\BEA
\L_{\tilde{\chi}^{\pm}}^{massa}
   &=& ig \LP[
                  v_1 \psi^2_{H_1}\la^+
                  + v_2\la^-\psi^1_{H_2}
           \RP]
          + \mu\; \psi^2_{H_1} \psi^1_{H_2}
        - M\,
            \la^{-}\la^{+}
             + h.c. 
             \,\ . \nonumber \\
             \label{chargino prop 1}
\EEA
Introduzindo a seguinte nota\c c\~ao
\BEA
   \psi^{+} \;\;=\;\; \LP( \BA{r} - i\la^{+} \\
     \psi^{1}_{H_{2}} \EA \RP) \,\ , \hspace{0.7cm}
   \psi^{-} \;\;=\;\; \LP( \BA{r} - i\la^{-} \\ \psi^{2}_{H_{1}} \EA \RP) \,\ ,
   \label{chargino prop 3}   \nonumber
\EEA
e
\BEA
     \Psi^{\pm} &=& \LP( \BA{c}   \psi^{+}  \\ \psi^{-}  \EA  \RP) \,\ ,  \nonumber
\EEA
Eq.(\ref{chargino prop 1}) adquire a seguinte forma
\BEA
  \L_{\tilde{\chi}^{\pm}}^{massa}
      &=&  \HA \LP( \Psi^{\pm} \RP)^{T} Y^{\pm} \Psi^{\pm} + h.c. \,\ .  \nonumber
\EEA
Onde
\BEA
      Y^{\pm} &=& \LP(  \BA{cc} 0  & X^{T} \\
                                X  & 0      \EA  \RP) \,\ ,
                                \label{chargino prop 6}
\EEA
com
\BEA
   X &=&   \LP(  \BA{cc}  M  & -\sqrt{2} M_{\mbox{w}}\sin{\b}   \\
                          -\sqrt{2} M_{\mbox{w}}\cos{\b}  &    \mu   \EA  \RP) \,\ .
                          \label{chargino prop 7}
\EEA
Agora, os auto-estados da massa pode ser definida por($i, j = 1,2$)
\BEA
     \chi^{+}_{i}  &=&  V_{ij} \psi^{+}_{j} \,\ ,
         \label{chargino prop 8} \SL
     \chi^{-}_{i}  &=&  U_{ij} \psi^{-}_{j} \,\ , \,\ i=1,2
         \label{chargino prop 9}
\EEA
onde U e V s\~ao matrizes unit\'arias, escolhida de maneira tal que
\BEA
   U X V^{-1}  &=& {\cal M}_{C} \,\ .
       \label{chargino prop 10}
\EEA
Onde $M^{\pm}_{D}$ \'e a matriz de massa do chargino. Desenvolvendo a 
equa\c c\~ao de autovalores, que \'e $\det[Y^{\pm}- \lambda I]=0$, 
acharemos
\BEA
(M_{W} \sin2 \theta_{W}+M \mu)^{2}-(2M^{2}_{W}+M^{2}+ \mu^{2}) 
\lambda^{2}+ \lambda^{4}=0 \,\ ,
\EEA
resolvendo a equa\c c\~ao acima para $\lambda^{2}$, encontraremos os 
seguintes autovalores
\BEA
M^{2}_{\tilde{\chi}}= \frac{1}{2}[(M^{2}+ \mu^{2}+2M_{W}^{2}) \pm 
\sqrt{(M^{2}+ \mu^{2}+2M^{2}_{W})^{2}-4(M \mu+M_{W} \sin2 \beta)^{2}}] 
\,\ . \nonumber \\
\EEA

Devo mencionar que at\'e aqui, $U$ e $V$ n\~ao s\~ao \'unicas. Isto reflete o fato 
que certas fases arbitr\'arias podem ser absorvidas na defini\c c\~ao dos campos 
f\'\i sicos. Os elementos $U_{ij}$ e $V_{ij}$ das matrizes de 
diagonaliza\c c\~ao podem ser expressas em termos dos par\^ametros $M$, $\mu$, e 
$\tan \beta$:
\begin{eqnarray}
  & & U_{12} = U_{21} = \frac{\theta_1}{\sqrt{2}}\,
    \sqrt{1 + \frac{M^2 - \mu^2 - 2\,m_W\cos 2\beta}{W}} \\[2mm]
  & & U_{22} = -U_{11} = \frac{\theta_2}{\sqrt{2}}\,
    \sqrt{1 - \frac{M^2 - \mu^2 - 2\,m_W\cos 2\beta}{W}} \\[2mm] 
  & & V_{21} = -V_{12} = \frac{\theta_3}{\sqrt{2}}\,
    \sqrt{1 + \frac{M^2 - \mu^2 + 2\,m_W\cos 2\beta}{W}} \\[2mm] 
  & & V_{22} = V_{11} = \frac{\theta_4}{\sqrt{2}}\,
    \sqrt{1 - \frac{M^2 - \mu^2 + 2\,m_W\cos 2\beta}{W}} 
\end{eqnarray}
com
\begin{equation}
  W = \sqrt{(M^2+\mu^2+2\,m_W^2)^2 - 4\,(M\!\cdot\!\mu -m_W^2\sin 2\beta)^2}
\end{equation}
e os fatores de sinais $\theta_i$, $i = 1\ldots4$, s\~ao 
\begin{equation}
  \{\theta_1,\,\theta_2,\theta_3,\theta_4\} = \; \left\{ \;
  \begin{array}{ll}
    \{ 1,\,\varepsilon_{\!_B},\,\varepsilon_{\!_A},\,1 \} & 
          \ldots\;\; \tan \beta > 1 \\[2mm]
    \{ \varepsilon_{\!_B},\,1,\,1,\,\varepsilon_{\!_A} \} & 
          \ldots\;\; \tan \beta < 1 \\    
  \end{array}\right.
\end{equation}
onde
\begin{equation}
  \varepsilon_{\!_A} = {\rm sign}(M\sin \beta + \mu\,\cos \beta),
  \hspace{6mm}
  \varepsilon_{\!_B} = {\rm sign}(M\cos \beta + \mu\,\sin \beta).
\end{equation}

Vamos escolher $U$ e $V$ tal que $M^{\pm}_{D}$ tenha apenas 
n\'umeros positivos e por conven\c c\~ao, escolheremos $\tilde{\chi}_{1}$ mais 
pesado que $\tilde{\chi}_{2}$, isto \'e 
$M^{2}_{\tilde{\chi}_{1}}>M^{2}_{\tilde{\chi}_{2}}$. Por simplicidade estamos 
assumindo que $M$ e $\mu$ s\~ao reais. Resolvendo o problema dos autovalores 
para $X^{\rm T} X$,
\begin{equation}
  {\cal M}_C^2 = {\rm diag}(m_{\chi{1}}^2,\,m_{\chi{2}}^2) = V\,X^{\rm T} X\,V^{-1}
\end{equation}
with
\begin{equation}
  V = \left( \begin{array}{rr}
    \cos\phi_{_1} & \sin\phi_{_1} \\ -\sin\phi_{_1} & \cos\phi_{_1}
    \end{array} \right)
\end{equation}
e a matriz $U$ \'e escrita da seguinte maneira
\begin{equation}
  U = \frac{1}{{\cal M}_C^{}}\;V\,X^{\rm T} = \left( \begin{array}{rr}
    \cos\phi_{_2} & \sin\phi_{_2} \\ -\sin\phi_{_2} & \cos\phi_{_2}
    \end{array} \right).
\end{equation}

Assim as massas do chargino s\~ao reais e positivas, \'e poss\'\i vel mostrar que 
as massas s\~ao
\BEA
     M_{\tilde{\chi}_{1}}^{2} &=& A+\sqrt{B} \,\ , \nn
     M_{\tilde{\chi}_{2}}^{2} &=& A-\sqrt{B} \,\ , \nonumber
\EEA
com
\BEA
   A &=& \HA \LP( M^{2}+\mu^{2}  \RP) +  M^{2}_{\mbox{w}} \,\ ,\nn
   B &=& \f{1}{4} \LP(M^{2}-\mu^{2} \RP)^{2}
          + M^{4}_{\mbox{w}}\cos^{2}\LP( 2\b \RP)
          + M^{2}_{\mbox{w}}\LP( M^{2}+\mu^{2}+2\mu M\sin\LP( 2\b \RP)\RP) \,\ .
     \nonumber
\EEA
Al\'em disso os espinores de duas componentes das Eqs.(\ref{chargino prop 8}) e 
(\ref{chargino prop 9}) pode ser expressas em termos de espinores de Dirac de
quatro componentes da seguinte maneira:
\BEA
     \tilde{\chi}_{i} &=& \LP(  \BA{c}  \chi^{+}_{i} \\
                                            \bar{\chi}^{-}_{i}  \EA  \RP)
                      \,\ ,\hspace{1cm}   i = 1,2 \,\ ,
            \label{chargino prop 11}
\EEA
e o conjugado de carga \'e
\BEA
\tilde{\chi}^{c}_{i} &=& \LP(  \BA{c}  \chi^{-}_{i} \\
                                            \bar{\chi}^{+}_{i}  \EA  \RP)
                      \,\ ,\hspace{1cm}   i = 1,2 \,\ ,
            \label{charginocc prop 11}
\EEA

\subsubsection{Mistura do Neutralino.}

Neutralinos~$\tilde{\chi}^{0}_{i}, (i=1,\ldots, 4$), surgem devido a mistura 
dos gauginos neutros $\lambda^{3}_{A}$ e $\lambda_{B}$ e dos higgsinos 
neutros $\tilde{H}^{0}_{1}$ e $\tilde{H}^{0}_{2}$. Os neutralinos s\~ao 
descritos por espinores de Majorana; contudo, se 2 neutralinos s\~ao 
degenerados em massa, eles podem se combinar em um espinor de Dirac.

O termo de massa do gaugino neutro, ver 
Eq.(\ref{The Soft SUSY-Breaking Term prop 3}), 
\'e
\BEA
\L^{neutro}_{\tilde{V}}=- \frac{M}{2}(\lambda^{3}_{A} \lambda^{3}_{A}+ 
\bar{\lambda}^{3}_{A} \bar{\lambda}^{3}_{A})- \frac{M^{\prime}}{2}
(\lambda_{B} \lambda_{B}+ \bar{\lambda}_{B} \bar{\lambda}_{B}) \,\ ,
\EEA
o termo de massa do higgsino neutro, ver 
Eq.(\ref{higgsino massa}), 
tem a seguinte forma
\BEA
\L^{neutro}_{\tilde{H}}&=&- \frac{\mu}{2} \bar{\tilde{H}}^{n}_{1} 
\tilde{H}^{n}_{2}- \frac{\mu}{2} \bar{\tilde{H}}^{n}_{2} 
\tilde{H}^{n}_{1} \nonumber \\
&=& \mu (\psi^{1}_{H1} \psi^{2}_{H2}+ \bar{\psi}^{1}_{H1} 
\bar{\psi}^{2}_{H2}) \,\ .
\EEA
J\'a os termos de mistura que contribuem para a massa do neutralino 
s\~ao
\BEA
\L^{mistura}_{massa}&=& \frac{g}{\sqrt{2} \cos \theta_{W}} ( \bar{\tilde{Z}}L 
\tilde{H}^{n}_{1}H^{0}_{1}+ \bar{H}^{0}_{1} \bar{\tilde{H}}^{n}_{1}R 
\tilde{Z})- \frac{g}{\sqrt{2} \cos \theta_{W}}( \tilde{H}^{n}_{2}L 
\tilde{Z}H^{0}_{2}+ \bar{H}^{0}_{2} \bar{\tilde{Z}}R \tilde{H}^{n}_{2} 
\nonumber \\
&=& \frac{ig}{\sqrt{2} \cos \theta_{W}}( \bar{H}^{0}_{1} \psi^{1}_{H1} 
\lambda_{Z}- \bar{\lambda}_{Z} \bar{\psi}^{1}_{H1}H^{0}_{1})- 
\frac{ig}{\sqrt{2} \cos \theta_{W}}( \bar{H}^{0}_{2} \lambda_{Z} 
\psi^{2}_{H2}- \bar{\psi}^{2}_{H2} \bar{\lambda}_{Z}H^{0}_{2}) 
\,\ , \nonumber \\
\EEA
usando as Eqs.(\ref{svtl11}) podemos escrever
\BEA
\L^{mistura}_{massa}&=& \frac{ig}{\sqrt{2}}( \bar{H}^{0}_{1} \psi^{1}_{H1} 
\lambda^{3}_{A}- \bar{\lambda}^{3}_{A} \bar{\psi}^{2}_{H2}H^{0}_{2})- 
\frac{ig \sin \theta_{W}}{\sqrt{2} \cos \theta_{W}}(\bar{H}^{0}_{1} 
\psi^{1}_{H1} \lambda_{B}- \bar{\lambda}_{B} \bar{\psi}^{1}_{H1} 
H^{0}_{1}) \nonumber \\
&-& \frac{ig}{\sqrt{2}}( \bar{H}^{0}_{2} \lambda^{3}_{A} 
\psi^{2}_{H2}- \bar{\psi}^{2}_{H2} \bar{\lambda}^{3}_{A}H^{0}_{2})- 
\frac{ig \sin \theta_{W}}{\sqrt{2} \cos \theta_{W}}( \bar{H}^{0}_{2} 
\lambda_{B} \psi^{2}_{H2}- \bar{\psi}^{2}_{H2} \bar{\lambda}_{B}
H^{0}_{2}) \, . \nonumber \\
\EEA

Juntando todas estas pe\c cas teremos
\BEA
\L_{\tilde{\chi}^{0}}^{massa}&=& \frac{ig}{\sqrt{2}}(\bar{H}^{0}_{1} 
\psi^{1}_{H1} \lambda^{3}_{A}- \bar{H}^{0}_{2} \psi^{2}_{H2} 
\lambda^{3}_{A}- \bar{\lambda}^{3}_{A} \bar{\psi}^{1}_{H1}H^{0}_{1}+ 
\bar{\lambda}^{3}_{A} \bar{\psi}^{2}_{H2}H^{0}_{2}) \nonumber \\
&+& \frac{ig \sin \theta_{W}}{\sqrt{2} \cos \theta_{W}}(
\bar{H}^{0}_{1} \psi^{1}_{H1} \lambda_{B}- \bar{H}^{0}_{2} 
\psi^{2}_{H2}- \bar{\lambda}_{B} \bar{\psi}^{1}_{H1}H^{0}_{1}+ 
\bar{\lambda}_{B} \bar{\psi}^{2}_{H2}H^{0}_{2}) \nonumber \\
&+& \mu(\psi^{1}_{H1} \psi^{2}_{H2}+ \bar{\psi}^{1}_{H1} 
\bar{\psi}^{2}_{H2})- \frac{M}{2}(\lambda^{3}_{A} \lambda^{3}_{A}+ 
\bar{\lambda}^{3}_{A} \bar{\lambda}^{3}_{A})- \frac{M^{\prime}}{2} 
(\lambda_{B} \lambda_{B}+ \bar{\lambda}_{B} \bar{\lambda}_{B}) 
\,\ , \nonumber \\
\EEA
ap\'os a quebra da simetria de gauge 
\BEA
 \L_{\tilde{\chi}^{0}}^{massa}
   &=&
         \f{ig}{\sqrt{2}}\LP\{
               v_{1}\,\la^{3}_{A}\psi^{1}_{H1}
             - v_{2}\,\la^{3}_{A}\psi^{2}_{H2} \RP\}+
\f{ig \sin \theta_{W}}{\sqrt{2} \cos \theta_{W}}\LP\{
v_{1}\,\la_{B}\psi^{1}_{H1}-v_{2}\,\la_{B}\psi^{2}_{H2} \RP\}
             +\mu\,\psi^{1}_{H1}\psi^{2}_{H_2}
   \nmb
         - \HA M\,\la^{3}_{A}\la^{3}_{A}
         - \HA M^{\prime}\,\la_{B}\la_{B}+h.c. \,\ .\hspace{1cm}
               \label{Neutralino Mixing prop 2}
\EEA
Na base
\BEA
    \psi^{0} &=& \LP(  \BA{cccc}
         i\la^{3}_{A} & i\la_{B} &
     \psi^{1}_{H_{1}} & \psi^{2}_{H_{2}}  \EA  \RP)^{T} \,\ ,
            \label{Neutralino Mixing prop 3}
\EEA
Eq.(\ref{Neutralino Mixing prop 2}) pode ser escrita da seguinte maneira
\BEA
    \L_{\tilde{\chi}^{0}}^{mass}
       &=& \HA \, \LP(\psi^{0}\RP)^{T}Y^{0}\psi^{0} + h.c. \,\ ,
       \label{Neutralino Mixing prop 4}
\EEA
onde $Y^{0}$ \'e dada por
\BEA
   Y^{0} &=&
     \LP( \BA{cccc}
         M         & 0 & M_{Z} \sin \beta \cos \theta_{W} & 
-M_{Z} \cos \beta \cos \theta_{W} \\
0 & M^{\prime}& M_{Z} \sin \beta \sin \theta_{W} & 
M_{Z} \cos \beta \sin \theta_{W} \\
         M_{Z} \sin \beta \cos \theta_{W} & 
M_{Z} \sin \beta \cos \theta_{W} & 0 & \mu \\
         -M_{Z} \cos \beta \cos \theta_{W} & 
-M_{Z} \cos \beta \sin \theta_{W} & \mu & 0
\EA \RP) \,\ .\nn            \label{Neutralino Mixing prop 5}
\EEA
Repare que $Y^{0}$ \'e sim\'etrica, a mesma coisa ocorre com a natureza dos
neutralinos. Como consequ\^encia, apenas uma matriz unit\'aria $N$ \'e
necess\'aria para diagonalizar $Y^{0}$:
\BEA
     N^{*}Y^{0}N^{\dagger} &=&  M^{0}_{D} \,\ .
         \label{Neutralino Mixing prop 6}
\EEA
Onde $M^{0}_{D}$ \'e a matriz de massa diagonal do neutralino, e os 
auto-valores s\~ao arranjados tal que 
$|m_{\tilde{\chi}_{1}}|<|m_{\tilde{\chi}_{2}}|<|m_{\tilde{\chi}_{3}}|
<|m_{\tilde{\chi}_{4}}|$.

Como na se\c c\~ao anterior definiremos dois auto-estados de duas componentes
da seguinte maneira
\BEA
    \chi^{0}_{i} &=& N_{ij} \psi^{0}_{j} \,\ , \hspace{1cm} i,j =1,\ldots, 4 \,\ ,
       \label{Neutralino Mixing prop 7}
\EEA
A matriz $N_{ij}$, que \'e da seguinte forma
\BEA
N_{ij}= \LP(  \BA{cccc}  
\eta_{1}& 0& 0& 0 \\
0& \eta_{2}& 0& 0 \\
0& 0& \eta_{3}& 0 \\
0& 0& 0& \eta_{4}  \EA  \RP) \,\ .
\EEA
Esta matriz \'e introduzida para garantir a mudan\c ca de fase das 
part\'\i culas cujos valores se tornam negativos pela diagonaliza\c c\~ao 
Eq.(\ref{Neutralino Mixing prop 6}), e seus valores s\~ao
\BEA
\eta_{i}= \LP\{ \BA{lcr}
1 \,\ m_{\tilde{\eta}_{i}}>0 \,\ , \\
i \,\  m_{\tilde{\eta}_{i}}<0 \,\ ,
\EA \RP.
\EEA
e $m_{\tilde{\chi}_{i}}= \eta^{2}_{i}m_{\tilde{\eta}_{i}}$.

Neste caso podemos arrumar eles em um espinor de Majorana de quatro
componentes definidos por
\BEA
   \tilde{\chi}^{0}_{i}  &=& \LP(  \BA{c}  \chi^{0}_{i} \\
                                           \bar{\chi}^{0}_{i}  \EA  \RP) \,\ ,
                            \hspace{1cm} i= 1,\ldots, 4 \,\ .
                            \label{Neutralino Mixing prop 8}
\EEA

\section{Regras de Feynman}

Uma vez j\'a obtidas os auto-estados f\'\i sicos e suas respectivas massas 
iremos agora derivar as regras de Feynman do {\bf MSSM}.

\subsection{Regras Higgs-F\'ermion}

Usando Eqs(\ref{eeHint}), (\ref{estado H1}) e (\ref{estado H2}) obteremos as 
seguintes regras

\noindent{\bf $H^{0}_{1}(H^{0}_{2},H^{0}_{3})$-F\'ermion-F\'ermion}\par

\BEA
\L_{eeH^{0}_{i}}&=&f \bar{\Psi}(e)L \Psi(e)H^{0}_{1}+f \bar{\Psi}(e)R \Psi(e)
\bar{H}^{0}_{1} = \frac{g m_{e} \cos \alpha}{\sqrt{2}M_{W} \cos \beta} 
\left( \bar{\Psi}(e)L \Psi(e)+\bar{\Psi}(e)R \Psi(e) 
\right) H^{0}_{1} \nonumber \\
&-& \frac{g m_{e} \sin \alpha}{\sqrt{2}M_{W} \cos \beta} 
\left( \bar{\Psi}(e)L \Psi(e)+\bar{\Psi}(e)R \Psi(e) 
\right)H^{0}_{2} \nonumber \\
&+& 
\frac{g m_{e} \sin \beta}{\sqrt{2}M_{W} \cos \beta} 
\left( \bar{\Psi}(e)L \Psi(e)-\bar{\Psi}(e)R \Psi(e) 
\right)H^{0}_{3} \nonumber \\
&=& \frac{g m_{e} \cos \alpha}{\sqrt{2}M_{W} \cos \beta} 
\bar{\Psi}(e) \Psi(e)H^{0}_{1}- 
\frac{g m_{e} \sin \alpha}{\sqrt{2}M_{W} \cos \beta} 
\bar{\Psi}(e) \Psi(e)H^{0}_{2} \nonumber \\
&-&
\frac{igm_{e}}{\sqrt{2}M_{W}} \tan \beta \bar{\Psi}(e) \gamma_{5} \Psi(e)
H^{0}_{3} \,\ .
\EEA

\noindent{\bf $H^{\pm}$-F\'ermion-F\'ermion}\par

\BEA
\L_{eeHC}&=&f \bar{\Psi}(e)L \Psi(\nu)H^{-}_{1}+f \bar{\Psi}(\nu)R \Psi(e)
\bar{H}^{-}_{1} \nonumber \\
&=& \frac{g m_{e} \sin \beta}{\sqrt{2}M_{W} \cos \beta} 
(\bar{\Psi}(\nu)R \Psi(e)H^{+}+\bar{\Psi}(e)L \Psi(e))H^{-}) \nonumber \\
&=&\frac{g}{\sqrt{2}M_{W}} m_{e} \tan \beta ( 
\bar{\Psi}(\nu)R \Psi(e)H^{+}+h.c.) \,\ .
\EEA

\noindent{\bf Goldstone-F\'ermion-F\'ermion}
\BEA
\L_{eeG}&=&f \bar{\Psi}(e)L \Psi(e)H^{0}_{1}+f \bar{\Psi}(e)R \Psi(e)
\bar{H}^{0}_{1} \nonumber \\
&+&f \bar{\Psi}(e)L \Psi(\nu)H^{-}_{1}+f \bar{\Psi}(\nu)R \Psi(e)
\bar{H}^{-}_{1} \nonumber \\
&=& \frac{igm_{e}}{\sqrt{2}M_{W}} (\bar{\Psi}(e)L \Psi(e)-
\bar{\Psi}(e)R \Psi(e))G^{0} \nonumber \\
&+& \frac{gm_{e}}{\sqrt{2}M_{W}} 
(\bar{\Psi}(\nu)R \Psi(e)G^{+}+\bar{\Psi}(e)L \Psi(e))G^{-}) \nonumber \\
&=& \frac{-igm_{e}}{\sqrt{2}M_{W}} (\bar{\Psi}(e) \Psi(e)G^{0}+ 
\frac{gm_{e}}{\sqrt{2}M_{W}} 
(\bar{\Psi}(\nu)R \Psi(e)G^{+}+h.c.) \,\ . \nonumber \\
\EEA

\acknowledgments 
Este estudo foi financiado pela Funda\c{c}\~ao de Amparo \`a Pesquisa
do Estado de S\~ao Paulo (FAPESP). Agrade\c co tamb\'em ao Professor N. 
Berkovits por ter me ensinado o formalismo de supercampos e aos Professores 
V. Pleitez e J. C. Montero por todo o incentivo a realizar este estudo bem como 
por todos os preciosos ensinamentos, e aos Professores M. Capdequi-Peyran\`ere,
M. Manna e G. Moultaka e a Profa. Maria C. Tijero. Agradeco tamb\'em a todos 
os meus colegas, principalmente a Carlos Tello Echevarria, Ricardo Martin 
Bentin Zacarias, Jos\'e Nemecio Acosta Jara, Juan Segundo Valverde Salvador, 
Te\'ofilo Vargas Auccalla, O. P. Ravinez e C. A. de S. Pires  pelo 
companherismo.

\newcommand{\PRD}{Phys. Rev. }
\newcommand{\PRL}{Phys. Rev. Lett. }
\newcommand{\NP}{Nucl. Phys. }
\newcommand{\PL}{Phys. Lett. }
\newcommand{\NC}{Riv. Nuovo Cimento }

\newcommand{\BI}[1]{\bibitem{#1}}

\newpage

\begin{center}
\begin{picture}(300,100)(0,0)
\DashCArc(150,50)(40,0,180){4}
\ArrowArc(150,50)(40,180,360)
\ArrowLine(50,50)(110,50)
\put(70,20){Y}
\Vertex(110,50){2}
\ArrowLine(190,50)(250,50)
\put(220,20){Y}
\Vertex(190,50){2}
\end{picture}
\label{Fig.1}
\end{center}

\begin{center}
\begin{picture}(300,100)(0,0)
\DashCArc(150,50)(40,0,180){4}
\DashCArc(150,50)(40,180,360){4}
\DashLine(50,10)(250,10){4}
\put(150,0){$ \lambda$}
\Vertex(150,10){2}
\end{picture}
\label{Fig.2a}
\end{center}

\begin{center}
\begin{picture}(300,100)(0,0)
\ArrowArc(150,50)(40,0,180)
\ArrowArc(150,50)(40,180,360)
\DashLine(50,50)(110,50){4}
\put(70,20){Y}
\Vertex(110,50){2}
\ArrowLine(190,50)(250,50)
\put(220,20){Y}
\Vertex(190,50){2}
\end{picture}
\label{Fig.2b}
\end{center}

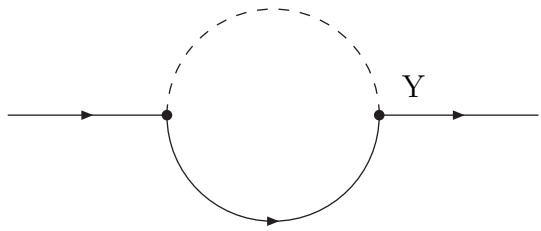
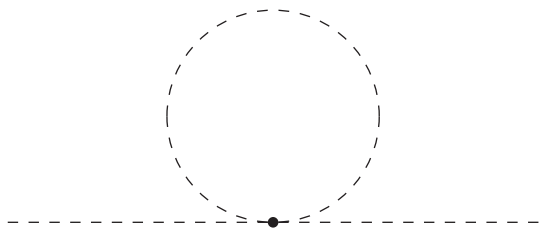
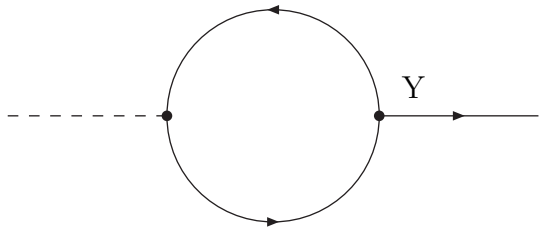
\end{document}